\documentclass[12pt,preprint]{emulateapj}

\begin{document}
\title{UV Luminosity Functions at redshifts $z\sim4$ to $z\sim10$: 10000 Galaxies from HST Legacy Fields\altaffilmark{1,2}}
\author{R. J. Bouwens\altaffilmark{3,4}, G. D. Illingworth\altaffilmark{4},
  P. A. Oesch\altaffilmark{4,5}, 
  M. Trenti\altaffilmark{6}, I. Labb{\'e}\altaffilmark{3},
  L. Bradley\altaffilmark{7}, M. Carollo\altaffilmark{8}, P.G. van
  Dokkum\altaffilmark{5}, V. Gonzalez\altaffilmark{4,9},
  B. Holwerda\altaffilmark{3}, M. Franx\altaffilmark{3},
  L. Spitler\altaffilmark{10,11}, R. Smit\altaffilmark{3},
  D. Magee\altaffilmark{4}}

\altaffiltext{1}{Based on observations made with the NASA/ESA Hubble
  Space Telescope, which is operated by the Association of
  Universities for Research in Astronomy, Inc., under NASA contract
  NAS 5-26555.}  
\altaffiltext{2}{Based on observations obtained with
  MegaPrime/MegaCam, a joint project of CFHT and CEA/IRFU, at the
  Canada-France-Hawaii Telescope (CFHT) which is operated by the
  National Research Council (NRC) of Canada, the Institut National des
  Science de l'Univers of the Centre National de la Recherche
  Scientifique (CNRS) of France, and the University of Hawaii. This
  work is based in part on data products produced at Terapix available
  at the Canadian Astronomy Data Centre as part of the
  Canada-France-Hawaii Telescope Legacy Survey, a collaborative
  project of NRC and CNRS.}
\altaffiltext{3}{Leiden Observatory, Leiden University, NL-2300 RA Leiden, Netherlands}
\altaffiltext{4}{UCO/Lick Observatory, University of California, Santa Cruz, CA 95064}
\altaffiltext{5}{Department of Astronomy, Yale University, New Haven, CT 06520}
\altaffiltext{6}{Institute of Astronomy, University of Cambridge, Madingley Road, Cambridge CB3 0HA, UK}
\altaffiltext{7}{Space Telescope Science Institute}
\altaffiltext{8}{Institute for Astronomy, ETH Zurich, 8092 Zurich, Switzerland; poesch@phys.ethz.ch}
\altaffiltext{9}{University of California, Riverside, CA 92521, USA}
\altaffiltext{10}{Australian Astronomical Observatory, PO Box 296 Epping, NSW 1710 Australia}
\altaffiltext{11}{Department of Physics \& Astronomy, Macquarie University, Sydney, NSW 2109, Australia}

\begin{abstract}
The remarkable HST datasets from the CANDELS, HUDF09, HUDF12, ERS, and
BoRG/HIPPIES programs have allowed us to map the evolution of the
rest-frame UV luminosity function from $z\sim 10$ to $z\sim 4$.  We
develop new color criteria that more optimally utilize the full
wavelength coverage from the optical, near-IR, and mid-IR observations
over our search fields, while simultaneously minimizing the
incompleteness and eliminating redshift gaps.  We have identified
5859, 3001, 857, 481, 217, and 6 galaxy candidates at $z\sim 4$,
$z\sim 5$, $z\sim 6$, $z\sim 7$, $z\sim 8$, and $z\sim 10$,
respectively from the $\sim$1000 arcmin$^{2}$ area covered by these
datasets.  This sample of $>$10000 galaxy candidates at $z\geq4$ is by
far the largest assembled to date with HST. The selection of
$z\sim4$-8 candidates over the five CANDELS fields allows us to assess
the cosmic variance; the largest variations are at $z\geq7$.  Our new
LF determinations at $z\sim4$ and $z\sim5$ span a 6-mag baseline and
reach to $-$16 AB mag.  These determinations agree well with previous
estimates, but the larger samples and volumes probed here result in a
more reliable sampling of $>L^*$ galaxies and allow us to re-assess
the form of the UV LFs.  Our new LF results strengthen our earlier
findings to $3.4\sigma$ significance for a steeper faint-end slope of
the $UV$ LF at $z>4$, with $\alpha$ evolving from
$\alpha=-1.64\pm0.04$ at $z\sim4$ to $\alpha= -2.06\pm0.13$ at
$z\sim7$ (and $\alpha=-2.02\pm0.23$ at $z\sim8$), consistent with that
expected from the evolution of the halo mass function.  We find less
evolution in the characteristic magnitude $M^*$ from $z\sim7$ to
$z\sim4$; the observed evolution in the LF is now largely represented
by changes in $\phi^*$.  No evidence for a non-Schechter-like form to
the $z\sim4$-8 LFs is found.  A simple conditional luminosity function
model based on halo growth and evolution in the M/L ratio $(\propto
(1+z)^{-1.5})$ of halos provides a good representation of the observed
evolution.
\end{abstract}

\section{Introduction}

Arguably the most fundamental and important observable for galaxy
studies in the early universe is the luminosity function.  The
luminosity function (LF) gives us the volume density of galaxies as a
function of their luminosity.  By comparing the luminosity function
with the halo mass function -- both in shape and normalization -- we
can gain insight into the efficiency of star formation as a function
of halo mass and cosmic time (e.g., van den Bosch et al.\ 2003; Vale
\& Ostriker 2004; Moster et al.\ 2010; Behroozi et al.\ 2013; Birrer
et al.\ 2014).  These comparisons then provide us with insight into
the halo mass scales where gas cooling is most efficient, where
feedback from AGN or SNe starts to become important, and how these
processes vary with cosmic time.  In the rest-frame $UV$, the
luminosity of galaxies strongly correlates with the star formation
rates for all but the most dust-obscured galaxies (e.g., Wang \&
Heckman 1996; Adelberger \& Steidel 2000; Martin et al.\ 2005).
Establishing the $UV$ LF at high redshift is also essential for
assessing the impact of galaxies on the reionization of the universe
(e.g., Bunker et al.\ 2004; Yan \& Windhorst 2004; Oesch et al.\ 2009;
Bouwens et al.\ 2012; Kuhlen \& Faucher-Gigu{\'e}re 2012; Robertson et
al.\ 2013).

\begin{deluxetable*}{ccccccccccccccc}
\tablewidth{0cm}
\tabletypesize{\footnotesize}
\tablecaption{Observational Data Utilized in Deriving the $z\sim4$-10 LFs.\tablenotemark{*}\label{tab:obsdata}}

\tablehead{
\colhead{} & \colhead{Area} & \colhead{Redshift} & \multicolumn{8}{c}{$5\sigma$ Depth (\# of orbits for HST, \# of hours for IRAC)\tablenotemark{a}}\\
\colhead{Field}  & \colhead{(arcmin$^2$)} & \colhead{Sel. Range} & \colhead{u\tablenotemark{b}} & \colhead{B\tablenotemark{b}} & \colhead{$B_{435}$} & \colhead{g\tablenotemark{b}} & \colhead{V\tablenotemark{b}} & \colhead{$V_{606}$} & \colhead{r\tablenotemark{b}} & \colhead{$i_{775}$} & \colhead{i\tablenotemark{b}} & \colhead{$I_{814}$} }
\startdata
XDF\tablenotemark{d}  & 4.7 & 4-10 & --- & --- & 29.6\tablenotemark{e} & --- & --- & 30.0\tablenotemark{e} & --- & 29.8\tablenotemark{e} & --- & 28.7 \\
                      &     &      &      &     & (56) &                     & & (56) & & (144) & & (16) \\
HUDF09-1              & 4.7 & 4-10 &  --- & --- & --- & --- & --- & 28.6 & --- & 28.5 & --- & --- \\
                      &     &      &      &     &      &     &     & (10)  &    & (23) &     & \\
HUDF09-2              & 4.7 & 4-10 &  --- & --- & 28.3 & --- & --- & 29.3 & --- & 28.8 & --- & 28.3 \\
                      &     &      &      &     & (10) &     &     & (32)  &    & (46) &     & (144) \\
CANDELS-GS/      & 64.5 & 4-10 &  --- & --- & 27.7 & --- & --- & 28.0 & --- & 27.5 & --- & 28.0 \\
 $~~$DEEP         &     &      &      &     & (3)  &     &     & (3)  &    & (3.5) &     & ($>$12) \\
CANDELS-GS/      & 34.2 & 4-10 &  --- & --- & 27.7 & --- & --- & 28.0 & --- & 27.5 & --- & 27.0 \\
 $~~$WIDE         &     &      &      &     & (3)  &     &     & (3)  &    & (3.5) &     & ($\sim$2)\\
ERS                  & 40.5 & 4-10 &  --- & --- & 27.5 & --- & --- & 27.7 & --- & 27.2 & --- & 27.6 \\
                 &       &     &      &     & (3)  &     &     & (3)  &    & (3.5) &     & ($\sim$4)\\
CANDELS-GN/      & 62.9 & 4-10 &  --- & --- & 27.5 & --- & --- & 27.7 & --- & 27.3 & --- & 27.9 \\
 $~~$DEEP         &     &      &      &     & (3)  &     &     & (3)  &    & (3.5) &     & ($>$12) \\
CANDELS-GN/      & 60.9 & 4-10 &  --- & --- & 27.5 & --- & --- & 27.7 & --- & 27.2 & --- & 27.0 \\
 $~~$WIDE         &     &      &      &     & (3)  &     &     & (3)  &    & (3.5) &     & ($\sim$2)\\
CANDELS-       & 151.2 & 5-10 & 25.5 & 28.0 & --- & --- & 27.7 & 27.2 & 27.5 & --- & 27.4 & 27.2 \\
 $~~$UDS              &       &      &      &      &     &     &      & ($\sim$1.5) &  & &  & ($\sim$3) \\
CANDELS-      & 151.9 & 5-10 & 27.8 & 28.0 & --- & 28.0 & 27.0 & 27.2 & 27.9 & --- & 27.8 & 27.2 \\
$~~$COSMOS            &       &      &      &      &     &     &      & ($\sim$1.5) &  & &  & ($\sim$4) \\
CANDELS-         & 150.7 & 5-10 & 27.4 & --- & --- & 27.9 & --- & 27.6 & 27.6 & --- & 27.5 & 27.6 \\
$~~$EGS               &       &      &      &      &     &     &      & ($\sim$2.5) &  & &  & ($\sim$4) \\
BoRG/$~$            & 218.3   & 8 & --- & --- & --- & --- & --- & 27.0-$~$ & --- & --- & --- & --- \\
$~$HIPPIES\tablenotemark{g} &   &  &  &  & & & & $~$28.7 &  &  &  \\
\\
 & z\tablenotemark{b} & $z_{850}$ & Y\tablenotemark{b} & $Y_{098}/Y_{105}$ & J\tablenotemark{b} & $J_{125}$ & $JH_{140}$ & H\tablenotemark{b} & $H_{160}$ & $K_s$\tablenotemark{b} & $3.6\mu$m\tablenotemark{c} & $4.5\mu$m\tablenotemark{c}\\
\tableline
XDF\tablenotemark{b} & --- & 29.2\tablenotemark{c} & --- & 29.7 & --- & 29.3 & 29.3 & --- & 29.4 & --- & 26.5 & 26.5\\
                     &     & (170) & & (100) & & (40) & (30) & & (85) & & (130) & (130) \\
HUDF09-1             & --- & 28.4 & --- & 28.3 & --- & 28.5 & 26.3\tablenotemark{f} & --- & 28.3 & --- & 26.4 & 26.4\\
                     &     & (71) & & (8) & & (12) & (0.3) & & (13) & & (80) & (80)\\
HUDF09-2             & --- & 28.8 & --- & 28.6 & --- & 28.9 & 26.3\tablenotemark{f} & --- & 28.7 & --- & 26.5 & 26.5\\
                     &     & (89) & & (11) & & (18) & (0.3) & & (19) & & (130) & (130)\\
CANDELS-GS/      & --- & 27.3 & --- & 27.5 & --- & 27.8 & 26.3\tablenotemark{f} & --- & 27.5 & --- & 26.1 & 25.9\\
$~~$Deep             &     & ($\sim$15) & & (3) & & (4) & (0.3) & & (4) & & (50) & (50) \\
CANDELS-GS/      & --- & 27.1 & --- & 27.0 & --- & 27.1 & 26.3\tablenotemark{f} & --- & 26.8 & --- & 26.1 & 25.9\\
$~~$Wide               &     & ($\sim$15) & & (1) & & (0.7) & (0.3) & & (1.3) & & (50) & (50)\\
ERS                  & --- & 27.1 & --- & 27.0 & --- & 27.6 & 26.4\tablenotemark{f} & --- & 27.4 & --- & 26.1 & 25.9\\
                     &     & ($\sim$15) & & (2) & & (2) & (0.3) & & (2) & & (50) & (50)\\
CANDELS-GN/      & --- & 27.3 & --- & 27.3 & --- & 27.7 & 26.3\tablenotemark{f} & --- & 27.5 & --- & 26.1 & 25.9\\
$~~$Deep                &     & ($\sim$15) & & (3) & & (4) & (0.3) & & (4) & & (50) & (50)\\
CANDELS-GN/      & --- & 27.2 & --- & 26.7 & --- & 26.8 & 26.2\tablenotemark{f} & --- & 26.7 & --- & 26.1 & 25.9\\
$~~$Wide                &     & ($\sim$15) & & (1) & & (0.7) & (0.3) & & (1.3) & & (50) & (50)\\
CANDELS-          & 26.2 & --- & 26.0 & --- & --- & 26.6 & 26.3\tablenotemark{f} & --- & 26.8 & 25.5 & 25.5 & 25.3 \\
$~~$UDS               &     &   & &    & & (0.6) & (0.3) & & (1.3) & & (12) & (12)\\
CANDELS-       & 26.5 & --- & 26.1 & --- & 25.4 & 26.6 & 26.3\tablenotemark{f} & 25.0 & 26.8 & 25.3 & 25.4 & 25.2\\
$~~$COSMOS             &     &   & &    & & (0.6) & (0.3) & & (1.3) & & (12) & (12)\\
CANDELS-          & 26.1 & --- & --- & --- & --- & 26.6 & 26.3\tablenotemark{f} & --- & 26.9 & 24.1 & 25.5 & 25.3\\
$~~$EGS                &     &   & &    & & (0.6) & (0.3) & & (1.3) & & (12) & (12)\\
BoRG/$~$             & --- & --- & --- & 26.5-$~$ & --- & 26.5-$~$ & --- & --- & 26.3-$~$ & --- & --- & --- \\
$~$HIPPIES\tablenotemark{g} & & &  & $~$28.2 &   & $~$28.4 &   &  & $~$28.1
\enddata
\tablenotetext{*}{More details on the observational data we use for
  each of these search fields is provided in Appendix A.}
\tablenotetext{a}{The $5\sigma$ depths for the HST observations are
  computed based on the median flux uncertainties (after correction to
  total) for the faintest 20\% of sources in our fields.  While these
  depths are shallower than one computes from the noise in
  $0.35''$-diameter apertures (and not extrapolating to the total
  flux), the depths we quote here are reflective of that
  achieved for real sources.}
\tablenotetext{b}{Indicates ground-based observations from
  Subaru/Suprime-Cam, CFHT/Megacam, CFHT/Megacam, HAWK-I, VISTA, and
  CFHT/WIRCam in the $BgVriz$, $ugriyz$, $u$, $YK_s$, $YJHK_s$, and
  $K_s$ bands, respectively.  The $5\sigma$ depths for the
  ground-based observations are derived from the noise fluctuations in
  1.2$''$-diameter apertures (after correction to total).  These
  apertures are almost identical in size to those chosen by Skelton et
  al.\ (2014) to perform photometry on sources over the CANDELS
  fields.}
\tablenotetext{c}{The $5\sigma$ depths for the Spitzer/IRAC
  observations are derived in 2.0$''$-diameter apertures (after correction to total).}
\tablenotetext{d}{The XDF refers to the 4.7 arcmin$^2$ region over the
  HUDF with ultra-deep near-IR observations from the HUDF09 and HUDF12
  programs (Illingworth et al.\ 2013).  It includes all ACS and
  WFC3/IR observations acquired over this region for the 10-year
  period 2002 to 2012.}
\tablenotetext{e}{The present XDF reduction (Illingworth et al.\ 2013)
  is typically $\sim$0.2 mag deeper than the original reduction of the
  HUDF ACS data provided by Beckwith et al.\ (2006).}
\tablenotetext{f}{The $JH_{140}$ observations are from the 3D-HST and
  GO-11600 (PI: Weiner) programs.}
\tablenotetext{g}{Only the highest quality (longer exposure)
  BoRG/HIPPIES fields (and similar programs) are considered in our
  analysis (see Appendix A.2).  For inclusion, we require search
  fields to have an average exposure time in the $J_{125}$ and
  $H_{160}$ bands of at least 1200 seconds and with longer exposure
  times in the optical $V_{606}+V_{600}$ bands than the average
  exposure time in the near-infrared $J_{125}+H_{160}$ observations.}
\end{deluxetable*}

Attempts to map out the evolution of the luminosity function of
galaxies in the high-redshift universe has a long history, beginning
with the discovery of Lyman-break galaxies at $z\sim3$ (Steidel et
al.\ 1996) and work on the Hubble Deep Field North (e.g., Madau et
al.\ 1996; Sawicki et al.\ 1997).  One of the most important early
results on the LF at high redshift were the $z\sim3$ and $z\sim4$
determinations by Steidel et al.\ (1999), based on a wide-area (0.23
degree$^2$) photometric selection and spectroscopic follow-up
campaign.  Steidel et al.\ (1999) derived essentially identical LFs
for galaxies at both $z\sim3$ and $z\sim4$, pointing towards a broader
peak in the star formation history extending out to $z\sim4$, finding
no evidence for the large decline that Madau et al.\ (1996) had
reported between $z\sim3$ and $z\sim4$.

Following upon these early results, there was a push to measure the
$UV$ LF to $z\sim5$ and higher (e.g., Dickinson 2000; Ouchi et
al.\ 2004; Lehnert \& Bremer 2003).  However, it was not until the
installation of the Advanced Camera for Surveys (Ford et al.\ 2003) on
the Hubble Space Telescope in 2002 that the first substantial
explorations of the $UV$ LF at $z\sim6$ began.  Importantly, the HST
ACS instrument enabled astronomers to obtain deep, wide-area imaging
in the $z_{850}$ band, allowing for the efficient selection of
galaxies at $z\sim6$ (Stanway et al.\ 2003; Bouwens et al.\ 2003b;
Dickinson et al.\ 2004).  Based on $z\sim6$ searches and the large HST
data sets from the wide-area GOODS and ultra-deep HUDF data sets, the
overall evolution of the $UV$ LF was quantified to $z\sim6$ (Bouwens
et al.\ 2004a; Bunker et al.\ 2004; Yan \& Windhorst 2004; Bouwens et
al.\ 2006; Beckwith et al.\ 2006).  The first quantification of the
evolution of the $UV$ LF with fits to all three Schechter parameters
was by Bouwens et al.\ (2006) and suggested a brightening of the
characteristic luminosity with cosmic time.  Most follow-up studies
supported this conclusion (Bouwens et al.\ 2007; McLure et al.\ 2009;
Su et al.\ 2011: though Beckwith et al.\ 2006 favored a simple
$\phi^*$ evolution model with no evolution in $\alpha$ or $M^*$).

The next significant advance in our knowledge of the $UV$ LF at high
redshift came with the installation of the Wide Field Camera 3 (WFC3)
and its near-IR camera WFC3/IR on the Hubble Space Telescope.  The
excellent sensitivity, field of view, and spatial resolution of this
camera allowed us to survey the sky $\sim$40$\times$ more efficiently
in the near-IR than with the earlier generation IR instrument NICMOS.
The high efficiency of WFC3/IR enabled the identification of
$\sim$200-500 galaxies at $z\sim7$-8 (e.g., Wilkins et al.\ 2010;
Bouwens et al.\ 2011; Oesch et al.\ 2012; Grazian et al.\ 2012;
Finkelstein et al.\ 2012; Yan et al.\ 2012; McLure et al.\ 2013;
Schenker et al.\ 2013; Lorenzoni et al.\ 2013; Schmidt et al.\ 2014),
whereas only $\sim$20 were known before (Bouwens et al.\ 2008, 2010b;
Oesch et al.\ 2009; Ouchi et al.\ 2009b).  While initial determinations
of the $UV$ LF at $z\sim7$-8 appeared consistent with a continued
evolution in the characteristic luminosity to fainter values (e.g.,
Bouwens et al.\ 2010a; Lorenzoni et al.\ 2011), the inclusion of
wider-area data in these determinations quickly made it clear that
some of the evolution in the LF was in the volume density $\phi^*$
(e.g., Ouchi et al.\ 2009b; Castellano et al.\ 2010; Bouwens et
al.\ 2011b; Bradley et al.\ 2012; McLure et al.\ 2013) and in the
faint-end slope $\alpha$ (Bouwens et al.\ 2011b; Bradley et al.\ 2012;
Schenker et al.\ 2013; McLure et al.\ 2013).

With the recent completion of the wide-area CANDELS program (Grogin et
al.\ 2011; Koekemoer et al.\ 2011) and availability of even deeper
optical+near-IR observations over the HUDF from the XDF/UDF12 data set
(Illingworth et al.\ 2013; Ellis et al.\ 2013), there are several
reasons to revisit determinations of the $UV$ LF not just at
$z\sim7$-10, but over the entire range $z\sim10$ to $z\sim4$ to more
precisely study the evolution.  First, the addition of especially deep
WFC3/IR observations to legacy fields with deep ACS observations
allows for an improved determination of the $UV$ LF at $z\sim5$-6 due
to the $\sim$1-mag greater depths of the $UV$ LF probed at $z\sim5$-6
by the WFC3/IR near-IR observations relative to the original
$z_{850}$-band observations.  The gains at $z\sim6$ are even more
significant, as the new WFC3/IR data make it possible (1) to perform a
standard two-color selection of $z\sim6$ galaxies and (2) to measure
their $UV$ luminosities at the same rest-frame wavelengths as with
other samples.  Bouwens et al.\ (2012a) already made use of the
initial observations over the CANDELS GOODS-South to provide such a
determination of the $z\sim6$ LF, but the depth and area of the
current data sets allow us to significantly improve upon this early
analysis.

Second, the availability of WFC3/IR observations over legacy fields
like GOODS or the HUDF can also significantly improve the redshift
completeness of Lyman-break-like selections at $z\sim4$, $z\sim5$, and
$z\sim6$, while keeping the overall contamination levels to a minimum
(as we will illustrate in \S3 of this paper).  Improving the overall
completeness and redshift coverage of Lyman-break-like selections is
important, since it will allow us to leverage the full search volume,
thereby reducing the sensitivity of the high-redshift results to
large-scale structure variations and shot noise (from small number
statistics).

Finally, the current area covered by the wide-area CANDELS program now
is in excess of 750 arcmin$^2$ in total area, or $\sim$0.2 square
degrees, over 5 independent pointings on the sky.  The total area
available at present goes significantly beyond the CANDELS-GS,
CANDELS-UDS, ERS, and BoRG fields that have been used for many
previous LF determinations at $z\sim7$-10 (e.g., Bouwens et al.\ 2011;
Oesch et al.\ 2012; Bradley et al.\ 2012; Yan et al.\ 2012; Grazian et
al.\ 2012; Lorenzoni et al.\ 2013; McLure et al.\ 2013; Schenker et
al.\ 2013).  While use of the full CANDELS area can be more
challenging due to a lack of deep HST data at $\sim$0.9-1.1$\mu$m over
the UDS, COSMOS, and EGS areas, the effective selection of $z\sim5$-10
galaxies is nevertheless possible, leveraging the available
ground-based observations, as we demonstrate in \S3 and \S4 (albeit
with some intercontamination between the CANDELS-EGS $z\sim7$ and
$z\sim8$ samples due to the lack of deep $Y$-band data).

Of course, there have been a significant number of studies on the $UV$
LF at $z\sim4$-7 over even wider survey areas than available over
CANDELS, e.g., van der Burg et al.\ (2010) and Willott et al.\ (2013)
at $z\sim3$-5 and $z\sim6$ from the $\sim$4 deg$^2$ Canada France
Hawaii Telescope (CFHT) Legacy Survey deep field observations, Ouchi
et al.\ (2009b) at $z\sim7$ from Subaru observations of the Subaru
Deep Field (Kashikawa et al.\ 2004) and GOODS North (Giavalisco et
al.\ 2004a), and Bowler et al.\ (2014) at $z\sim7$ from the UltraVISTA
and UDS programs.  While each of these surveys also provide
constraints on the volume density of the bright rare sources, these
programs generally lack high-spatial-resolution data on their
candidates, making the rejection of low-mass stars from these survey
fields more difficult.  In addition, integration of the results from
wide-area fields with deeper, narrower fields can be particularly
challenging, as any systematic differences in the procedure for
measuring magnitudes or estimating volume densities can result in
significant errors on the measured shape of the LF (e.g., see
Figure~\ref{fig:oldi} from Appendix F.2 for an illustration of the
impact that small systematics can have).

Controlling for cosmic variance is especially important given the
substantial variations in the volume density of luminous sources
observed field to field.  The use of independent sightlines -- as
implemented in the CANDELS program -- is remarkably effective in
reducing the impact of cosmic variance on our results.  In fact, we
would expect the results from the 0.2 degree$^2$ search area available
over the 5 CANDELS fields to be reasonably competitive with the 1.5
deg$^2$ UltraVISTA field (McCracken et al.\ 2012), as far as
large-scale structure uncertainties are concerned.  While the
uncertainties on the 5 CANDELS fields are formally expected to be
$\sim$1.6$\times$ larger,\footnote{Using the Trenti \& Stiavelli
  (2008) ``cosmic variance calculator,'' a $z=5.8\pm0.5$ redshift
  selection window for each sample, galaxies with an intrinsic volume
  density of $4\times10^{-4}$ Mpc$^{-3}$, and 5 independent
  20$'$$\times$7.5$'$ CANDELS survey fields, we estimate a total
  uncertainty of 10\% on the volume density of galaxies over the
  entire CANDELS program from ``cosmic variance.''  Repeating this
  calculation over the 90$'$$\times$60$'$ survey area from UltraVISTA
  yields $\sim$7\%.}  CANDELS usefully allows for a measurement of the
field-to-field variations and hence uncertainties due to large-scale
structure (which is especially valuable if factor of $\sim$1.8
variations in the volume density of bright $z\gtrsim6$ galaxies are
present on square-degree scales: Bowler et al.\ 2015).  Of course,
very wide-area ground-based surveys can also make use of multiple
search fields, both to estimate the uncertainties arising from
large-scale structure and as a further control on cosmic variance
(e.g., Ouchi et al.\ 2009; Willott et al.\ 2013; Bowler et al.\ 2014,
2015), and can also benefit from smaller shot noise uncertainties (if
the goal is the extreme bright end of the LF).

The purpose of the present work is to provide for a comprehensive and
self-consistent determination of the $UV$ LFs at $z\sim4$, $z\sim5$,
$z\sim6$, $z\sim7$, $z\sim8$, and $z\sim10$ using essentially all of
the deep, wide-area observations available from HST over five
independent lines of sight on the sky and including the full data sets
from the CANDELS, ERS, and HUDF09+12/XDF programs.  The deepest,
highest-quality regions within the BoRG/HIPPIES program (relevant for
selecting $z\sim8$ galaxies) are also considered.  In deriving the
present LFs, we use essentially the same procedures, as previously
utilized in Bouwens et al.\ (2007) and Bouwens et al.\ (2011).  Great
care is taken to minimize the impact of systematic biases on our
results.  Where possible, extensive use of deep ground-based
observations over our search fields is made to ensure the best
possible constraints on the redshifts of the sources.  A full
consideration of the available Spitzer/IRAC SEDS (Ashby et al.\ 2013),
Spitzer/IRAC GOODS (Dickinson et al.\ 2004), and IRAC Ultra Deep Field
2010 (IUDF10: Labb{\'e} et al.\ 2013) observations over our fields are
made in setting constraints on the LF at $z\sim10$ (see Oesch et
al.\ 2014).

For consistency with previous work, we find it convenient to quote
results in terms of the luminosity $L_{z=3}^{*}$ Steidel et
al.\ (1999) derived at $z\sim3$, i.e., $M_{1700,AB}=-21.07$.  We refer
to the HST F435W, F606W, F600LP, F775W, F814W, F850LP, F098M, F105W,
F125W, F140W, and F160W bands as $B_{435}$, $V_{606}$, $V_{600}$,
$i_{775}$, $I_{814}$, $z_{850}$, $Y_{098}$, $Y_{105}$, $J_{125}$,
$JH_{140}$, and $H_{160}$, respectively, for simplicity.  Where
necessary, we assume $\Omega_0 = 0.3$, $\Omega_{\Lambda} = 0.7$, and
$H_0 = 70\,\textrm{km/s/Mpc}$.  All magnitudes are in the AB system
(Oke \& Gunn 1983).

\begin{figure*}
\epsscale{1.12}
\plotone{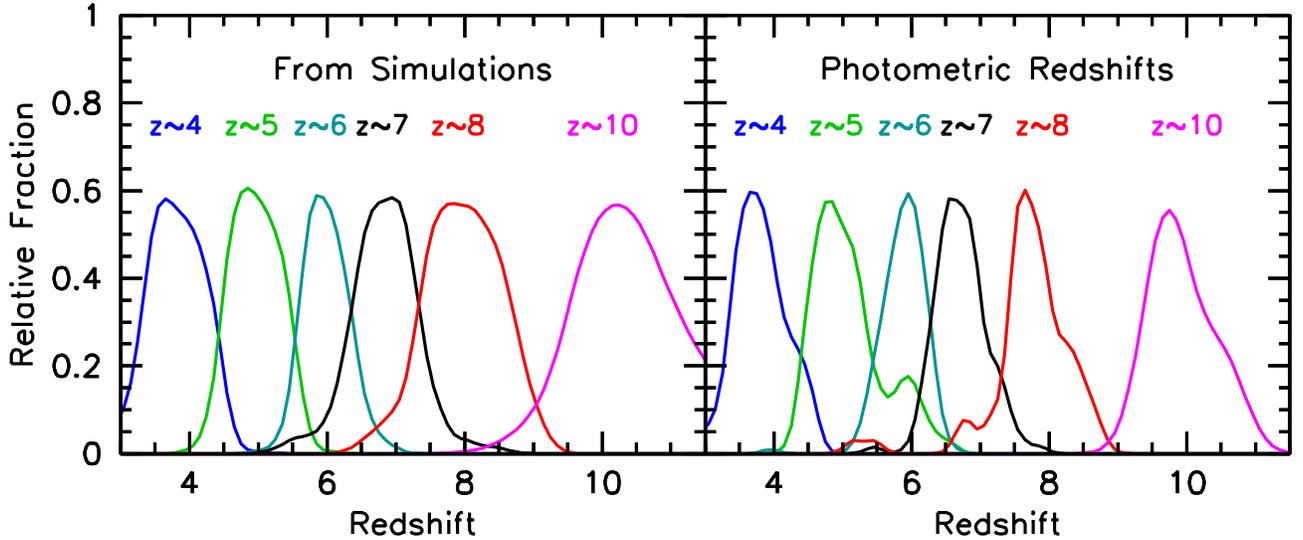}
\caption{(\textit{left}) The expected redshift distributions for our
  $z\sim4$, $z\sim5$, $z\sim6$, $z\sim7$, $z\sim8$, and $z\sim10$
  samples from the XDF using the Monte-Carlo simulations described in
  \S4.1.  The mean redshifts for these samples are 3.8, 4.9, 5.9, 6.8,
  7.9, and 10.4, respectively.  These simulations demonstrate the
  effectiveness of our selection criteria in isolating galaxies within
  fixed redshift ranges.  Each selection window is smoothed by a
  normal distribution with scatter $\sigma_z \sim 0.2$.
  (\textit{right}) Redshift distribution we recover for sources in our
  $z\sim4$, $z\sim5$, $z\sim6$, $z\sim7$, $z\sim8$, and $z\sim10$
  samples using the EAZY photometric redshift code (with similar
  smoothing as in the left panel).  Our color-color selections
  segregate sources by redshift in a very similar manner to what one
  would find selecting sources according to their best-fit photometric
  redshift estimate (e.g., McLure et al.\ 2010; Finkelstein et
  al.\ 2012; Bradley et al.\ 2014).\label{fig:zdist}}
\end{figure*}

\section{Observational Data Sets}

The present determinations of the $UV$ LFs at $z\sim4$-10 make use of
all the ultra-deep, wide-area observations obtained as part of the
HUDF09+HUDF12, ERS, and CANDELS programs, in conjunction with archival
HST observations over these fields.  The pure parallel observations
from the BoRG/HIPPIES programs are also utilized.  A summary of all
the deep, wide-area data sets used in the present study is provided in
Table~\ref{tab:obsdata}, along with the redshift ranges of the sources
we can select in these data sets.  The 5$\sigma$ depths reported in
Table~\ref{tab:obsdata} are based on the median uncertainties in the
total fluxes, as found for the faintest 20\% of sources identified as
part of a data set (total fluxes are derived using the procedures
described in \S3.1).

Except for the reduced HST data made publicly available by the BoRG
team through the Mikulski archive for Space
Telscopes,\footnote{http://archive.stsci.edu/prepds/borg/} we
rereduced all of these data using the ACS GTO pipeline \textsc{apsis}
(Blakeslee et al.\ 2003) and our WFC3/IR pipeline \textsc{wfc3red.py}
(Magee et al.\ 2011).  All fields were reduced and analyzed at a
$0.03''$-pixel scale, except the CANDELS UDS/COSMOS/EGS fields (where
the pixel scale was $0.06''$) or BoRG/HIPPIES pure-parallel data sets
(where the pixel scale was $0.08''$ for the reductions we utilized
from Bradley et al.\ 2012 or $0.06''$ where we carried out our own
reductions).

\textit{XDF}: Our deepest search field (reaching to $\sim$30 mag at
$5\sigma$) is located over the particularly deep 4.7 arcmin$^2$
WFC3/IR pointing defined by the HUDF09 and HUDF12 programs within the
HUDF (Beckwith et al.\ 2006) and takes full advantage of the entire
XDF data set (Illingworth et al.\ 2012) incorporating all ACS and
WFC3/IR observations ever taken over the HUDF (reaching $\sim$0.2 mag
deeper than the original optical HUDF: Beckwith et al.\ 2006).

\textit{HUDF09-Ps Fields}: Our second and third deepest search fields
are the two deep $\sim$4.7 arcmin$^2$ WFC3/IR pointings HUDF09-1 and
HUDF09-2 defined by the HUDF09 program (Bouwens et al.\ 2011).
Ultra-deep ACS observations in the $V_{606}i_{775}z_{850}$ bands are
available over these fields from the HUDF05, HUDF09, HUDF12, and other
programs (Oesch et al.\ 2007; Bouwens et al.\ 2011; Ellis et
al.\ 2013).  Deep $B_{435}$ observations are available over the
HUDF09-2 field.

\textit{CANDELS GOODS-North (GN) + CANDELS GOODS-South (GS) Fields}:
We also make use of both the deep and intermediate depth observations
that exist over the GN and GS fields from the CANDELS program (Grogin
et al.\ 2011).  These observations probe $\sim$1.5-2.5 mag shallower
than our deepest field, the XDF, but cover $\sim$30$\times$ more area.
Deep ACS $B_{435}V_{606}i_{775}z_{850}$ observations are available
over the entire CANDELS-GN, with the deep regions are covered with
especially sensitive HST ACS $I_{814}$ observations ($\gtrsim$0.5 mag
deeper than in the $i_{775}$ band).  Our reductions of these
observations include the full set of SNe search and follow-up
observations associated with the Riess et al.\ (2007) programs.
Shallow observations in the $JH_{140}$ band (0.3 orbits) are available
over most of this area as part of the 3D-HST (Brammer et al.\ 2012)
and AGHAST (Weiner et al.\ 2014) programs.  

\begin{figure*}
\epsscale{1.12}
\plotone{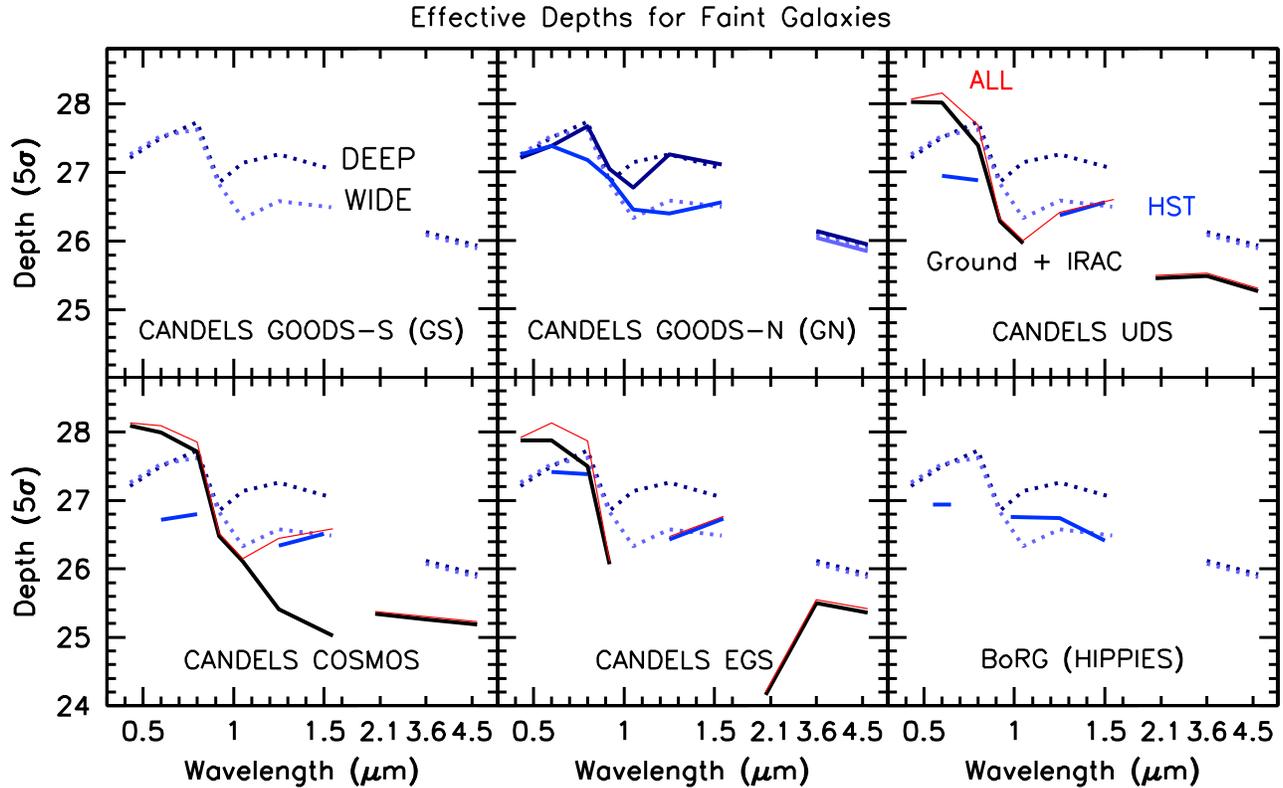}
\caption{An illustration of the $5\sigma$ depths of the various data
  sets used in this study (calculated based on the median $1\sigma$
  flux errors measured for all sources found between
  $H_{160,AB}\sim26$ and $H_{160,AB}\sim26.5$, after correcting each
  of these fluxes [Kron apertures for HST, $1.2''$-diameter aperture
    for ground-based, and $2''$-diameter apertures for Spitzer/IRAC
    observations] to total).  The upper leftmost panel shows the
  depths of the two shallower data sets available over the GOODS-S
  sightline, i.e., the CANDELS DEEP data set (dotted dark blue line)
  and the CANDELS WIDE data set (dotted blue line).  The other panels
  show the depths of the data available over the other four CANDELS
  fields and those BoRG/HIPPIES fields where $z\sim8$ candidates have
  been identified.  The blue lines indicate the depths available in
  the HST observations alone, while the red lines indicate the depths
  of all available observations, i.e., HST + ground-based.  The dark
  blue solid lines indicate the depths of the HST observations
  associated with the CANDELS DEEP GN program.  In 5 out of 6 cases
  that $z$ and $Y$ band observations exist over the CANDELS-UDS,
  CANDELS-COSMOS, and CANDELS-EGS fields, these data reach within 0.5
  mag of that available over the CANDELS-GS+GN fields.  As a result,
  current observations allow for the effective selection of galaxies
  at $z\sim6$, $z\sim7$, and $z\sim8$ over the CANDELS-UDS/COSMOS/EGS
  fields, if we limit ourselves to a somewhat brighter limit than we
  consider over CANDELS GN and GS (as we demonstrate from end-to-end
  simulations in \S4.1 and as shown in Figure~\ref{fig:zdistsel}).
\label{fig:depth}}
\end{figure*}

\textit{ERS Field}: Additional constraints on the prevalence of
intermediate luminosity $z\sim4$-10 galaxies is provided by the ACS
$B_{435}V_{606}i_{775}z_{850}$ and WFC3/IR $Y_{098}J_{125}H_{160}$
observations available as part of the $\sim$40 arcmin$^2$ Early
Release Science observations over GOODS South (Windhorst et
al.\ 2011).  

\textit{CANDELS-UDS, CANDELS-COSMOS, and CANDELS-EGS Fields}: Our
strongest constraint on the volume density of the brightest, most
luminous galaxies is provided by the $\sim$450 arcmin$^2$ search area
available over the CANDELS-UDS, CANDELS-EGS, and CANDELS-COSMOS data
sets (Grogin et al.\ 2011).  Essentially this entire area is covered
by moderately deep WFC3/IR $J_{125}H_{160}$ and ACS $V_{606}I_{814}$
observations.  Deep ground-based observations in both the optical and
near-IR from Subaru, CFHT, VLT, and VISTA largely fill out the
wavelength coverage available from HST so that it extends from
3500$\AA\,$ to 23000$\AA$, making it possible to select galaxies at
$z\sim5$, $z\sim6$, $z\sim7$, $z\sim8$, and $z\sim10$ and also ensure
that our selected samples are largely free of contamination by lower
redshift interlopers.

\textit{BoRG/HIPPIES Fields}: The $\sim$450 arcmin$^2$ wide-area
BoRG/HIPPIES data set (Trenti et al.\ 2011; Yan et al.\ 2011; Bradley
et al.\ 2012; Schmidt et al.\ 2014) effectively doubles the search
volume we have available to constrain the prevalence of the rarest,
brightest $z\sim8$ galaxies.  The data set features deep observations
in $J_{125}$ and $H_{160}$ bands (from $\sim$25.5 mag to $\sim$28.4
mag, 5$\sigma$), as well as observations in two bands blueward of the
break, $Y_{098}$/$Y_{105}$ and $V_{606}$/$V_{600}$.  The BoRG/HIPPIES
observations were obtained with HST in parallel with observations from
other science programs, providing for excellent controls on
large-scale structure uncertainties, due to the many independent areas
of the sky probed.  Here we make use of the highest-quality search
fields ($\sim$220 arcmin$^2$) taken as part of both the BoRG program
and similar data sets.  37 arcmin$^2$ of this search area derives from
the HIPPIES program.

With the exception of the BoRG/HIPPIES fields, all of our search
fields have deep Spitzer/IRAC observations available that can be used
to improve our search for $z\sim9$-10 galaxies and better distinguish
$z\leq7$ galaxies from $z\geq7$ galaxies.  Here we make use of the
Spitzer/IRAC observations from the GOODS (Dickinson et al.\ 2004),
SEDS (Ashby et al.\ 2013), IUDF (Labb{\'e} et al.\ 2013), and
S-CANDELS (PI Fazio: Oesch et al.\ 2014) data set over the CANDELS-GN
and GS, the IUDF data set over the HUDF/XDF and HUDF09-Ps fields, and
the SEDS data set over the CANDELS UDS/COSMOS/EGS fields.

The zeropoints for the ACS and WFC3/IR observations were set according
to the STScI zeropoint
calculator \footnote{\texttt{http://www.stsci.edu/hst/acs/analysis/zeropoints/zpt.py}}
and the WFC3/IR data handbook (Dressel et al.\ 2012).  These
zeropoints were corrected for foreground galaxy extinction based the
Schlafly \& Finkbeiner (2011) maps.

Additional details on the data sets or search fields utilized in this
study can be found in Appenidx A.

\section{Sample Selection}

\subsection{Photometry}

\subsubsection{HST Photometry}

As in our other recent work, we make use of the SExtractor (Bertin \&
Arnouts 1996) software in dual-image mode to construct the source
catalogs from which we will later select our high-redshift samples.
For the detection images, we utilize the square root of $\chi^2$ image
(Szalay et al.\ 1999: similar to a coadded image) constructed from all
available $Y_{098}Y_{105}J_{125}H_{160}$ WFC3/IR observations for our
$z\sim4$, $z\sim5$, $z\sim6$, and $z\sim7$ samples, the $J_{125}$ and
$H_{160}$-band observations for our $z\sim8$ samples, and the
$H_{160}$-band observations for our $z\sim10$ samples.  For the
$z\sim7$ and $z\sim8$ samples from the XDF data set, we also include
the deep $JH_{140}$-band observations in generating the $\chi^2$
image.

Color measurements are then made from the observations PSF-matched to
the $H_{160}$-band in small-scalable apertures derived adopting a Kron
(1980) parameter of 1.6.  The PSF matching is performed using a kernel
derived that when convolved with the tighter PSF matches the
$H_{160}$-band encircled energy distribution (\S2.2 of Bouwens et
al.\ 2014a).  We can obtain even higher S/N color measurements at
optical wavelengths for sources in our search fields by taking
advantage of the narrower PSF of the HST ACS observations.  Our
procedure is simply (1) to PSF match the ACS observations to the
$z_{850}$-band and (2) to do the photometry in an aperture that was
just 70\% the size of that used on the WFC3/IR data.  We arrived at
the 70\% scale factor by comparing the sizes of the scalable
Kron-style apertures derived for individual $z\sim4$-6 galaxies found
in HUDF+GOODS, if PSF-matching is done to the ACS $z_{850}$-band data
and to the WFC3/IR $H_{160}$-band data.  Higher S/N optical colors are
useful for measuring the amplitude of the Lyman Break in candidate
$z\sim4$, $z\sim5$, and $z\sim6$ galaxies.

The fluxes measured in the small-scalable apertures were then
corrected to total magnitudes in two steps.  In the first step, we
multiply the small aperture fluxes by the excess light found in a
larger scalable aperture (Kron factor of 2.5) relative to smaller
scalable aperture.  This estimate is made using the square root of
$\chi^2$ image.  Second, we correct for the light outside the large
scalable aperture and on the wings of the PSF using the standard
encircled energy distributions for point sources tabulated in Dressel
(2013) or Sirianni et al.\ (2005).  Figure~\ref{fig:compm13} from
Appendix H illustrates the typical size of the apertures we use
relative to the size of a source.  While the source included in
Figure~\ref{fig:compm13} is the one of the largest $z\sim7$ galaxies
known (i.e., the largest in the HUDF: Oesch et al.\ 2010; Ono et
al.\ 2013), this figure illustrates the usefulness of scalable
apertures.

\subsubsection{Photometry on Ground-Based Imaging Data}

In selecting our samples over the wide-area CANDELS-UDS,
CANDELS-COSMOS, and CANDELS-EGS fields, we also made use of the deep
optical and near-infrared ground-based data available over these same
areas of the sky from Subaru, CFHT, VLT, and VISTA (see Appendix A.1).
The optical observations reach as deep or deeper than the HST
observations and are important for excluding lower redshift
contaminants from the $z\sim5$-10 samples we construct from these
fields.  Moderately deep near-IR observations are available in the $Y$
band and are valuable for discriminating between $z\sim7$ and $z\sim8$
candidates in the CANDELS-UDS and CANDELS-COSMOS fields (Appendix
A.1).

A significant challenge in extracting photometry for sources from the
ground-based data was the broad PSF and therefore the occasional
blending of sources with nearby neighbors in the ground-based imaging
data.  To obtain accurate photometry of sources in the presence of
this blending, we made use of \textsc{Mophongo} (Labb{\'e} et
al.\ 2006, 2010a, 2010b, 2013) to do photometry on sources in our
fields.  Since this software has been presented more extensively in
other places, we only include a brief description here.

The most important step for doing photometry on faint sources
contaminated by light from neighboring sources is the removal of the
contaminating flux.  This is accomplished by using the deep WFC3/IR
$H_{160}$-band observations as a template to model the positions and
isolated flux profiles of the foreground sources.  These flux profiles
are then convolved to match the ground-based PSFs and then
simultaneously fit to the ground-based imaging data leaving only the
fluxes of the sources as unknowns.  The best-fit model is then used to
subtract the flux from neighboring sources and normal aperture
photometry is performed on sources in 1.2$''$-diameter apertures.  The
measured fluxes are then corrected to account for the light on the
wings of the ground-based PSFs.  Our correction of the measured flux
in 1.2$''$-diameter apertures to total makes use of the HST template
we have for each source (after convolution to match the ground-based
PSF).  The typical residuals we find in our registration of the
ground-based images to the HST observations were $\sim$0.04$''$.  The
CANDELS team use a similar approach in deriving photometry for the
CANDELS-UDS and CANDELS-GS fields (Galametz et al.\ 2013; Guo et
al.\ 2013).

\subsubsection{IRAC Photometry}

Deep Spitzer/IRAC imaging observations available over our search
fields provide essential constraints on the shape of source SEDs
redward of $1.6\mu$m for the $z\sim10$ searches we perform, allowing
us to distinguish $z\sim10$ star-forming galaxies from lower redshift
interlopers.  See Appendix A of Oesch et al.\ (2012a) for a discussion
of these contaminants.

Our procedure for performing photometry on the deep IRAC observations
(Labb{\'e} et al.\ 2006, 2010a, 2010b, 2013) is almost identical to
the approach we adopt for the deep ground-based observations
(\S3.1.2).  The positions and morphology of sources in the deep HST
observations are used to model and subtract contamination from
neighboring sources on candidate $z\sim10$ galaxies in our search
fields.  Photometry is then performed on the sources in
$2.0''$-diameter apertures, and the measured flux is corrected to
total based on the HST template we have for each source convolved to
match the Spitzer/IRAC PSF.

To ensure that the photometry we derive is robust, we compared the
fluxes we measure for individual sources with results using
$3''$-diameter apertures and find almost exactly the same measured
flux in the mean at both $3.6\mu$ and $4.5\mu$m ($\Delta m < 0.03$).

\begin{figure*}
\epsscale{1.03}
\plotone{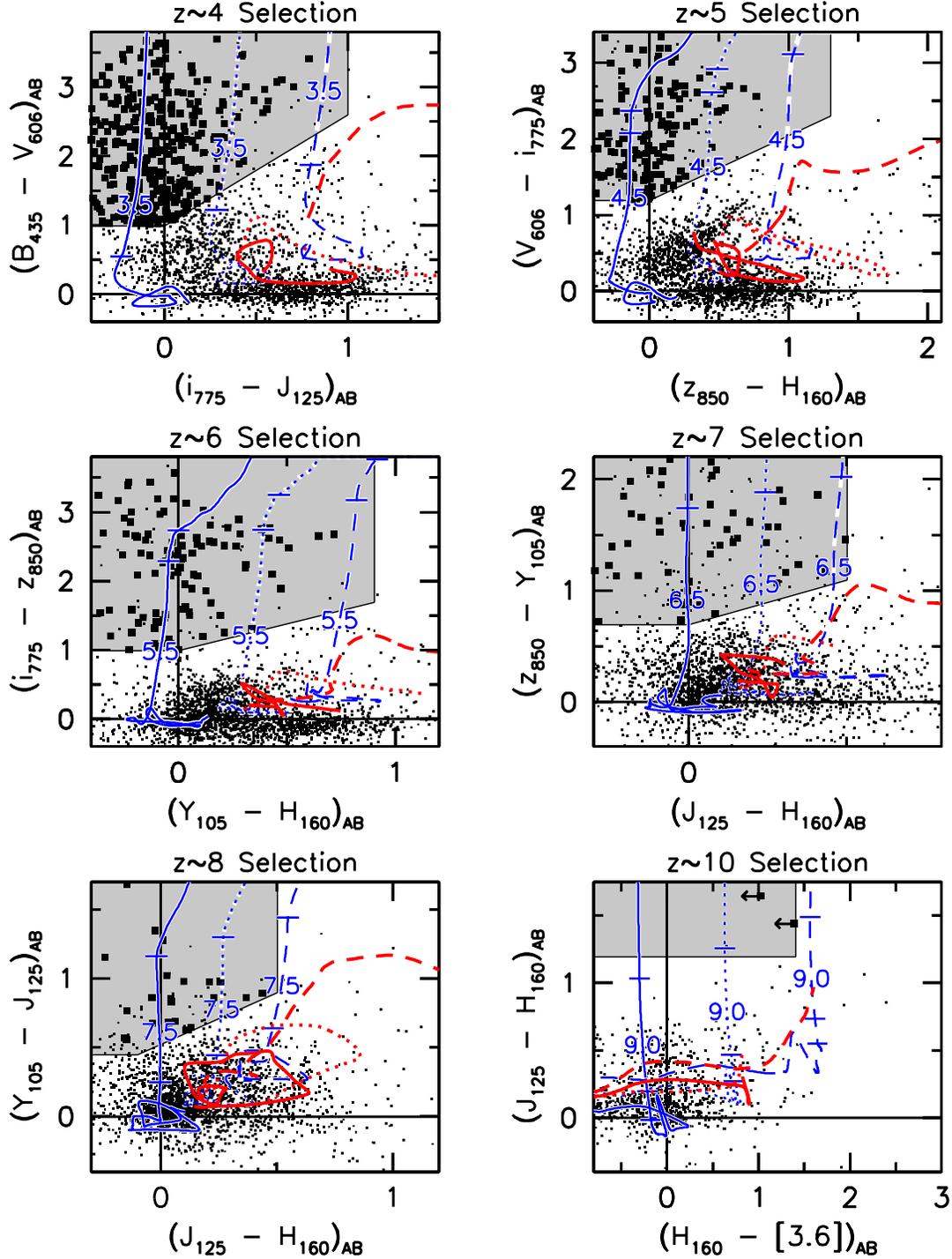}
\caption{Color-color selection criteria that we use to identify
  star-forming galaxies at $z\sim4$, $z\sim5$, $z\sim6$, $z\sim7$,
  $z\sim8$, and $z\sim10$ over the XDF, HUDF09-1, HUDF09-2,
  CANDELS-GN, and CANDELS-GS field (\S3.2.2).  The gray-shaded regions
  show the regions in color-color space where we select sources.  The
  solid, dashed, and dotted blue lines show the expected colors we
  would expect star-forming galaxies to have as a function of
  redshift, for $UV$-continuum slopes $\beta$ of $-2.3$, $-1.15$, and
  0, respectively (with hashes shown every $\Delta z = 0.5)$.  The red
  lines show the colors we would expect for various lower-redshift
  contaminants (using the SEDs from Coleman et al.\ 1980), again as a
  function of redshift.  The black dots show the colors of individual
  sources found in the XDF, while the large black squares indicate the
  colors of sources from the XDF identified as part of the relevant
  high-redshift selection.  The arrows indicate the $1\sigma$ upper
  limits on the $H_{160}-[3.6]$ colors for two $z\sim10$ candidates
  from the XDF.  Our criteria make use of the color formed from the
  two bands straddling the targeted Lyman Break and the color that
  best constrains the spectral slope redward of the break.  The
  criteria allow us to identify a relatively complete selection of
  star-forming galaxies at $z\gtrsim 3.3$, $z\gtrsim 4.5$, $z\gtrsim
  5.5$, $z\gtrsim6.4$, and $z\gtrsim 7.3$, and $z\gtrsim 9.5$.  To
  ensure a good redshift separation between these samples, we impose
  an upper redshift cut-off to each sample by also requiring that
  sources \textit{not} satisfy the selection criteria of the sample
  just above it in redshift.  In addition to the two-color criteria
  shown here, we also require that sources be undetected in the
  available HST observations blueward of the break, both on a
  passband-by-passband basis and in terms of a $\chi^2$ stack of all
  the fluxes blueward of the break (\S3.2.2).\label{fig:sel1}}
\end{figure*}

\begin{deluxetable*}{cccc}
\tablewidth{0cm}
\tabletypesize{\footnotesize}
\tablecaption{Criteria Utilized in Selecting our $z\sim4$-10 Samples\tablenotemark{*}\label{tab:selcrit}}
\tablehead{
\colhead{} & \multicolumn{3}{c}{Data Set} \\
\colhead{Sample} & \colhead{XDF, HUDF09-Ps} & \colhead{} & \colhead{CANDELS-UDS}\\
\colhead{$<$$z$$>$} & \colhead{CANDELS-GS+GN} & \colhead{ERS, BoRG/HIPPIES\tablenotemark{$\dagger$}} & \colhead{COSMOS,EGS}} \\
\startdata
4 & $(B_{435}$$-$$V_{606}$$>$1)$\wedge$($i_{775}$$-$$J_{125}$$<$1)$\wedge$ & $(B_{435}$$-$$V_{606}$$>$1)$\wedge$($i_{775}$$-$$J_{125}$$<$1)$\wedge$ & \\
  & $(B_{435}$$-$$V_{606}$$>$1.6($i_{775}$$-$$J_{125}$)+1)$\wedge$ & $(B_{435}$$-$$V_{606}$$>$1.6($i_{775}$$-$$J_{125}$)+1)$\wedge$ & \\
  & [not in $z\sim5$ selection] & [not in $z\sim5$ selection]\\\\
5 & $(V_{606}$$-$$i_{775}$$>$1.2)$\wedge$$(z_{850}$$-$$H_{160}$$<$1.3)$\wedge$ & $(V_{606}$$-$$i_{775}$$>$1.2)$\wedge$$(z_{850}$$-$$H_{160}$$<$1.3)$\wedge$ & $((V_{606}$$-$$I_{814}$$>$1.3)$\wedge$$(I_{814}$$-$$H_{160}$$<$1.25)$\wedge$\\
  & $(V_{606}$$-$$i_{775}$$>$0.8($z_{850}$$-$$H_{160})$$+$1.2)$\wedge$ & $(V_{606}$$-$$i_{775}$$>$0.8($z_{850}$$-$$H_{160})$$+$1.2)$\wedge$ & $(V_{606}$$-$$I_{814}$$>$$0.72(I_{814}$$-$$H_{160})$$+$$1.3)$$\wedge$ \\
  & [$z\sim5$ non-detection criterion]\tablenotemark{a}$\wedge$ & [$z\sim5$ non-detection criterion]\tablenotemark{a}$\wedge$ & $(f_{u}$$/$$ef_{u}$$<$$2.5)$$\wedge$ \\
  & [not in $z\sim6$ selection] & [not in $z\sim6$ selection] & ($4.2$$<$$z_{phot}$$<$5.5)$\wedge$($J_{125}$$<$$26.7$))$\vee$\tablenotemark{c}\\
  & & & [other LBGs with $4.2$$<$$z_{phot}$$<$5.5]\tablenotemark{b}\\\\
6 & $(i_{775}$$-$$z_{850}$$>$1.0)$\wedge$($Y_{105}$$-$$H_{160}$$<$1.0)$\wedge$ & $(i_{775}$$-$$z_{850}$$>$1.0)$\wedge$($Y_{098}$$-$$H_{160}$$<$1.0)$\wedge$ & ($I_{814}$$-$$J_{125}$$>$0.8)$\wedge$$(J_{125}$$-$$H_{160}$$<$0.4)$\wedge$ \\
  & $(i_{775}$$-$$z_{850}$$>$0.78($Y_{105}$$-$$H_{160}$)$+$1.0)$\wedge$ 
  & $(i_{775}$$-$$z_{850}$$>$0.6($Y_{098}$$-$$H_{160}$)$+$1.0)$\wedge$ 
  & ($I_{814}$$-$$J_{125}$$>$2($J_{125}$$-$$H_{160}$)$+$0.8)$\wedge$ \\
  & [$z\sim6$ non-detection criterion]\tablenotemark{a}$\wedge$ & [$z\sim6$ non-detection criterion]\tablenotemark{a}$\wedge$ & $(f_{ubg}$$/$$ef_{ubg}$$<$ 2.5)$\wedge$ \\
  & [not in $z\sim7$ selection] & [not in $z\sim7$ selection] & ($5.5$$<$$z_{phot}$$<$6.3)$\wedge$($J_{125}$$<$$26.7$))$\vee$\tablenotemark{c} \\
  & & & [other LBGs with $5.5$$<$$z_{phot}$$<$6.3]\tablenotemark{b}\\\\
7 & $(z_{850}$$-$$Y_{105}$$>$0.7)$\wedge$$(J_{125}$$-$$H_{160}$$<$0.45)$\wedge$ & $(z_{850}$$-$$Y_{098}$$>$1.3)$\wedge$ $(J_{125}$$-$$H_{160}$$<$0.5)$\wedge$ & $(I_{814}$$-$$J_{125}$$>$2.2)$\wedge$$(J_{125}$$-$$H_{160}$$<$0.4)$\wedge$\\
  & $(z_{850}$$-$$Y_{105}$$>$0.8($J_{125}$$-$$H_{160}$)+0.7)$\wedge$ & $(z_{850}$$-$$J_{125}$$>$0.8($J_{125}$$-$$H_{160}$)$+$0.7)$\wedge$ & $(I_{814}$$-$$J_{125}$$>$2($J_{125}$$-$$H_{160}$)$+$2.2)$\wedge$\\
  & $((I_{814}$$-$$J_{125}$$>$1.0)$\vee$$(SN(I_{814})$$<$1.5))$\wedge$ & $((I_{814}$$-$$J_{125}$$>$1.0)$\vee$$(SN(I_{814})$$<$1.5))$\wedge$ & $(f_{ubgvri}$$/$$ef_{ubgvri}$$<$2.5)$\wedge$ \\
  & [$z\sim7$ non-detection criterion]\tablenotemark{a}$\wedge$ & [$z\sim7$ non-detection criterion]\tablenotemark{a}$\wedge$ & $(6.3 < z_{phot} < 7.3) \wedge$ \\
  & [not in $z\sim8$ selection] & [not in $z\sim8$ selection] & ($J_{125,AB}$$<$26.7)$\vee$\tablenotemark{c} \\
  & & & [other LBGs with $6.3$$<$$z_{phot}$$<$7.3]\tablenotemark{b}\\\\
8 & ($Y_{105}$$-$$J_{125}$$>$0.45)$\wedge$$(J_{125}$$-$$H_{160}$$<$0.5)$\wedge$ 
  & $(Y_{098}$$-$$J_{125}$$>$1.3)$\wedge$$(J_{125}$$-$$H_{160}$$<$0.5)$\wedge$  
  & $(I_{814}$$-$$J_{125}$$>$2.2)$\wedge$$(J_{125}$$-$$H_{160}$$<$0.4)$\wedge$\\
  & $(Y_{105}$$-$$J_{125}$$>$0.75($J_{125}$$-$$H_{160}$)$+$0.525)$\wedge$ 
  & $(Y_{098}$$-$$J_{125}$$>$0.75($J_{125}$$-$$H_{160}$)$+$1.3)$\wedge$
  & $(I_{814}$$-$$J_{125}$$>$2($J_{125}$$-$$H_{160}$)$+$2.2)$\wedge$\\
  & [$z\sim8$ non-detection criterion]\tablenotemark{a} & [$z\sim8$ non-detection criterion]\tablenotemark{a,d} & $(f_{ubgvri}$$/$$ef_{ubgvri}$$<$2.5)$\wedge$ \\
  & & & (7.3$<$$z_{phot}$$<$9.0)$\wedge$$(H_{160,AB}$$<$26.7)$\vee$\tablenotemark{c} \\
         & & & [other LBGs with $7.3$$<$$z_{phot}$$<$9.0]\tablenotemark{b}\\\\
10 & $(J_{125}$$-$$H_{160}$$>$1.2)$\wedge$ & ($J_{125}$$-$$H_{160}$$>$1.2)$\wedge$ & ($J_{125}$$-$$H_{160}$$>$1.2)$\wedge$ \\
  & $((H_{160}$$-$[3.6]$<$1.4)$\vee$ & (($H_{160}$$-$$[3.6]$$<$1.4)$\vee$ & (($H_{160}$$-$[3.6]$<$1.4) $\vee$ \\
  & $(\textrm{S/N}([3.6])$$<$2))$\wedge$ & $(\textrm{S/N}([3.6])$$<$2))$\wedge$ & $(\textrm{S/N}([3.6])$$<$2))$\wedge$ \\
  & [$z\sim10$ non-detection criterion]\tablenotemark{a} & [$z\sim10$ non-detection criterion]\tablenotemark{a} & $(f_{ubgvriz}$$/$$ef_{ubgvriz}$$<$2.5)$\wedge$ \\
  & & & ($\chi_{V,I}^2$$<$2) \\\\
All & [Stellarity Criterion]\tablenotemark{e} & [Stellarity Criterion]\tablenotemark{e} & [Stellarity Criterion]\tablenotemark{e} \\
  & $(\chi_{Y+J+JH+H}^2 > 25)$\tablenotemark{f} & $(\chi_{Y+J+JH+H}^2 > 25)$\tablenotemark{f} & $(\chi_{Y+J+JH+H}^2 > 25)$\tablenotemark{f}
\enddata
\tablenotetext{*}{Throughout this table, $\wedge$ and $\vee$ represent
  the logical \textbf{AND} and \textbf{OR} symbols, respectively, and SN
  represents the signal to noise.  $\chi^2$ statistic is as defined in
  \S3.2 (see also Bouwens et al.\ 2011b).  In the application of these
  criteria, flux in the dropout band is set equal to the $1\sigma$
  upper limit in cases of a non-detection.}
\tablenotetext{$\dagger$}{The BoRG/HIPPIES data set is only used in searches
  for $z\sim8$ galaxies.}
\tablenotetext{a}{The optical non-detection criteria are as follows:
  $(SN(B)<2)$ [$z\sim5$], $(SN(B)<2)\wedge((V_{606}-z_{850} >
  2.7)\vee(SN(V)<2))$ [$z\sim6$],
  $(SN(B)<2)\wedge(SN(V)<2)\wedge(SN(i)<2)\wedge(\chi_{bvi}^2 < 3)$
  [$z\sim7$],
  $(SN(B)<2)\wedge(SN(V)<2)\wedge(SN(i)<2)\wedge(SN(I)<2)\wedge
  (\chi_{b,v,i,I}^2 < 3)$ [$z\sim8$], and $(\chi_{b,v,i,I,z,Y}^2 < 3)
  \wedge (SN(B)<2)\wedge(SN(V)<2)\wedge (SN(i)<2)\wedge(SN(I)<2)\wedge
  (SN(z)<2)\wedge(SN(Y)<2)$ [$z\sim10$].  For our $z\sim7$-10
  selections, we also require that the optical $\chi^2$ be less than 4
  and 3 in fixed $0.35''$-diameter and 0.2$''$-diameter apertures, respectively.  We
  also impose a stricter optical non-detection criterion for the
  faintest sources in each of our selections (i.e., where the total
  detection significance as defined by $\chi_{Y,J,JH,H}^2<64$).  These
  criteria are $SN(B)<1$ ($z\sim5$), $(SN(B)<1)\wedge
  ((V_{606}-z_{850}>2.3)\vee(\chi_{B,V}^2 <2))$ ($z\sim6$),
  $\chi_{B,V,i}^2 < 2$ ($z\sim7$), and $\chi_{B,V,i,I}^2<2$
  ($z\sim8$).}
\tablenotetext{b}{We also include sources in our $z\sim5$, $z\sim6$,
  $z\sim7$, and $z\sim8$ selections, respectively, if they satisfy any
  of our $z\sim5$, $z\sim6$, and $z\sim7$-8 LBG criteria and the
  photometric redshifts we estimate for the sources are $4.2<z<5.5$,
  $5.5<z<6.3$, $6.3<z<7.3$, and $7.3<z<9.0$, respectively, with a
  total measured magnitude of $J_{125,AB}<26.7$, $J_{125,AB}<26.7$,
  $J_{125,AB}<26.7$, and $H_{160,AB}<26.7$.  See \S3.2.3.}
\tablenotetext{c}{While we select sources to $\sim$26.7 mag, we only
  include sources brightward of 26.5 mag in our LF determinations.}
\tablenotetext{d}{We required sources identified
  within the BoRG/HIPPIES data set to satisfy an even more stringent
  optical non-detection criteria ($SN(V)<1.5$) to effectively exclude
  all low-redshift interlopers from our selection.}
\tablenotetext{e}{We require that the measured stellarity of sources
  (as measured in the detection image) be less than 0.9 to exclude
  stars from our samples (0 = extended source and 1 = point source).
  We also exclude particularly compact sources, with detection-image
  stellarities less than 0.9 if its HST + ground-based + Spitzer
  photometry is significantly better fit with a stellar SED than a
  $z\geq3$ galaxy ($\Delta \chi^2 > 2$) and the measured stellarity in
  either the $J_{125}$ or $H_{160}$ band is at least 0.8.  The
  stellarity requirement is only imposed within 1 magnitude of the
  detection limit of the sample, i.e., 26.5 mag for the CANDELS/Wide
  data sets, 27.0 mag for the CANDELS/DEEP data sets, 28.0 mag for the
  HUDF09-1+HUDF09-2 data sets, and 28.5 mag for the XDF data set.}
\tablenotetext{f}{Even more stringent requirements are made on the
  detection significance of sources in data sets shallower than the
  XDF.  Candidates are required to have a total signal-to-noise in the
  $Y_{105}$, $J_{125}$, $JH_{140}$, and $H_{160}$ bands of 5.5 in the
  HUDF09-Ps and CANDELS data set, and 6.0 in the BoRG/HIPPIES data
  set.  For $z\sim8$ and $z\sim10$ selections, only the
  $J_{125}JH_{140}H_{160}$ and $JH_{140}H_{160}$ fluxes, respectively,
  are used in assessing the detection significance of candidate
  sources.  $z\sim10$ candidates over the CANDELS-UDS/COSMOS/EGS
  fields are required to have a root-mean square S/N of 2.0 in the
  Spitzer/IRAC $3.6\mu$m and $4.5\mu$m imaging to ensure they are real.}
\end{deluxetable*}

\subsection{Source Selection}

\subsubsection{Lyman-Break Selection Criteria}

As in previous work, we construct the bulk of our high-redshift
samples using two color Lyman-break-like criteria.  Substantial
spectroscopic follow-up work has shown that this approach is quite
effective at identifying large samples of star-forming galaxies at
$z\gtrsim 3$ (Steidel et al.\ 1999; Bunker et al.\ 2003; Dow-Hygelund
et al.\ 2007; Popesso et al.\ 2009; Vanzella et al.\ 2009; Stark et
al.\ 2010).

Lyman-break samples typically take advantage of three pieces of
information in identifying probable sources at high redshift: (1)
color information from two adjacent passbands necessary to locate the
position and measure the amplitude of the Lyman break, (2) color
information redward of the break needed to define the intrinsic color
of the source (thereby distinguishing the selected high-redshift
sources from intrinsically-red galaxies), and (3) evidence that
sources show essentially no flux blueward of the spectral break.

Our selection is constructed to take advantage of all three pieces of
information and to do so in a suitably optimal manner, within the
context of simple color criteria.  The most noteworthy gains can be
achieved by taking advantage of the additional wavelength leverage
provided by the deep near-IR and mid-IR observations for constraining
the intrinsic colors of candidate sources.  This allows us to go
beyond what is possible from the Lyman-break-like selection utilized
in Giavalisco et al.\ (2004b) and Bouwens et al.\ (2007).  Obviously,
the color which provides us with the most significant leverage in
probing the intrinsic colors of the sources are those we would use to
provide optimal measurements of the spectral slope $\beta$ (e.g., we
use the same $i_{775}-J_{125}$ color below in constructing our color
criterion for the $z\sim4$ selection as would be optimal for deriving
$\beta$ for $z\sim4$ galaxies: Bouwens et al.\ 2012; Bouwens et
al.\ 2014a).

While one could consider selecting $z\sim4$-10 samples based on the
best-fit photometric redshift or redshift likelihood contours (e.g.,
McLure et al.\ 2010; Finkelstein et al.\ 2012; Bradley et al.\ 2014:
see Figure~\ref{fig:zdist} [right]), Lyman-break selection procedures
can be simpler to apply and offer a slight advantage in terms of
operational transparency.  This makes our selection procedure easier
to reproduce by both theorists and observers, as follow-up studies by
Shimizu et al.\ (2013), Lorenzoni et al.\ (2013), and Schenker et
al.\ (2013) utilizing our color criteria all illustrate.

Despite the present procedural choice, photometric redshift techniques
also work quite well, particularly when used with a well-calibrated
prior or as refinements to the redshift estimate, as direct
comparisons between LF determinations conducted using Lyman-break-like
selection criteria (e.g., Schenker et al.\ 2013) and
photometric-redshift selection criteria (e.g., McLure et al.\ 2013)
illustrate.  Indeed, we will be utilizing photometric redshift
techniques in \S3.2 to redistribute sources across our
CANDELS-UDS/COSMOS/EGS $z\sim5$, $z\sim6$, $z\sim7$, and $z\sim8$
samples based on our best estimate redshifts from the HST +
ground-based + Spitzer/IRAC observations.

\subsubsection{XDF, HUDF09-1, HUDF09-2, CANDELS-GS, CANDELS-GN, ERS, 
BoRG/HIPPIES}

In this section, we describe the selection criteria we employ for data
sets with deep observations in the $Y$-band with HST, i.e., the XDF,
HUDF09-1, HUDF09-2, CANDELS-GS, CANDELS-GN, ERS, and the BoRG/HIPPIES
fields.

We have constructed two-colour selection criteria so that the
lower-redshift boundary is approximately the same for sources
independent of their spectral slope.  For those areas where the
$Y$-band observations are available in the $Y_{105}$-band filter, we
use one set of criteria, while for those areas where the $Y_{098}$ is
available, we employ an alternate set of selection criteria.  The
specific color criteria we have developed are presented in
Table~\ref{tab:selcrit}.

The new color criteria we have developed are not directly comparable
to those previously developed to work with optical/ACS observations
(Giavalisco et al.\ 2004b; Bouwens et al.\ 2007), though we remark
that the $B_{435}-V_{606}$, $V_{606}-i_{775}$, $i_{775}-z_{850}$ color
criteria that we utilize (to identify the existence of a Lyman-break)
are almost identical to previous criteria.  Our color criteria are
most similar in spirit to the $z=4$ criteria previously developed by
Castellano et al.\ (2012), though Castellano et al.\ (2012) use a
$V_{606}-H_{160}$ color to quantify the color of galaxies redward of
the break rather than an $i_{775}-J_{125}$ color.  The advantage of
using the $i_{775}-J_{125}$ colors over the $V_{606}-H_{160}$ colors
is the cleaner measurement it provides of the slope of the
$UV$-continuum for candidate $z\sim4$ galaxies (though the wavelength
leverage it provides is less).

The $z\sim7$ and $z\sim8$ color criteria we utilize here are very
similar to the criteria we had previously applied in Bouwens et
al.\ (2011b) to the HUDF and ERS data sets.  See Figures 2 and 3 and
Figures 6 and 7 from Bouwens et al.\ (2011b).  The $z\sim8$ selection
criteria we employ over the ERS+BoRG+HIPPIES fields utilizes a less
stringent $Y_{098}-J_{125}>1.3$ cut than the $Y_{098}-J_{125}>1.75$
cut utilized in the standard BoRG search (e.g., Bradley et al.\ 2012),
making our selection slightly more susceptible to contamination by
low-mass stars.  However, such sources should be largely excluded by
stellarity criterion we discuss below (see also \S3.5.5).

Finally, our $z\sim10$ selection criteria are identical to those
previously presented by Bouwens et al.\ (2011), Oesch et al.\ (2012a),
and Oesch et al.\ (2014).

In applying these criteria, we set the flux in the dropout band to be
equal to the $1\sigma$ upper limit in cases of a non-detection.

In isolation, the color criteria we present in Table~\ref{tab:selcrit}
would allow for the selection of sources at least one unit higher in
redshift than our desired high-redshift boundaries for these
selections (e.g., our $z\sim4$ selection criteria could allow us to
select sources from $z\sim3.5$ to $z\sim5.5$).  Fortunately, we can
impose a high-redshift boundary for each of our selections by
explicitly requiring that sources \textit{not} satisfy the selection
criteria for the sample just above it in redshift.  This ensures that
our selections are both essentially complete and disjoint from one
another.

To keep contamination from lower redshift sources to a minimum, we
require that sources in our $z\sim5$ and $z\sim6$ selections be
undetected ($<2\sigma$) in $B_{435}$-band imaging data for our fields,
if it is available.  For our $z\sim6$ selections, we require the
$V_{606}-z_{850}$ color to be redder than 2.6 or for sources to be
undetected ($<2\sigma$) in the $V_{606}$-band imaging data (similar to
Bouwens et al.\ 2006).  For our $z\sim7$-10 selections, we calculate
an optical ``$\chi^2$'' for each candidate source (Bouwens et
al.\ 2011), as $\chi_{opt} ^2 = \Sigma_{i} \textrm{SGN}(f_{i})
(f_{i}/\sigma_{i})^2$ where $f_{i}$ is the flux in band $i$ in a
consistent aperture, $\sigma_i$ is the uncertainty in this flux, and
SGN($f_{i}$) is equal to 1 if $f_{i}>0$ and $-1$ if $f_{i}<0$.  The
$B_{435}V_{606}i_{775}$ flux measurements (where available) were used
in calculating $\chi_{opt}^2$ for our $z\sim7$ selections, while the
$B_{435}V_{606}i_{775}I_{814}$ and
$B_{435}V_{606}i_{775}I_{814}z_{850}Y_{105}$ observations were used in
computing $\chi_{opt} ^2$ for our $z\sim8$ and $z\sim10$ selections,
respectively.  $\chi_{opt}^2$ is computed on the basis of the flux
measurements in small-scalable apertures; any candidate with a
measured $\chi_{opt}$ in excess of 3 is excluded from our selections.

For our highest redshift selections, i.e., $z\sim7$-10, we also
computed a $\chi_{opt}^2$ for sources in 0.35$''$-diameter apertures
and especially small $0.2''$-diameter apertures (before PSF smoothing
to preserve S/N) and required sources to be less than 3 and 4
respectively.  An even lower threshold of 2 for $\chi_{opt}^2$ was
used in selecting $z\sim7$-8 sources over the HUDF09-1 field, due to
the lack of $B_{435}$-band observations over that field.  Finally, for
the faintest $z\sim5$-8 candidates in each of our selections with a
coadded significance of the detections in the $Y_{098}$, $Y_{105}$,
$J_{125}$, $JH_{140}$, and $H_{160}$ bands less than 8 (i.e.,
$\chi_{Y,J,JH,H}^2<64$), we used the even more stricter requirements
on the flux in the optical bands listed in footnote a of
Table~\ref{tab:selcrit}.

\begin{figure}
\epsscale{0.7} \plotone{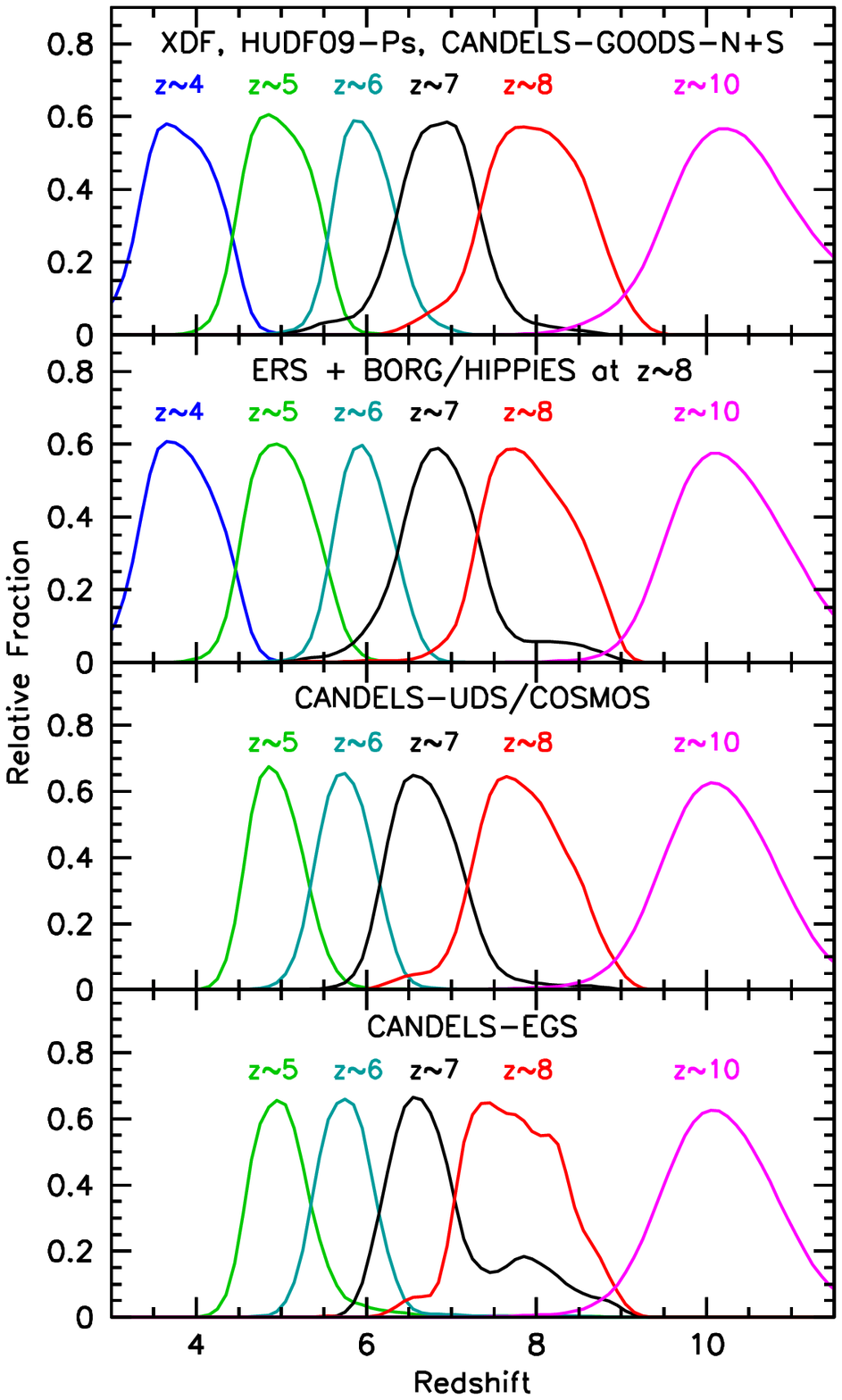}
\caption{The expected redshift distributions for our samples of
  $z\sim4$, $z\sim5$, $z\sim6$, $z\sim7$, $z\sim8$, and $z\sim10$
  galaxies selected from the XDF+HUDF09-Ps+CANDELS-GN+GS fields with
  the $B_{435}V_{606}i_{775}z_{850}Y_{105}J_{125}H_{160}$ filter set
  (\textit{upper panel}: see \S3.2.2 for selection procedure), from
  the ERS data set with the
  $B_{435}V_{606}i_{775}z_{850}Y_{098}J_{125}H_{160}$ filter set
  (\textit{middle panel}: see \S3.2.2 for selection procedure), and
  from the CANDELS-UDS+COSMOS+EGS data set with the
  $V_{606}I_{814}J_{125}H_{160}$ filter set augmented by ground-based
  data (\textit{lower panel: see \S3.2.3 for selection procedure}).
  Each selection window is smoothed by a normal distribution with
  scatter $\sigma_z \sim 0.2$.  We derived the redshift distributions
  for the $z\sim4$-10 samples shown in all four panels using the full
  end-to-end Monte simulations described in \S4.1 (the redshift
  distribution for the faintest sources from the CANDELS
  UDS/COSMOS/EGS fields [i.e., within $\sim$0.5 mag of the limit] have
  a width that is only $\sigma_{z}$$\sim$0.1 greater than what is
  shown here.)  For sources in the CANDELS EGS data set, the
  Spitzer/IRAC photometry is used to help discriminate between
  $z\sim7$ and $z\sim8$ galaxies (as $z<7$ galaxies are known to have
  bluer $3.6\mu$-$4.5\mu$m colors than $z>7$ given the strong high EW
  of [OIII]+H$\beta$: Labb{\'e} et al.\ 2013; Stark et al.\ 2013; Smit
  et al.\ 2014; Ono et al.\ 2012; Finkelstein et al.\ 2013; Laporte et
  al.\ 2014).  The redshift distribution for the $z\sim8$ BoRG/HIPPIES
  samples should be quite similar to our $z\sim8$ ERS samples, but is
  based on the $V_{606}Y_{098}J_{125}H_{160}$ or
  $V_{600}Y_{098}J_{125}H_{160}$ filters alone
  (Table~\ref{tab:selcrit}).
\label{fig:zdistsel}}
\end{figure}

For our deepest field the XDF, sources are required to be detected at
$5\sigma$ in a $\chi^2$ stack of all the HST observations redward of
the break (in a fixed 0.36$''$-diameter aperture).  This is to ensure
source reality.  For sources over the deep HUDF09-1 and HUDF09-2
fields and the wider-area CANDELS and ERS fields, we require sources
be detected at $5.5\sigma$.  For sources over the BoRG/HIPPIES fields,
we require sources to be detected at $6\sigma$.  Our use of more
stringent criteria for our shallower fields is quite reasonable, given
the much smaller number of exposures in these data and therefore noise
that is less Gaussian in its characteristics (e.g., see Schmidt et
al.\ 2014).\footnote{While we could increase the total number of
  sources in our selections somewhat by searching for sources at lower
  significance levels, these sources are not of substantial value for
  current LF determinations, given the considerable uncertainties in
  correcting for both the incompleteness and contamination expected
  for such samples.}

\begin{deluxetable*}{ccccccc}
\tablewidth{0pt}
\tablecolumns{11}
\tabletypesize{\footnotesize}
\tablecaption{A complete list of the sources included in our $z\sim 4$, $z\sim 5$, $z\sim 6$, $z\sim 7$, $z\sim 8$, and $z\sim 10$ samples.\tablenotemark{*}\label{tab:catalog}}
\tablehead{
\colhead{ID} & \colhead{R.A.} & \colhead{Dec} & \colhead{$m_{AB}$} & \colhead{Sample\tablenotemark{a}} & \colhead{Data Set\tablenotemark{b}} & \colhead{$z_{phot}$\tablenotemark{c,d}}}
\startdata
XDFB-2384848214 & 03:32:38.49 & $-$27:48:21.4 & 27.77 & 4 & 1 &  3.49 \\
XDFB-2384248186 & 03:32:38.42 & $-$27:48:18.7 & 29.18 & 4 & 1 &  3.82 \\
XDFB-2376648168 & 03:32:37.66 & $-$27:48:16.9 & 28.61 & 4 & 1 &  4.01 \\
XDFB-2385948162 & 03:32:38.60 & $-$27:48:16.2 & 28.04 & 4 & 1 &  4.16 \\
XDFB-2382548139 & 03:32:38.26 & $-$27:48:13.9 & 28.18 & 4 & 1 &  4.37 \\
XDFB-2394448134 & 03:32:39.45 & $-$27:48:13.4 & 26.40 & 4 & 1 &  3.58 \\
XDFB-2381448127 & 03:32:38.14 & $-$27:48:12.7 & 28.58 & 4 & 1 &  3.68 \\
XDFB-2390248129 & 03:32:39.03 & $-$27:48:13.0 & 27.99 & 4 & 1 &  3.91 \\
XDFB-2379348121 & 03:32:37.93 & $-$27:48:12.1 & 27.45 & 4 & 1 &  4.11 \\
XDFB-2378848108 & 03:32:37.88 & $-$27:48:10.9 & 30.13 & 4 & 1 &  3.72 
\enddata
\tablenotetext{*}{Table~\ref{tab:catalog} is published in its entirety
  in the electronic edition of the Astrophysical Journal.  A portion
  is shown here for guidance regarding its form and content.}
\tablenotetext{a}{The mean redshift of the sample in which the source
  was included for the purposes of deriving LFs.}
\tablenotetext{b}{The data set from which the source was selected: 1 =
  HUDF/XDF, 2 = HUDF09-1, 3 = HUDF09-2, 4 = ERS, 5 = CANDELS-GS, 6 =
  CANDELS-GN, 7 = CANDELS-UDS, 8 = CANDELS-COSMOS, 9 = CANDELS-EGS,
  and 10 = BoRG/HIPPIES or other pure-parallel programs.}
\tablenotetext{c}{Most likely redshift in the range $z=2.5$-11 as derived 
  using the EAZY photometric redshift code (Brammer et al.\ 2008) using 
  the same templates as discussed in \S3.2.3.}
\tablenotetext{d}{``*'' indicates that for a flat redshift prior, the
  EAZY photometric redshift code (Brammer et al.\ 2008) estimates that
  this source shows at least a 68\% probability for having a redshift
  significantly lower than the nominal low-redshift limit for a
  sample, i.e., $z<2.5$, $z<3.5$, $z<4.4$, $z<5.4$, $z<6.3$, and $z<8$
  for candidate $z\sim4$, $z\sim5$, $z\sim6$, $z\sim7$, $z\sim8$, and
  $z\sim10$ galaxies, respectively.}
\end{deluxetable*}

For sources that are at least 1 magnitude brightward of the nominal
detection limit for our samples (i.e., 26.5 mag for the CANDELS/Wide
data sets, 27.0 mag for the CANDELS/DEEP data sets, 28.0 mag for the
HUDF09-1+HUDF09-2 data sets, and 28.5 mag for the XDF data set), the
SExtractor stellarity parameter for sources (from SExtractor detection
image) is required to be less than 0.9 (where 0 corresponds to very
extended sources and 1 corresponds to point sources).  We also exclude
particularly compact sources, with measured stellarities (from
detection image) greater than 0.5 if its HST photometry was
significantly better fit to a stellar SED than a $z\geq3$ galaxy
($\Delta \chi^2 > 2$) and if the measured stellarity in either the
$J_{125}$ or $H_{160}$ image is greater than 0.8.  The templates we
use for our stellar SED fits are from the SpeX prism library of
low-mass stars (Burgasser et al.\ 2004) extended to $5\mu$m using the
derived spectral types and the known $J$-[3.6] or $J$-[4.5] colors of
these spectral types (Patten et al.\ 2006; Kirkpatrick et al.\ 2011).

Finally, a careful visual inspection was performed on all of the
candidate $z\sim4$-10 galaxies that otherwise satisfy our selection
criteria to exclude obvious artifacts (e.g., diffraction spikes,
spurious ``sources'' on the wings of ellipticals) or any sources that
seemed likely to be associated with bright foreground
sources.\footnote{We note the exclusion of two bright
  ($H_{160,AB}\sim25$) $z\sim8$ candidates identified over the
  BoRG/HIPPIES data set from our selection as a result of these
  concerns (at positions $\alpha, \delta$ 22:02:50.00, 18:51:00.2 and
  08:35:13.13, 24:55:38.1).}  We also verified that none of the
sources in our selection were previously included in the catalogs of
candidate low-mass stars from Holwerda et al.\ (2014a) or were
associated with SNe identified during the CANDELS observations (Rodney
et al.\ 2014).

\subsubsection{CANDELS-UDS, CANDELS-COSMOS, CANDELS-EGS Fields}

Because of the lack of deep HST imaging in $B_{435}$, $z_{850}$, or
$Y_{098}$/$Y_{105}$-band over the CANDELS-UDS, CANDELS-COSMOS, and
CANDELS-EGS fields (Table~\ref{tab:obsdata} and
Figure~\ref{fig:depth}), it is not possible to select $z\sim5$-8
galaxies over those fields using the same color criteria as we
utilized over our primary search fields (i.e., the XDF, CANDELS-GN,
and CANDELS-GS).

Our procedure for selecting our samples of $z\sim5$, $z\sim6$,
$z\sim7$ and $z\sim8$ galaxies over these is therefore more involved
and makes significant use of the ground-based observations.  We
describe our procedure in the paragraphs that follow.  The first step
was to identify all those sources that plausibly corresponded to
star-forming galaxies at $z\sim5$-8 through the systematic selection
of Lyman-break-like galaxies at $z\sim5$, $z\sim6$, and $z\sim7$-8.
The criteria we used to do this preselection is presented in Appendix
B.

In the second step, we obtained photometry on each of these sources in
deep ground-based Subaru+CFHT+VLT+VISTA + Spitzer/IRAC observations
that are available over our search fields.  We then used the
\textit{EAZY} photometric redshift code (Brammer et al.\ 2008) to
estimate redshifts for all the sources.  The photometry utilized in
deriving the photometric redshifts included flux measurements from the
HST $V_{606}I_{814}J_{125}JH_{140}H_{160}$ + Subaru-SuprimeCam
$BgVriz$ + CFHT/Megacam $ugriyz$ + UltraVISTA $YJHK_s$ data sets for
the CANDELS COSMOS field, HST $V_{606}I_{814}J_{125}JH_{140}H_{160}$ +
Subaru-SuprimeCam $BVriz$ + CFHT/Megacam $u$ + UKIRT/WFCAM $K_s$ +
VLT/HAWKI/HUGS $YK_s$ data sets for the CANDELS UDS field, and the HST
$V_{606}I_{814}J_{125}JH_{140}H_{160}$ + CFHT/Megacam $ugriyz$ +
CFHT/WIRCam $K_s$ + Spitzer/IRAC 3.6$\mu$m+4.5$\mu$m data sets for the
CANDELS EGS field.  No consideration of the Spitzer/IRAC photometry is
made for sources over the CANDELS-UDS and CANDELS-COSMOS fields due to
the availability of deep $Y$-band observations to distinguish $z\sim7$
sources from $z\sim8$ sources.\footnote{In addition, this exclusion of
  the Spitzer/IRAC data in this selection allowed us to avoid
  introducing any coupling between redshift and the Spitzer/IRAC
  properties of our sources for future analyses.}

Sources with photometric redshifts in the range $z=4.2$-5.5,
$z=5.5$-6.3, $z=6.3$-7.3, and $z=7.3$-9.0 were tentatively assigned to
our $z\sim5$, $z\sim6$, $z\sim7$, and $z\sim8$ selections,
respectively.  These redshift ranges were chosen to ensure a good
match with the mean redshifts for the color selections defined in
\S3.2.2.  Our photometric redshift fitting is conducted using the
EAZY\_v1.0 template set supplemented by SED templates from the Galaxy
Evolutionary Synthesis Models (GALEV: Kotulla et al.\ 2009).  Nebular
continuum and emission lines were added to the later templates using
the Anders \& Fritze-v. Alvensleben (2003) prescription, a $0.2
Z_{\odot}$ metallicity, and a rest-frame EW for H$\alpha$ of
1300\AA.\footnote{While the rest-frame EW we assume for $H\alpha$ for
  our adapted GALEV templates is larger than the $\sim$500-600$\AA$ EW
  typical for many $z\gtrsim5$ galaxies (e.g., Shim et al.\ 2011;
  Stark et al.\ 2013), such templates have been included to give the
  EAZY photometric code (which can consider arbitrary linear
  combinations of SED templates) the flexibility to accurately model
  the SEDs of galaxies with very strong line emission.  These
  templates effectively counterbalance our use of the standard
  template set, where the impact of line emission is minimal.}

We only included galaxies in our $z\sim6$, $z\sim7$, and $z\sim8$
samples brightward of $J_{125}=26.7$ mag ($z=6$-7) and $H_{160}=26.7$
mag ($z=8$) in our samples as a whole.  However, only sources
brightward of 26.5 mag are used in our LF determinations (\S4).  This
was to ensure good redshift separation, given the limited depth of the
$I_{814}$-band observations and ground-based $z$ and $Y$-band
observations (Figure~\ref{fig:depth}).  As we demonstrate with the
simulations in \S4.1 (illustrated in Figure~\ref{fig:zdistsel}), we
can effectively split sources into different redshift subsamples to
26.5 mag.

To ensure that each of these candidate $z\sim5$-8 galaxies was robust,
we required that each of these sources show $<$2.5$\sigma$ detection
blueward of the break.  To this end, inverse-variance-weighted fluxes
were derived for each source blueward of the Lyman break.  Included in
the inverse-variance-weighted measurements for the samples in brackets
below were the CFHT Megacam $u$ and Subaru Suprime-Cam $B$
[CANDELS-UDS $z\sim5$], CFHT $u$ and Subaru $B$ [CANDELS-UDS
  $z\sim6$], CFHT $u$ and Subaru $BVr$ [CANDELS-UDS $z\sim7$], CFHT
$u$ and Subaru $BVri$ [CANDELS-UDS $z\sim8$], CFHT MegaCam $u$ [COSMOS
  $z\sim5$], Subaru $Bg$ and CFHT $ug$ [COSMOS $z\sim6$], Subaru
$BgVr$ and CFHT $ugr$ [COSMOS $z\sim7$], Subaru $BgVri$ and CFHT
$ugriy$ [COSMOS $z\sim8$], CFHT $u$ [EGS $z\sim5$], CFHT $ug$ [EGS
  $z\sim6$], CFHT $ugr$ [EGS $z\sim7$], and CFHT $ugriy$ [EGS
  $z\sim8$] flux measurements, respectively.  We also excluded
candidate $z\sim6$, $z\sim7$, and $z\sim8$ galaxies from our selection
where flux in the HST $V_{606}$, $V_{606}$, and $V_{606}+I_{814}$
bands was greater than 1.5$\sigma$.  Exclusion of sources with
detections blueward of the break only had a modest effect on the size
of the $z\sim5$, $z\sim6$, $z\sim 7$, and $z\sim8$ samples we derived
from the wide-area CANDELS fields (removing 2\%, 7\%, 8\%, and 21\% of
the sources from the $z\sim5$, $z\sim6$, $z\sim7$, and $z\sim8$
samples, respectively).\footnote{We also note the exclusion of a
  $z\sim5$ candidate at 10:00:13.93, 2:22:14.9, due to its showing far
  too much flux in the ground-based $B$ and $g$-band data (3-$4\sigma$
  discrepancy in both cases) to be a robust $z\sim5$ candidate.}

We used a similar strategy for excluding stars from our CANDELS-UDS,
COSMOS, and EGS fields, as what we utilize for selections over the
XDF, HUDF09-Ps, ERS, CANDELS-GN+GS, and the BoRG/HIPPIES fields
(\S3.2.2).  The only procedural difference with the present fields is
that we also make use of the ground-based + Spitzer/IRAC photometry we
obtain for sources in ascertaining whether their SEDs are more
consistent with that of a star or a $z\sim5$-8 galaxy.  Using
simulations where we added point-like sources to the real data with
input fluxes taken from random stars in the SpeX prism library of
late-type stars (Burgasser et al.\ 2004), we found that our schema was
successful at excluding 97\%, 97\%, and 94\% of $H_{160,AB}=26.0$-26.5
stars from our selection over the CANDELS-UDS, CANDELS-COSMOS, and
CANDELS-EGS fields, respectively (with late L and early T type stars
being the most challenging to exclude).

As a check on the fidelity of our $z\sim5$-8 samples, we also derived
fluxes for sources in our samples in larger $1.8''$-diameter apertures
than we used for our fiducial selection.  Coadding the fluxes of
sources blueward of the break while weighting by the inverse variance,
we found that 94\% of the sources in our samples remain undetected at
$<2.5\sigma$ even in larger $1.8''$-diameter aperture.  To interpret
these findings, we repeated this experiment on the mock images we
created in \S4.1 and found similar incompleteness levels, strongly
arguing that the slight detection rate we found for our $z\sim5$-8
samples in the larger apertures could be explained as resulting from
noise and imperfectly subtracted nearby neighbors.\footnote{To check
  the robustness of our flux measurements in the $Y$-band to the size
  of the high-redshift sources, we derived $Y$-band fluxes for the
  brightest $H<25.5$ sources in $1.8''$-diameter apertures for
  comparison with our smaller-aperture measurements.  Encouragingly
  enough, the $Y$-band fluxes we recovered were completely consistent
  (3$\pm$5\% lower) using the wider apertures as using our fiducial
  $1.2''$-diameter apertures.  This is not surprising, since
  \textsc{mophongo} accounts for the expected profile of the source in
  the ground-based observations in correcting the aperture
  measurements to total.}

We further stacked the optical $V_{606}$-band observations (blueward
of the break for $z\geq7$ galaxies) for all 107 $z\sim7$ and $z\sim 8$
candidates from the wide-area fields and found no detection
($<1\sigma$).  Similar stack results in the $I_{814}$ band for our
CANDELS-UDS/COSMOS/EGS $z\sim8$ samples yielded no detection.

Our selection of $z\sim10$ galaxies over these fields is very similar
to our selection of $z\sim10$ galaxies from HST fields with $Y$-band
imaging (\S3.2.2).  Again, we require that sources satisfy a
$J_{125}-H_{160}>1.2$ color cut, show a $6\sigma$ detection in the
$H_{160}$-band, be undetected in a stack of the optical/ACS data
($\chi_{opt}^2 < 3$), and also be detected at $\geq6\sigma$ in the
$H_{160}$ band.  However, we also require sources remain undetected
($<2\sigma$) in whatever $Y$-band observations were available over our
search fields (i.e., from HAWK-I and VISTA over the CANDELS-UDS and
CANDELS-COSMOS fields, respectively), that sources also remain
undetected ($<2.5\sigma$) in a stack of the optical ground-based
Subaru+CFHT observations available over each field, and that sources
be detected at $>2\sigma$ in the available $3.6\mu$m+$4.5\mu$m IRAC
imaging over the CANDELS fields from the SEDS program (Ashby et
al.\ 2013) to ensure source reality.

\begin{deluxetable*}{lrrrrrrr}
\tablewidth{0pt}
\tablecolumns{11}
\tabletypesize{\footnotesize}
\tablecaption{Total number of sources in our $z\sim4$, $z\sim5$, $z\sim6$, $z\sim7$, $z\sim8$, and $z\sim10$ samples used in deriving the present high-redshift LFs.\label{tab:sampnumbers}}
\tablehead{
\colhead{} & \colhead{Area} &
\colhead{$z\sim4$} & \colhead{$z\sim5$} & \colhead{$z\sim6$} & \colhead{$z\sim7$} & \colhead{$z\sim8$} & \colhead{$z\sim10$}\\
\colhead{Field} & \colhead{(arcmin$^2$)} &
\colhead{\#} & \colhead{\#} & \colhead{\#} & \colhead{\#} & \colhead{\#} & \colhead{\#}}
\startdata
HUDF/XDF & 4.7  & 357 & 153 & 97 & 57 & 30 & 2\\
HUDF09-1 & 4.7  & --- & 91 & 38 & 22 & 18 & 0\\
HUDF09-2 & 4.7  & 147 & 77 & 32 & 23 & 17 & 0\\
CANDELS-GS-DEEP & 64.5  & 1590 & 471 & 198 & 77 & 27 & 1\\
CANDELS-GS-WIDE & 34.2  & 451 & 117 & 43 & 5 & 3 & 0\\
ERS & 40.5  & 815 & 205 & 61 & 47 & 6 & 0\\
CANDELS-GN-DEEP & 68.3  & 1628 & 634 & 188 & 134 & 51 & 2\\
CANDELS-GN-WIDE & 65.4  & 871 & 282 & 69 & 39 & 18 & 1\\
CANDELS-UDS & 151.2  & --- & 270 & 33 & 18 & 6 & 0\\
CANDELS-COSMOS & 151.9  & --- & 320 & 48 & 15 & 9 & 0\\
CANDELS-EGS & 150.7  & --- & 381 & 50 & 44 & 9 & 0\\
BORG/HIPPIES & 218.3  & --- & --- & --- & --- & 23 & 0\\
Total & 959.1 & 5859 & 3001 & 857 & 481 & 217 & 6
\enddata
\end{deluxetable*}

\begin{figure*}
\epsscale{1.14}
\plotone{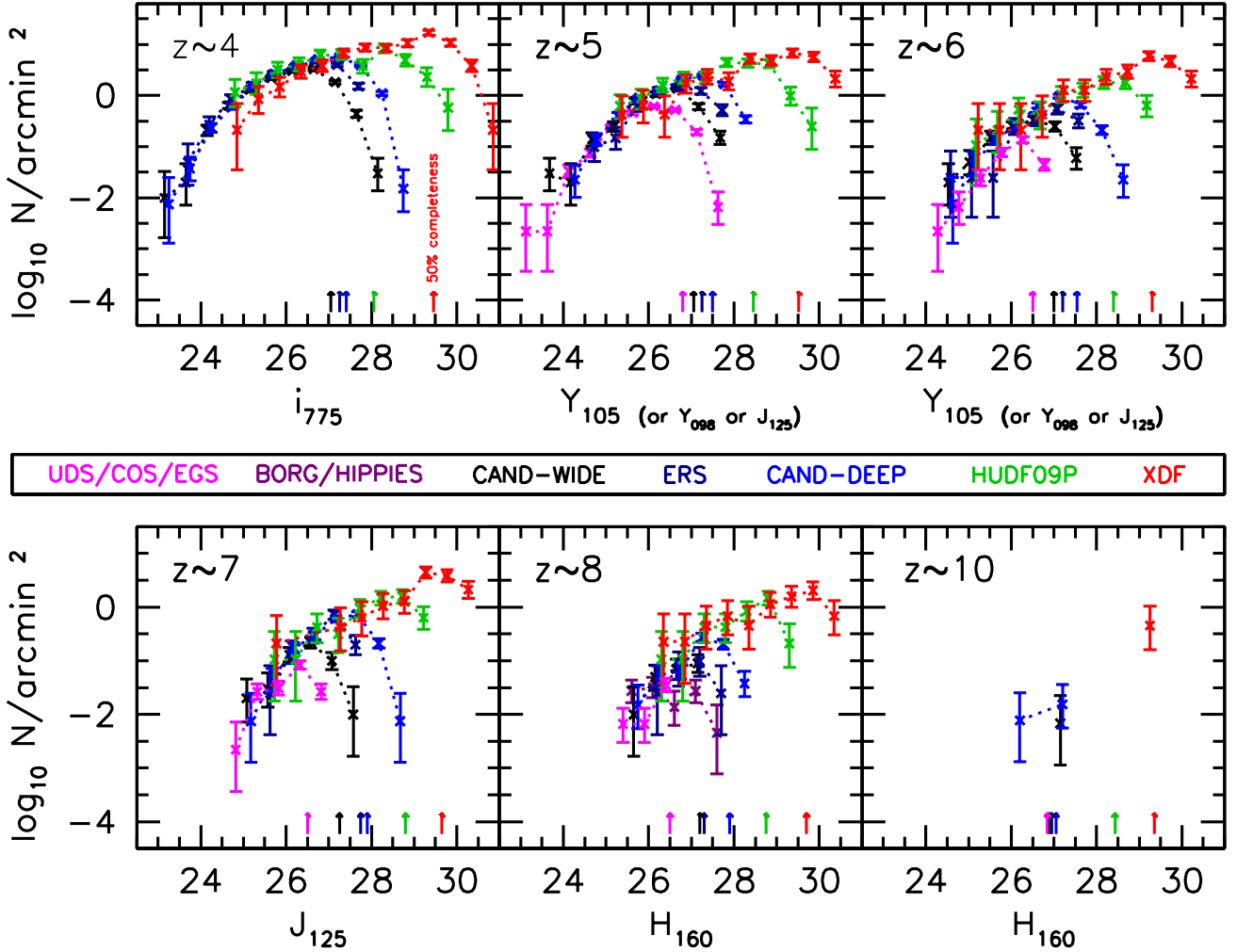}
\caption{Surface densities of candidate $z\sim4$, $z\sim5$, $z\sim6$,
  $z\sim7$, $z\sim8$, and $z\sim10$ galaxies for all the search fields
  considered in this analysis.  Shown are the results from the
  CANDELS-UDS/CANDELS-COSMOS/CANDELS-EGS fields (\textit{magenta
    points}), BoRG/HIPPIES (\textit{dark violet}), CANDELS-GN-WIDE and
  CANDELS-GS-WIDE (\textit{black points}), CANDELS-GN-DEEP and
  CANDELS-GS-DEEP (\textit{blue points}), HUDF09-1 and HUDF09-2 fields
  (\textit{green points}), and the XDF data set (\textit{red points}).
  Surface densities are presented as a function of the $i_{775}$,
  $Y_{105}$, $Y_{105}$, $J_{125}$, $H_{160}$, and $H_{160}$ band
  magnitudes that provide the best measure of the rest-frame $UV$ flux
  of galaxies at 1600$\AA$ for our $z\sim4$, $z\sim5$, $z\sim6$,
  $z\sim7$, $z\sim8$, and $z\sim10$ selections, respectively.  Surface
  densities for our $z\sim5$ and $z\sim6$ selections over the ERS and
  CANDELS-UDS/CANDELS-COSMOS/CANDELS-EGS fields are presented as a
  function of the $Y_{098}$ and $J_{125}$-band fluxes, respectively,
  due to the lack of deep $Y_{105}$-band coverage of these fields.
  The points have been offset horizontally from each other for clarity.
  The available HST + ground-based + Spitzer/IRAC observations allow
  for the selection of $z\sim5$, $z\sim6$, $z\sim7$, $z\sim8$, and
  $z\sim10$ galaxies from the wide-area CANDELS-UDS, CANDELS-COSMOS,
  and CANDELS-EGS fields.  The HST observations available over the
  BoRG/HIPPIES search fields are only particularly effective for
  selecting candidate $z\sim8$ galaxies.  The upward arrows at the
  bottom of each panel indicate the approximate magnitude where the
  efficiency of selecting galaxies at a specific redshift from some
  data set is just 50\% of the maximum efficiency.  The onset of
  incompleteness in our different samples is clearly seen in the
  observed decrease in surface density of sources near the magnitude
  limit.  With our selection volume estimates, we can correct for the
  increased incompleteness at fainter magnitudes.  We do not make use
  of the faintest sources in each search field, due to the large
  uncertainties in the completeness (and contamination) corrections.
  Table~\ref{tab:surfdens} from Appendix C provides these surface
  densities in tabular form.
\label{fig:surfdens}}
\end{figure*}

\subsection{Selection Results}

Applying the selection criteria from \S3.2 to XDF, HUDF09-Ps, ERS,
BoRG/HIPPIES, and CANDELS data sets results in 5859, 3001, 857, 481,
217, and 6 sources in our $z\sim4$, $z\sim5$, $z\sim6$, $z\sim7$,
$z\sim8$, and $z\sim10$ samples.  Our total $z\sim4$-10 sample
includes $\sim$10400 sources.  The individual number of high-redshift
candidates in each field is provided in Table~\ref{tab:sampnumbers}.

The surface density of galaxies we find in our different redshift
samples is presented in Figure~\ref{fig:surfdens} as a function of
magnitude.  While it is clear that some field-to-field variations
exist in the surface density of galaxies in our different samples
(e.g., $z\sim4$ galaxies in the HUDF seem to be underdense relative to
our other search fields), overall the surface density of galaxies as a
function of magnitude is fairly similar for each of our search fields,
over magnitude ranges where our search is largely complete.  We
discuss field-to-field variations in detail in \S4.5.  In
Table~\ref{tab:surfdens} from Appendix C, we tabulate the average
surface density of galaxies in our different samples as a function of
magnitude.

Our best estimate of the approximate redshift distribution for our
different high-redshift samples is shown in the left panel of
Figure~\ref{fig:zdist} and is based on the simulations we describe in
\S4.1 for the XDF, HUDF09-1, and HUDF09-2 fields.  The mean redshift
for galaxies in our $z\sim4$, $z\sim5$, $z\sim6$, $z\sim7$, and
$z\sim8$ samples is 3.8, 4.9, 5.9, 6.8, and 7.9.  From these
simulations, it is clear that our selection criteria are quite
effective in dividing high-redshift galaxies into discrete redshift
slices.  In the right panel of Figure~\ref{fig:zdist}, we also present
the redshift distributions we derive for our XDF, HUDF09-1, and
HUDF09-2 samples using the photometric redshift code EAZY (Brammer et
al.\ 2008).  Photometric redshifts are estimated based on the HST
photometry (for our $z\sim4$-8 samples) and HST+Spitzer photometry
(for our $z\sim10$ sample).  As is clear from the figure, our simple
color-color selections result in essentially the same subdivision of
sources by redshift, as one would find if one relied on a photometric
redshift code to do the selection.

We include our complete $z\sim4$, $z\sim5$, $z\sim6$, $z\sim7$,
$z\sim8$, and $z\sim10$ catalogs in Table~\ref{tab:catalog}, with
coordinates and rest-frame $UV$ luminosities.  We have also provided
our best estimate redshifts for each of the $z\sim5$-8 candidates we
identified over the CANDELS-UDS/COSMOS/EGS fields.  Photometric
redshift estimates are also provided for $z\sim4$-10 candidates over
XDF, HUDF09-Ps, ERS, CANDELS-GS/GN, and BoRG/HIPPIES by applying the
photometric redshift software EAZY (Brammer et al.\ 2008) and the
template set from \S3.3.2 to the HST photometry we have available for
these candidates.  To improve the accuracy of the photometric redshift
estimates for our $z\sim4$ CANDELS-GN+GS+ERS samples (where the lack
of photometric constraints blueward of the $B_{435}$ band can impact
the results), we have also incorporated the $U$-band photometry of
these candidates from KPNO (Capak et al\ 2004) and VLT/VIMOS (Nonino
et al.\ 2009) using \textsc{mophongo} in fixed 1.2$''$-diameter
apertures.

\subsection{Comparisons with Previous $z\sim4$-10 Samples}

The present compilation of $z\sim4$-10 galaxy candidates from the XDF,
HUDF09-1, HUDF09-2, the ERS, and the five CANDELS fields contains
$\sim$10400 $z\sim4$-10 candidates and is the largest such compilation
obtained to the present based on HST observations.  Previously, the
largest such samples of galaxies found in HST observations were
reported in Bouwens et al.\ (2007: 6714 sources over the range
$z=4$-6) and Bouwens et al.\ (2014a: 4004 sources over the range
$z=4$-8).

A substantial fraction ($\sim$30-70\%) of the sources from the current
catalogs appeared in previous wide-area selections.  2331, 586, and
206 of the $z\sim4$, $z\sim5$, and $z\sim6$ candidates (44\%, 34\%,
and 37\% of this sample, respectively) from our wide-area CANDELS+ERS
selections were previously reported by Bouwens et al.\ (2007).  For
$z\sim7$-8 selections over the CANDELS-GS, 59 and 28 of the $z\sim7$
and $z\sim8$ candidates (19\% and 27\% of this sample, respectively),
were previously reported by Bouwens et al.\ (2011), Oesch et
al.\ (2012), Grazian et al.\ (2012), Yan et al.\ (2012), Lorenzoni et
al.\ (2013), Schenker et al.\ (2013), and McLure et al.\ (2013).  22
of the present $z\sim7$ candidates over the CANDELS-UDS and
CANDELS-EGS fields (35\% of our sample) previously appeared in the
Grazian et al.\ (2012) or McLure et al.\ (2013).  The brightest
$z\sim7$ candidate we find in the CANDELS-UDS field is the well-known
``Himiko'' $z=6.595$ Ly$\alpha$-emitting galaxy previously reported by
Ouchi et al.\ (2009a).  The brightest 3 $z\sim6$ and brightest 2
$z\sim7$ galaxies from our CANDELS-COSMOS catalog were previously
identified by Willott et al.\ (2013) and Bowler et al.\ (2014),
respectively.

11 of the 23 $z\sim8$ candidates we identified over the BoRG/HIPPIES
fields and similar data sets (i.e., 48\%) were previously identified
as $z\sim8$ candidates by Bradley et al.\ (2012), McLure et
al.\ (2013), and Schmidt et al.\ (2014).  The reason our catalogs
include many $z\sim8$ candidates not included in the Bradley et
al.\ (2012) and Schmidt et al.\ (2014) compilation is our use of one
additional data set not previously considered (i.e., a parallel field
outside of Abell 1689) and our selecting sources with slightly weaker
$Y_{098}-J_{125}$ breaks and slightly redder $J_{125}-H_{160}$ colors
(consistent with our $z\sim8$ selection from the ERS data set).  While
excluding these sources may allow Bradley et al.\ (2012) and Schmidt
et al.\ (2014) to identify a marginally cleaner selection of $z\sim8$
galaxies, Bradley et al.\ (2012) and Schmidt et al.\ (2014)
potentially miss a modest fraction of the luminous $z\sim8$ galaxies
over the BoRG/HIPPIES search fields (i.e., those having significantly
redder $J_{125}-H_{160}$ colors than would be selected by their
criteria).\footnote{A good fraction of the brightest $z\sim6$-8
  sources would have $\beta$'s of $-1.6$ (Bouwens et al.\ 2012;
  Finkelstein et al.\ 2012; Willott et al.\ 2013; Bouwens et
  al.\ 2014a), which is redder than would be selected by the Bradley
  et al.\ (2012) and Schmidt et al.\ (2014) criteria.  Our selection
  criteria are effective in identifying $z\sim8$ galaxies with
  $\beta$'s as blue as 0 (corresponding to $J_{125}-H_{160}$ colors of
  0.5).}

For fainter $z\sim4$-8 samples from the XDF, HUDF09-1, and HUDF09-2
data sets, our samples again show very good overlap.  209, 139, 92,
75, and 45 of the present sample of $z\sim4$, $z\sim5$, $z\sim6$,
$z\sim7$, and $z\sim8$ candidates (41\%, 43\%, 55\%, 72\%, and 71\% of
this sample, respectively) were previously reported by Bouwens et
al.\ (2007), Wilkins et al.\ (2010), Bouwens et al.\ (2011), Schenker
et al.\ (2013), and McLure et al.\ (2013).  The reason the current
selection contains many sources that were not previously found by
Bouwens et al.\ (2007) is due to our ability to probe to greater
depths with WFC3/IR than was possible using the HST/ACS optical camera
alone and deeper optical observations now available over the XDF/HUDF
and HUDF09-2 fields.

The present $z>6$ sample is thus far the most comprehensive in the
literature, including some 698 $z\sim7$-8 high-quality candidates
based on all search fields.

The present $z\sim10$ sample contains 6 candidates in total and is
almost identical to the Oesch et al.\ (2014) $z\sim10$ sample, with 1
$z\sim10$ candidate over the CANDELS-GS field, 3 $z\sim10$ candidates
over the CANDELS-GN field, and 2 $z\sim10$ candidates over the XDF
data set.  One of the 6 $z\sim10$ candidates from the present
$z\sim10$ sample (XDFyj-40248004) was classified as a $z\sim9$
candidate in Oesch et al.\ (2013b).  The earlier analyses of Ellis et
al.\ (2013) and Oesch et al.\ (2013b) had only identified one
plausible $z\sim10$ candidate each,\footnote{Oesch et al.\ (2013b)
  demonstrated that one of the two $z\sim9.5$ candidates reported by
  Ellis et al.\ (2013), i.e., HUDF12-4106-7304, is significantly
  boosted by a diffraction spike and therefore cannot be considered a
  reliable candidate.}  while McLure et al.\ (2013) did not identify
any $z\sim10$ candidates over our search fields.\footnote{While we
  would have expected McLure et al.\ (2013) to have identified both of
  the plausible $z\sim9$-10 candidates Oesch et al.\ (2014) identified
  over the CANDELS-GS field, the apertures McLure et al.\ (2013) used
  on these sources could have easily included optical flux from
  neighboring sources (as occurred for Oesch et al.\ 2012a: see
  Appendix A of Oesch et al.\ 2014), resulting in McLure et
  al.\ (2013) excluding them from their ``robust'' $z>6.5$ candidate
  list.}

\subsection{Contamination}

We carefully considered many possible sources of contamination for our
$z\sim4$-10 samples.  Potential contaminants include stellar sources,
time-variable events like supernovae, spurious sources, extreme
emission-line galaxies, and photometric scatter.  We discuss possible
contamination by each of these sources in the subsections that follow.

\subsubsection{Stars}

One potential contaminant for our samples is from stars in our own
galaxy, particularly very low-mass stars.  It is now well established
that low-mass stars have very similar colors to those of $z\sim6$-7
galaxies and hence could be a meaningful contaminant, if one does not
have information on the spatial profile of galaxies (Stanway et
al.\ 2003; Bouwens et al.\ 2006; Ouchi et al.\ 2009b; Tilvi et
al.\ 2013).  Since we explicitly exclude points sources from our
selection, i.e., sources with a SExtractor stellarity index greater
than 0.9 (where 0 and 1 correspond to an extended and point source,
respectively) and an apparent magnitude at least one magnitude
brighter than the limit, we would expect contamination from stellar
sources to be somewhat limited.  Bouwens et al.\ 2006 found the
SExtractor stellarity parameter to be very effective in distinguishing
point sources from extended sources, for sources with sufficiently
high signal-to-noise (i.e., $>$10).  

Near the detection limit of our samples, a small level of
contamination is expected, given that we no longer attempt to remove
point sources at such low S/Ns.  We estimated this contamination by
deriving the number counts for all point-like sources in the CANDELS
fields (stellarity $>$0.9) which would satisfy our selection criteria
if placed near the selection limit of surveys.  We identified $\sim$25
stars over the magnitude range $21<H_{160,AB}<26$ per CANDELS field
which could contaminate our $z\sim4$-10 selections, with no especially
significant increase in the surface density of such sources from
$H_{160,AB}\sim 21$ to $H_{160,AB}\sim 26$ (similar to that found by
Pirzkal et al.\ 2009).  This is equivalent to a surface density of
$\sim$0.04 arcmin$^{-2}$ mag$^{-1}$, which is within a factor of two
of the surface density of low-mass stars (M4 and later) found by
Pirzkal et al.\ (2009) and Holwerda et al.\ (2014a), i.e., 0.09
arcmin$^{-2}$ mag$^{-1}$ and 0.11 arcmin$^{-2}$ mag$^{-1}$,
respectively.  Extrapolating the observed counts to beyond the limit
where we explicitly reject point-like sources (e.g., 27 mag for
CANDELS DEEP), we estimate a contamination rate of $\leq$2, 5, and
$\leq$2 sources per field for our $z\sim5$, $z\sim6$, and $z\sim7$
samples from the GN+GS fields, $<$1 contaminant for our XDF and
HUDF09-Ps samples, and $\sim$1 contaminant over the BoRG/HIPPIES
program.  This works out to surface densities of potential stellar
contaminants of $\lesssim$0.02, $\sim$0.05, and $\lesssim$0.02
arcmin$^{-2}$, respectively, for our $z\sim5$, $z\sim6$, and $z\sim7$
samples.

Finally, it is also possible that our samples include a small number
of contaminant stars even at brighter magnitudes where we exclude
pointlike sources or compact sources that significantly prefer a
stellar SED.  Using simulations similar to those described in \S4.1
(but for point-like sources with SEDs randomly drawn from the SpeX
library), we estimate that our samples would contain at most 2 such
contaminant stars per CANDELS field to $\sim$27 mag.  Overall, this
works out to a contamination rate of $<$1\% for our $z\sim4$
selections and $<$2\% for our $z\sim5$-8 samples.

\subsubsection{Transient Sources or Supernovae}

Another potential source of contamination for our high-redshift
samples are time-variable events like supernovae.  Such events could
contaminate our samples if observations of sources at bluer and redder
wavelengths did not take place over the same time frame and such
sources only became bright during observations in the redder bands.
Circumstances could then conspire to make such a SNe look like a
high-redshift star-forming galaxies with a prominent Lyman break, if
the SNe was sufficiently separated from its host galaxy that it could
be identified as a distinct source.

Fortunately, we can easily see from simple arguments that such
contaminants will be of negligible importance for our probes.  Our
explicit exclusion of pointlike sources at bright magnitudes and known
SNe events (e.g., Rodney et al.\ 2014) should guarantee that all but
the faintest SNe's make it in our sample, i.e., $\gtrsim$27 mag (where
we no longer exclude point sources).  Furthermore, for the CANDELS
WIDE fields where the various epochs of optical and near-IR
observations were acquired almost simultaneously (i.e., CANDELS UDS,
CANDELS COSMOS, and $\sim$50\% of CANDELS EGS), the contamination rate
will be negligible, as the two epochs are taken with a $\sim$50-day
time scale which is short relative to $\sim$100-day decay time for
most SNe events.  Contamination from SNe over the CANDELS DEEP regions
should be similarly low.  Due to the long $\sim$16-month observational
baseline, most of the pixels associated with a SNe brighter than
$\sim$27 mag would be rejected during the reduction of the WFC3/IR
data itself (or if temporarily brighter than 25 mag identified as a
SNe by the CANDELS SNe search team: Rodney et al.\ 2014).

The only scenario where SNe would likely contaminate our selection is
if the SNe were likely fading at the time of the first WFC3/IR
observations over the ERS, CANDELS-GN+GS WIDE, or deep field
observations and hence beyond our $\sim$26.5-mag limit for rejecting
point-like sources over those fields.  If we use the approximate SNe
rate of 0.03 SNe arcmin$^{-2}$ derived by Riess et al.\ (2007) per
40-day period from the GOODS SNe program, use the fact that only
$\sim$40\% of SNe would be sufficiently separated from their host
galaxy to be identified as a SNe (Strolger et al.\ 2004; Bouwens et
al.\ 2008), we estimate that at most 2 $z\sim7$ galaxies from our
program could correspond to SNe.  In addition, the lack of any overlap
between published SNe events (e.g., Rodney et al.\ 2014) and current
$z\sim4$-10 catalogs (\S3.3) provides us with further evidence that
the contamination is small.

\subsubsection{Lower-Redshift Galaxies}

Are there significant numbers of lower redshift galaxies in our
high-redshift samples?  For such galaxies to exist in our samples in
large numbers, they would need to have similar colors to $z\sim4$-10
galaxies, showing a deep spectral break, blue colors redward of the
break, and relatively small sizes.  It is not clear what such objects
would be, but low-mass, moderate-age, Balmer-break galaxies in the
$z\sim1$-3 universe are one possibility (e.g., Wilkins et al.\ 2010),
as would intermediate redshift galaxies with extreme-emission lines
(see \S3.5.4).  Dust-reddened intermediate-redshift sources would have
far too red colors redward of the break to be included in our
high-redshift samples.

Whatever the nature of intermediate-redshift contaminants, they are
unlikely to be present in our high-redshift samples, except in very
small numbers.  Perhaps, the most compelling argument for this can be
obtained by stacking the flux information in our high-redshift
samples.  If our samples were significantly contaminated by
lower-redshift galaxies, one would expect the stacks of the optical
data to show significant detections in the bluest bands.  However,
deep stacks of our $z\sim6$, $z\sim7$, and $z\sim8$ samples show
absolutely no flux in the $B_{435}$, $B_{435}V_{606}$, and
$B_{435}V_{606}i_{775}$ bands, respectively, consistent with our
high-redshift samples being almost exclusively composed of
high-redshift galaxies.  In addition, the spectroscopic follow-up done
on high-redshift samples reveal very small numbers of lower redshift
contaminants (e.g., Vanzella et al.\ 2009; Stark et al.\ 2010).

\subsubsection{Extreme Emission-Line Galaxies}

Another potential contaminant of our high-redshift samples are from
so-called extreme emission-line galaxies, where a significant fraction
of the flux from a galaxy is concentrated into a small number of very
high-equivalent-width emission lines (van der Wel et al.\ 2011; Atek
et al.\ 2011).  These emission lines can cause intermediate-redshift
sources to show apparent spectral breaks between adjacent bands,
mimicking the appearance of high-redshift Lyman-break galaxies (Atek
et al.\ 2011).  Fortunately, this is not expected to be a huge concern
for our selections except perhaps near the detection limit of our
samples due to the fact that extreme emission-line galaxies typically
show spectral slopes $\beta$ of $\sim-2$ (van der Wel et al.\ 2011)
over a wide wavelength range.  Such sources would therefore be easily
excluded in most cases from our high-redshift selections based on the
deep optical data that exist over our search fields.  The only
possible exception to this is if these sources also show substantial
amounts of dust reddening as may be present in the extreme [OIII]
emission-line galaxy identified by Brammer et al.\ (2013: see also
Brammer et al.\ 2012) at $z=1.6$ and also the $z\sim2$/$z\sim12$
candidate UDFj-39546284 (Bouwens et al.\ 2011; Ellis et al.\ 2013;
Brammer et al.\ 2013; Bouwens et al.\ 2013; Capak et al.\ 2013).

We can approximately quantify the contamination from these sources to
$z\sim4$-8 samples by creating a mock catalog of EELGs with the
observed surface densities on the sky (1 arcmin$^{-2}$: Atek et
al.\ 2011), $Y_{105}-J_{125}$ colors ($\sim$0.4 mag), $J_{125}$-band
magnitudes ($J_{125}\sim23$-27), and spectral slopes $\lambda^{\beta}$
(where $\beta$ ranges from $-1$ to $-2.3$: Atek et al.\ 2011; van der
Wel et al.\ 2011), and then adding noise.  Given the red
$Y_{105}-J_{125}$ colors of the known population of EELGs (and blue
$J_{125}-H_{160}$ colors), they would predominantly act as
contaminants for our $z\sim8$ selections.  Of the 959 EELGs expected
to be present over our 959-arcmin$^2$ search area, our simulations
suggest just 1 of these EELGs would make it into our overall $z\sim8$
sample.  However, we will not include that in our contamination
corrections since EELGs naturally contribute to the input sample of
galaxies used in the ``photometric scattering'' simulations described
below (\S3.5.5) and are therefore already implicitly corrected for.

\subsubsection{Establishing the contamination from low-redshift galaxies by adding noise to real data}

In general, the most important source of contamination for
high-redshift selections is from lower-redshift galaxies scattering
into our color selection windows due to the impact of noise.  As in
some earlier work (Bouwens et al.\ 2006; Bouwens et al.\ 2007), we
estimate the impact of such contamination by repeatedly adding noise
to the imaging data from the deepest fields, creating catalogs, and
then attempting to reselect sources from these fields in exactly the
same manner as the real observations.  Sources which are found with
the same selection criteria as our real searches in the degraded data
but which show detections blueward of the break in the original
observations are classified as contaminants.

The availability of deep imaging data with similar filter coverage as
the wider area observations makes it possible to use this procedure on
our wide-area CANDELS-GN, CANDELS-GS, and HUDF09-Ps samples.
Estimating the contamination rate by adding noise to real observations
is superior to making these estimates based on photometric catalogs,
since it allows one to inspect the results and exclude sources that
are obvious artifacts or consist of obviously overlapping galaxies..
This approach also provides a more direct and robust estimate of the
contamination rate than relying on the redshift likelihood
distributions from the photometric redshift approach (e.g., McLure et
al.\ 2013) due to the dependence on an uncertain redshift prior.  We
refer the interested reader to Appendix A from Bouwens et al.\ 2007
and Bouwens et al.\ (2006) for an earlier extensive application of
such simulations.

For the selection of sources from the XDF, it is not possible to make
use of such a procedure given the lack of an imaging data set with
deeper observations.  Nevertheless, we can estimate the likely
contamination by using brighter, higher S/N sources in the XDF to
model contamination in fainter sources.  In detail, we shift all
sources in the XDF $\sim$1 mag fainter in all passbands, add noise to
match that seen in the XDF, and then attempt to reselect these sources
using the same selection criteria as we use with the XDF (similar to
the procedure used in Bouwens et al.\ 2008; Ouchi et al.\ 2009b;
Wilkins et al.\ 2011).

In total, we consider degradation experiments for all six of our
Lyman-break selections, involving eight different combinations of
field depths, i.e., from XDF to HUDF09-1, XDF to HUDF09-2, XDF to
CANDELS-DEEP, XDF to CANDELS-WIDE, from HUDF09-1 to CANDELS-DEEP, from
HUDF09-1 to CANDELS-WIDE, from HUDF09-2 to CANDELS-DEEP, and from
HUDF09-2 to CANDELS-WIDE.  For each depth combination, ten different
realizations of the noise were considered to minimize the dependence
of the results upon a particular noise realization.

Using this procedure and ignoring sources brightward of the faintest
0.5 mag of each sample, we estimate a contamination rate of 2$\pm$1\%,
3$\pm$1\%, 6$\pm$2\%, 10$\pm$3\%, and 8$\pm$2\% for our $z\sim4$,
$z\sim5$, $z\sim6$, $z\sim7$, and $z\sim8$ selections, respectively.
At the faint end of each of our selections (within 0.5 mag of the
$5$-6$\sigma$ selection limit), the contamination rates we estimate
are approximately $2\times$ higher than this, but we do not make use
of such sources in the determinations of our LFs due to the larger
uncertainties in their completeness and contamination rates.  The
contamination rates in the HUDF09-2 and CANDELS-WIDE fields tend to be
lower, due to the greater sensitivities of the optical observations
relative to the near-IR observations.  The uncertainties on these
contamination rate estimates are typically $\sim$30\%, due to the
rather limited number of input objects (i.e., from the XDF and
HUDF09-Ps fields) used in these simulations and which contribute
meaningfully to the contamination rate.

For our CANDELS-UDS, CANDELS-COSMOS, CANDELS-EGS, and ERS wide-area
samples, we estimate the contamination rate using the complete
photometric catalog from the XDF.  We first derive model SEDs for each
source from our XDF catalogs using EAZY.  All those sources without
clear $\geq$3$\sigma$ detections in the $B_{435}$-band are excluded
(since such sources could be potentially at high redshift).  We then
add noise to the photometry of individual sources to match the noise
seen in the real data and then run the EAZY photometric redshift
software, while excluding those sources detected at $>2.5\sigma$
blueward of the break.  The contamination rates we find over the wide
fields from photometric scatter is just $2\%$ for $z\sim5$ candidates
and $1\%$ for $z\sim6$-8 candidates.

For our $z\sim8$ selection over the BoRG/HIPPIES program, we estimate
the contamination rate by using the same selection criteria on the
$V_{606}Y_{098}J_{125}H_{160}$ observations over the ERS data set and
then comparing the selected sources with our actual $z\sim8$ sample
from the ERS data set.  Applying the BoRG criteria to the HST
observations over the ERS field, we identify 8 candidate $z\sim8$
galaxies.  6 of these 8 candidates are likely to correspond to
$z\sim8$ galaxies, as they were previously selected using the full HST
observations (\S3.2.3).  The other 2 candidates show modest flux in
the other optical bands and therefore are unlikely $z\sim8$ galaxies.
These tests suggest a 25\% contamination rate for our BoRG/HIPPIES
selection, similar to what Bradley et al.\ (2012) adopt for the
contamination rate of their BoRG selection.  As a check on this
estimate, we also estimated the number of contaminants in the
wide-area BoRG/HIPPIES fields using almost identical simulations to
that perfomed above on the CANDELS-UDS/COSMOS/EGS fields.  The
contamination rate we recovered ($20\pm8$\%) was quite similar to that
derived from the ERS data set above; we will therefore assume a
contamination rate of 25\% for our $z\sim8$ BoRG/HIPPIES selection in
deriving our LF results.

\subsubsection{Spurious Sources}

Spurious sources also represent a potentially important contaminant
for high-redshift selections if there are significant non-Gaussian
artifacts in the data one is using to identify sources or one selects
sources of low enough significance.  To guard against contamination by
spurious sources, we require sources be detected at $5\sigma$
significance in our deepest data set the XDF, at $5.5\sigma$
significance in our HUDF09-1, HUDF09-2, CANDELS, and ERS search
fields, and $6\sigma$ significance in BoRG/HIPPIES.  Since almost all
of our sources (99.7\%) are detected at $>3\sigma$ in at least two
passbands, it is extraordinarily unlikely that a meaningful fraction
(i.e., $>$0.3\%) of our high-redshift samples is composed of spurious
sources.  Based on the number of single-band $3\sigma$ detections, we
estimate the likely spurious fraction to be $<$0.3\%.

\subsubsection{Summary}

We estimate a total contamination level of just $\sim$2\%, $\sim$3\%,
$\sim$6\%, $\sim$7\%, and $\sim$10\% for all but the faintest sources
in our $z\sim4$, $z\sim5$, $z\sim6$, $z\sim7$, and $z\sim8$ samples,
respectively.  The most significant source of contamination for our
high-redshift samples is due to the effect of noise in perturbing the
photometry of lower-redshift galaxies so that they satisfy our
high-redshift selection criteria, but stars also contribute at a low
level ($\sim$2\%).  Similar results are found in other recent
selections of sources in the high redshift universe (e.g., Giavalisco
et al.\ 2004b; Bouwens et al.\ 2006, 2007, 2011; Wilkins et al.\ 2011;
Schenker et al.\ 2013).

\begin{deluxetable*}{lclclc}
\tablewidth{0pt}
\tabletypesize{\footnotesize}
\tablecaption{Stepwise Determination of the rest-frame $UV$ LF at $z\sim4$, $z\sim5$, $z\sim6$, $z\sim7$, $z\sim8$, and $z\sim10$ using the SWML method (\S4.1).\label{tab:swlf}}
\tablehead{
\colhead{$M_{1600,AB}$\tablenotemark{a}} & \colhead{$\phi_k$ (Mpc$^{-3}$ mag$^{-1}$)} & \colhead{$M_{1600,AB}$\tablenotemark{a}} & \colhead{$\phi_k$ (Mpc$^{-3}$ mag$^{-1}$)} & \colhead{$M_{1600,AB}$\tablenotemark{a}} & \colhead{$\phi_k$ (Mpc$^{-3}$ mag$^{-1}$)}}
\startdata
\multicolumn{2}{c}{$z\sim4$ galaxies}\ & \multicolumn{2}{c}{$z\sim6$ galaxies} & \multicolumn{2}{c}{$z\sim8$ galaxies}\\
$-$22.69 & 0.000003$\pm$0.000004 & $-$22.52 & 0.000002$\pm$0.000002 & $-$22.87 & $<$0.000002\tablenotemark{b}\\
$-$22.19 & 0.000015$\pm$0.000009 & $-$22.02 & 0.000015$\pm$0.000006 & $-$22.37 & $<$0.000002\tablenotemark{b}\\
$-$21.69 & 0.000134$\pm$0.000023 & $-$21.52 & 0.000053$\pm$0.000012 & $-$21.87 & 0.000005$\pm$0.000003\\
$-$21.19 & 0.000393$\pm$0.000040 & $-$21.02 & 0.000176$\pm$0.000025 & $-$21.37 & 0.000013$\pm$0.000005\\
$-$20.69 & 0.000678$\pm$0.000063 & $-$20.52 & 0.000320$\pm$0.000041 & $-$20.87 & 0.000058$\pm$0.000015\\
$-$20.19 & 0.001696$\pm$0.000113 & $-$20.02 & 0.000698$\pm$0.000083 & $-$20.37 & 0.000060$\pm$0.000026\\
$-$19.69 & 0.002475$\pm$0.000185 & $-$19.52 & 0.001246$\pm$0.000137 & $-$19.87 & 0.000331$\pm$0.000104\\
$-$19.19 & 0.002984$\pm$0.000255 & $-$18.77 & 0.001900$\pm$0.000320 & $-$19.37 & 0.000533$\pm$0.000226\\
$-$18.69 & 0.005352$\pm$0.000446 & $-$17.77 & 0.006680$\pm$0.001380 & $-$18.62 & 0.001060$\pm$0.000340\\
$-$18.19 & 0.006865$\pm$0.001043 & $-$16.77 & 0.013640$\pm$0.004200 & $-$17.62 & 0.002740$\pm$0.001040\\
$-$17.69 & 0.010473$\pm$0.002229 & \multicolumn{2}{c}{$z\sim7$ galaxies\tablenotemark{c}} & \multicolumn{2}{c}{$z\sim10$ galaxies}\\
$-$16.94 & 0.024580$\pm$0.003500 & $-$22.66 & $<$0.000002\tablenotemark{b} & $-$22.23 & $<$0.000001\tablenotemark{b}\\
$-$15.94 & 0.025080$\pm$0.007860 & $-$22.16 & 0.000001$\pm$0.000002 & $-$21.23 & 0.000001$\pm$0.000001\\
\multicolumn{2}{c}{$z\sim5$ galaxies}\ & $-$21.66 & 0.000033$\pm$0.000009 & $-$20.23 & 0.000010$\pm$0.000005\\
$-$23.11 & 0.000002$\pm$0.000002 & $-$21.16 & 0.000048$\pm$0.000015 & $-$19.23 & $<$0.000049\tablenotemark{b}\\
$-$22.61 & 0.000006$\pm$0.000003 & $-$20.66 & 0.000193$\pm$0.000034 & $-$18.23 & 0.000266$\pm$0.000171\\
$-$22.11 & 0.000034$\pm$0.000008 & $-$20.16 & 0.000309$\pm$0.000061 &  & \\
$-$21.61 & 0.000101$\pm$0.000014 & $-$19.66 & 0.000654$\pm$0.000100 &  & \\
$-$21.11 & 0.000265$\pm$0.000025 & $-$19.16 & 0.000907$\pm$0.000177 &  & \\
$-$20.61 & 0.000676$\pm$0.000046 & $-$18.66 & 0.001717$\pm$0.000478 &  & \\
$-$20.11 & 0.001029$\pm$0.000067 & $-$17.91 & 0.005840$\pm$0.001460 &  & \\
$-$19.61 & 0.001329$\pm$0.000094 & $-$16.91 & 0.008500$\pm$0.002940 &  & \\
$-$19.11 & 0.002085$\pm$0.000171 &  &  &  & \\
$-$18.36 & 0.004460$\pm$0.000540 &  &  &  & \\
$-$17.36 & 0.008600$\pm$0.001760 &  &  &  & \\
$-$16.36 & 0.024400$\pm$0.007160 &  &  &  &
\enddata
\tablenotetext{a}{Derived at a rest-frame wavelength of 1600\AA.}
\tablenotetext{b}{Upper limits are $1\sigma$.}

\tablenotetext{c}{The CANDELS-EGS field contains a much larger number
  of luminous ($M_{UV,AB}<-21.41$) $z\sim7$ galaxy candidates than the
  other CANDELS fields (7 vs. 1, 2, 3, and 4) and may represent an
  extreme overdensity.  Therefore, as an alternative to the present
  determination, we also provide a stepwise determination of the
  $z\sim7$ LF in Table~\ref{tab:swlfegs} from Appendix E, which excludes
  the CANDELS-EGS data set.}
\end{deluxetable*}

\section{Luminosity Function Results}

In this section, we make use of our large, comprehensive samples of
$z\sim4$-10 galaxies we selected from the XDF+ERS+CANDELS+BoRG/HIPPIES
data sets to obtain the best available determinations of the UV LFs at
these redshifts.  In constructing the present LFs, we make use of
essentially the same procedures as we previously utilized in Bouwens
et al.\ (2007) and Bouwens et al.\ (2011).

We first derive the LFs in the usual non-parametric stepwise way
(\S4.1), and then in terms of the Schechter parameters (\S4.2).  In
\S4.3, we compare our LF results with previous results from our team.
In \S4.4, we use our large samples of galaxies at both higher and
lower luminosities to derive the shape of the $UV$ LF and attempt to
ascertain whether it is well represented by a Schechter function.  In
\S4.5, we quantify variations in the volume density of $z\sim4$-8
galaxies themselves across the five CANDELS fields.  Finally, in
\S4.6, we use our search results across the full CANDELS, ERS, XDF,
HUDF09-Ps data set to set constraints on the $UV$ LF at $z\sim10$.

\subsection{SWML Determinations}

We first consider a simple stepwise (binned) determination of the $UV$
LFs at $z\sim4$-8.  The baseline approach in the literature for these
type of determinations is to use the stepwise maximum-likelihood
(SWML) approach of Efstathiou et al.\ (1988).  With this approach, the
goal is to find the maximum likelihood LF shape which best reproduces
the available constraints.  Since the focus with this approach is in
reproducing the \textit{shape} of the LF, this approach is largely
robust against field-to-field variations in the normalization of the
luminosity function and hence large-scale structure effects.

\begin{deluxetable}{ccccc}
\tablewidth{0pt}
\tabletypesize{\footnotesize}
\tablecaption{
STY79 Determinations of the Schechter Parameters for the rest-frame $UV$ LFs
at $z\sim4$, $z\sim5$, $z\sim6$, $z\sim7$, $z\sim8$, and $z\sim10$ (\S4.2).\label{tab:lfparm}}
\tablehead{
\colhead{Dropout} & \colhead{} & \colhead{} & \colhead{$\phi^*$ $(10^{-3}$} & \colhead{} \\
\colhead{Sample} & \colhead{$<z>$} &
\colhead{$M_{UV} ^{*}$\tablenotemark{a}} & \colhead{Mpc$^{-3}$)} & \colhead{$\alpha$}}
\startdata
\multicolumn{5}{c}{Reddy \& Steidel 2009}\\
$U$ & 3.0 & $-20.97\pm0.14$ & $1.71\pm0.53$ & $-1.73\pm0.13$\\
\\
\multicolumn{5}{c}{XDF+HUDF09-Ps+CANDELS-GN+GS+ERS}\\
$B$ & 3.8 & $-20.88\pm0.08$ & $1.97_{-0.29}^{+0.34}$ & $-1.64\pm0.04$\\ 
$V$ & 4.9 & $-21.10\pm0.15$ & $0.79_{-0.18}^{+0.23}$ & $-1.76\pm0.06$\\
$i$ & 5.9 & $-21.10\pm0.24$ & $0.39_{-0.14}^{+0.21} $ & $-1.90\pm0.10$\\
$z$ & 6.8 & $-20.61\pm0.31$ & $0.46_{-0.21}^{+0.38} $ & $-1.98\pm0.15$\\
$Y$ & 7.9 & $-20.19\pm0.42$ & $0.44_{-0.24}^{+0.52} $ & $-1.81\pm0.27$\\
$J$ & 10.4 & $-20.92$ (fixed) & $0.013_{-0.005}^{+0.007}$ & $-2.27$ (fixed) \\
\multicolumn{5}{c}{All Fields (excluding CANDELS-EGS)\tablenotemark{b}}\\
$z$ & 6.8 & $-20.77\pm0.28$ & $0.34_{-0.14}^{+0.24}$ & $-2.03\pm0.13$ \\
$Y$ & 7.9 & $-20.21\pm0.33$ & $0.45_{-0.21}^{+0.42}$ & $-1.83\pm0.25$ \\
\multicolumn{5}{c}{All Fields}\\
$B$ & 3.8 & $-20.88\pm0.08$ & $1.97_{-0.29}^{+0.34}$ & $-1.64\pm0.04$\\ 
$V$ & 4.9 & $-21.17\pm0.12$ & $0.74_{-0.14}^{+0.18}$ & $-1.76\pm0.05$\\
$i$ & 5.9 & $-20.94\pm0.20$ & $0.50_{-0.16}^{+0.22}$ & $-1.87\pm0.10$\\
$z$ & 6.8 & $-20.87\pm0.26$ & $0.29_{-0.12}^{+0.21}$ & $-2.06\pm0.13$\\
$Y$ & 7.9 & $-20.63\pm0.36$ & $0.21_{-0.11}^{+0.23}$ & $-2.02\pm0.23$\\
$J$ & 10.4 & $-20.92$ (fixed) & $0.008_{-0.003}^{+0.004} $ & $-2.27$ (fixed)
\enddata
\tablenotetext{a}{Derived at a rest-frame wavelength of 1600\AA.}
\tablenotetext{b}{While our simulation results
  (Figure~\ref{fig:zdistsel}) suggest that it is possible to identify
  $z\sim7$ and $z\sim8$ galaxies using the available observations over
  the CANDELS EGS field (albeit with some intercontamination between
  $z\sim7$ and $z\sim8$ samples), the lack of deep $Y$-band
  observations over this search field make the results slightly less
  robust than over the other CANDELS fields.  Our quantification of
  the stepwise LFs at $z\sim7$ and $z\sim8$ from all fields (but
  excluding the CANDELS-EGS data set) is presented in
  Table~\ref{tab:swlfegs} from Appendix E.}
\end{deluxetable}

\begin{figure}
\epsscale{1.15}
\plotone{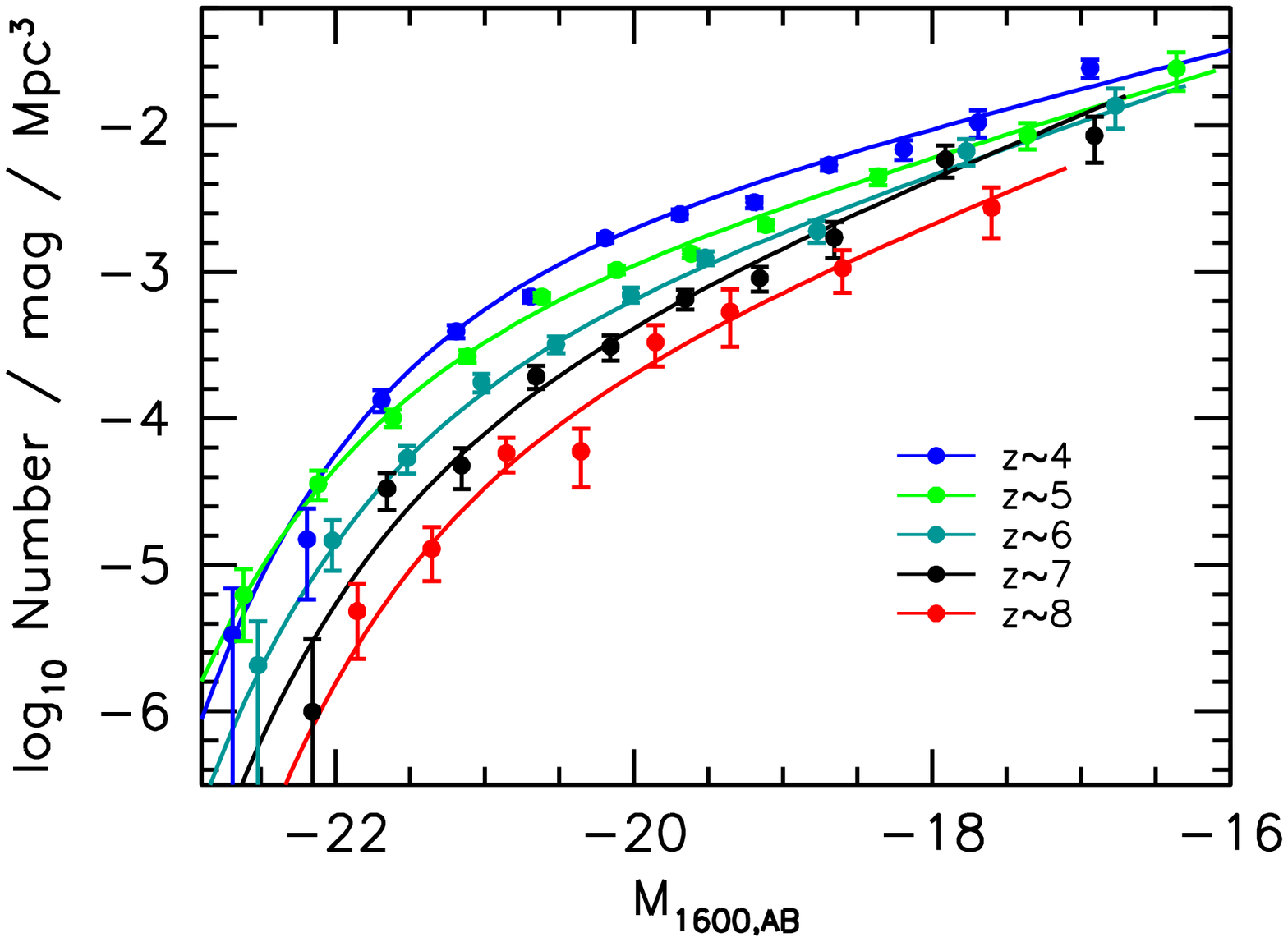}
\caption{SWML determinations of the $UV$ LFs at $z\sim4$ (\textit{blue
    solid circles}), $z\sim5$ (\textit{green solid circles}), $z\sim6$
  (\textit{light blue solid circles}), $z\sim7$ (\textit{black
    circles}), and $z\sim8$ (\textit{red solid circles}).  Also shown
  are independently-derived Schechter fits to the LFs using the STY
  procedure (see \S4.2).  The $UV$ LFs we have derived from the
  complete CANDELS+ERS+XDF+HUDF09 data sets show clear evidence for
  the build-up of galaxies from $z\sim8$ to $z\sim4$.  Note the
  appreciable numbers of luminous galaxies at $z\sim6$, $z\sim7$ and
  $z\sim8$.\label{fig:lfall}}
\end{figure}

\begin{figure*}
\epsscale{1.2}
\plotone{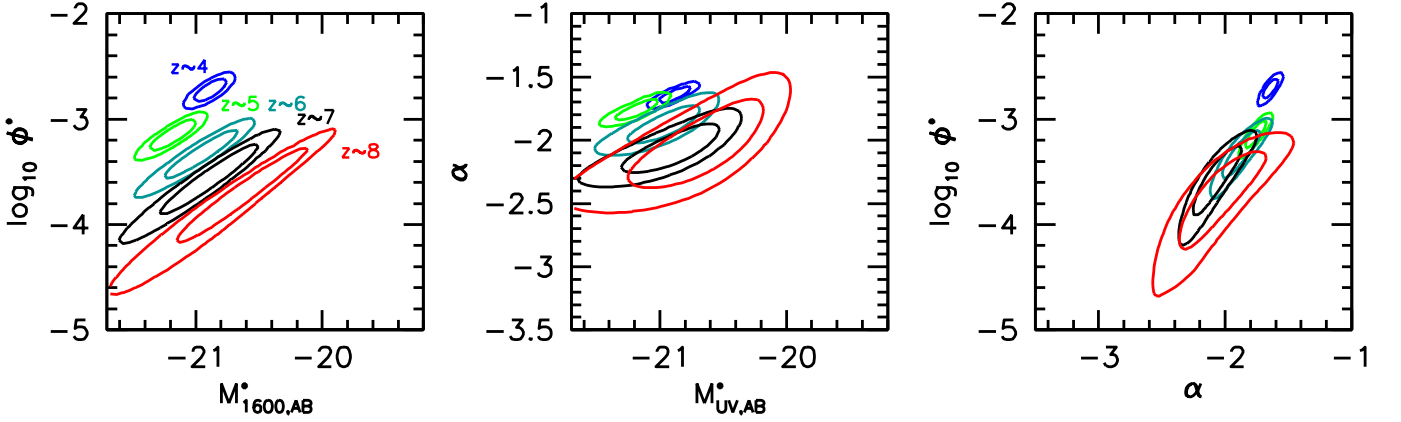}
\caption{The 68\% and 95\% confidence intervals on the Schechter
  parameters $M^*$, $\phi^*$, and $\alpha$ we derive for the $UV$ LFs
  at $z\sim4$ (\textit{dark blue contours}), $z\sim5$ (\textit{green
    contours}), $z\sim6$ (\textit{blue contours}), $z\sim7$
  (\textit{black contours}), and $z\sim8$ (\textit{red contours})
  using an STY-like procedure (\S4.2).  These confidence intervals
  show evidence for an evolution in the faint-end slope $\alpha$ and
  $\phi^*$ with redshift.  Evolution in both $\phi^*$ and $\alpha$
  looks very similar to an evolution in the characteristic luminosity
  $M^*$ (previously proposed by Bouwens et al.\ 2007 and Bouwens et
  al.\ 2008) with cosmic time, except at the bright end of the LF (see
  Figure~\ref{fig:modelcomp}).\label{fig:mlcontours}}
\end{figure*}

As in Bouwens et al.\ (2007) and Bouwens et al.\ (2011), we can write
the stepwise LF $\phi_k$ as $\Sigma \phi_k W(M-M_k)$ where $k$ is an
index running over the magnitude bins, where $M_k$ corresponds to the
absolute magnitude at the center of each bin, where
\begin{equation}
W(x) = 
\begin{array}{cc} 
0, & x < -0.25\\
1, & -0.25 < x < 0.25\\
0, & x > 0.25.
\end{array}
\end{equation}
and where $x$ gives the position within a magnitude bin (for a 0.5-mag
binning scheme).  The goal then is to find the LF which maximizes the
likelihood of reproducing the observed source counts over our various
search fields.  The likelihood $\cal L$ can be expressed analytically
as
\begin{equation}
{\cal L}=\Pi_{field} \Pi_i p(m_i)
\label{eq:ml}
\end{equation}
where
\begin{equation}
p(m_i) = \left (\frac{n_{expected,i}}{\Sigma_j n_{expected,j}} \right )^{n_{observed,i}}
\end{equation}
where the above products runs over the different search fields and
magnitude interval $i$ used in the LF determinations, $n_{expected,i}$
is the expected number of sources in magnitude interval $i$ for a
given LF, and $n_{observed,i}$ is the observed number of sources in
magnitude interval $i$.  The quantity $n_{observed,i}$ is derived
using the apparent magnitude of the source closest to 1600\AA, which
occurs in the $i_{775}$ band for sources in our $z\sim4$ samples, in
the $Y_{105}$ band for our $z\sim5$ and $z\sim6$
samples,\footnote{Even though the $z_{850}$-band magnitude of sources
  in our $z\sim5$ sample is nominally closer to 1600\AA$~$rest frame,
  we elected to use the $Y_{105}$ band flux due to the greater overall
  depth of these data in many of our data sets (particularly the XDF)}
in the $J_{125}$ band for our $z\sim7$ samples, and in the $H_{160}$
band for our $z\sim8$ samples.  For the ERS and CANDELS UDS/COSMOS/EGS
wide-area samples where no $Y_{105}$ coverage is available, we make
use of the $Y_{098}$-band and $J_{125}$ magnitudes, respectively,
instead for our $z\sim5$ and $z\sim6$ samples.  We apply a small
correction to the apparent magnitude of individual sources (typically
$\lesssim$0.1 mag) so that it corresponds to an effective rest-frame
wavelength 1600\AA.  The correction we apply is based on the biweight
mean $\beta$ Bouwens et al.\ (2014a) derive for galaxies with a given
absolute magnitude and redshift.  The quantity $n_{observed,i}$ is
also corrected for contamination using the simulations we describe in
\S3.5.5.

Similar to our previous work, we compute the number of sources
expected in a given magnitude interval $i$ assuming a model LF as
\begin{equation}
n_{expected,i} = \Sigma _{j} \phi_j V_{i,j}
\label{eq:numcountg}
\end{equation}
where $V_{i,j}$ is the effective volume over which one could expect to
find a source of absolute magnitude $j$ in the observed magnitude
interval $i$.  We estimate $V_{i,j}$ for a given search field using an
extensive suite of Monte-Carlo simulations where we add sources with
an absolute magnitude $j$ to the different search fields and then see
if we select a source with apparent magnitude $i$.  The $V_{i,j}$
factors implicitly correct for flux-boosting type effects that are
important near the detection limits of our samples, whereby faint
sources scatter to brighter apparent fluxes and thus into our samples.

Computing the relevant $V_{i,j}$'s for all of our samples and search
fields required our running an extensive suite of Monte-Carlo
simulations.  In these simulations, large numbers of artificial
sources were inserted into the input data (typically $\sim$50
arcmin$^{-2}$ in each simulation).  Catalogs were then constructed
from the data and sources selected.  To ensure that the candidate
galaxies in these simulations had realistic sizes and morphologies, we
randomly selected similar-luminosity $z\sim4$ galaxies from the Hubble
Ultra-Deep Field to use as a template to model the two-dimensional
spatial profile for individual sources.  We assigned each galaxy in
our simulations a $UV$ color using the $\beta$ vs. $M_{UV}$
determinations of Bouwens et al.\ (2014a), with an intrinsic scatter
in $\beta$ varying from 0.35 brightward of $-20$ mag to 0.20 faintward
of $-20$ mag.  This matches the intrinsic scatter in $\beta$ measured
for brightest $z\sim4$-5 sources by Bouwens et al.\ (2009, 2012) and
Castellano et al.\ (2012), as well as the decreased scatter in $\beta$
for the faintest sources (Rogers et al.\ 2014).  Finally, the
templates were artificially redshifted to the redshift in the catalog
using our well-tested ``cloning'' software (Bouwens et al.\ 1998;
Bouwens et al.\ 2003a) and inserted these sources into the real
observations.  In projecting galaxies to higher redshift, we scaled
source size approximately as $(1+z)^{-1.2}$ to match that seen in the
observations (Oesch et al.\ 2010a; Grazian et al.\ 2012; Ono et
al.\ 2013; Holwerda et al.\ 2014b; Kawamata et al.\ 2014).  We
verified through a series of careful comparisons that the source sizes
we utilized were similar to those in the real observations, both as a
function of redshift and luminosity (Appendix D).

In calculating the effective selection volumes over the CANDELS-UDS,
COSMOS, and EGS search areas, we also constructed fully simulated
images of our mock sources in the ground-based and Spitzer/IRAC
observations, adding these sources to the real observations, and
extracting their fluxes using the same photometric procedure as we
applied to the real observations.  Finally, we made use of the full
set of flux information we were able to derive for the mock sources
(HST+ground-based+Spitzer/IRAC) to estimate photometric redshifts for
these sources and hence determine whether sources fell within our
redshift selection windows.  As with the real observations, mock
sources were excluded from the selection, if they were detected at
$>$2.5$\sigma$ significance in passbands blueward of the break.  We
note that in producing simulated IRAC images for the mock sources, we
assume a rest-frame EW of 300\AA$\,$for H$\alpha$+[NII] emission and
500\AA$\,$for [OIII]+H$\beta$ emission over the entire range $z=4$-9,
a flat rest-frame optical color, and a $H_{160}$-optical continuum
color of 0.2-0.3 mag, to match the observational results of Shim et
al.\ (2011), Stark et al.\ (2013), Gonz{\'a}lez et al.\ (2012, 2014),
Labb{\'e} et al.\ (2013), Smit et al.\ (2014a,b), and Oesch et
al.\ (2013a).

\begin{figure*}
\epsscale{0.73} \plotone{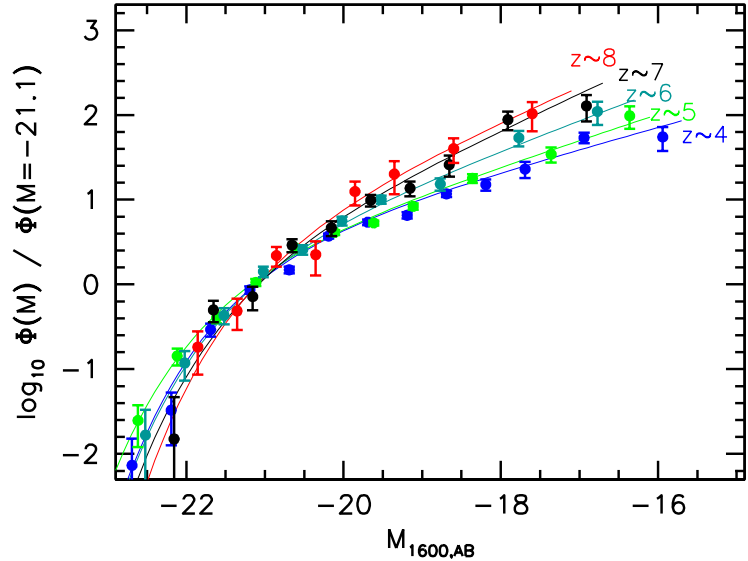}
\caption{$UV$ LFs at $z\sim4$, $z\sim5$, $z\sim6$, $z\sim7$, and
  $z\sim8$ renormalized to have approximately the same volume density
  at $\sim-21.1$ mag (\S4.2).  There is strong evidence for an evolution in the
  effective slope of the $UV$ LFs with redshift.  The effective slope
  of the LF is considerably steeper at $z\sim7$ and $z\sim8$ than it
  is at $z\sim4$-5.  \label{fig:shapelf}}
\end{figure*}

After deriving the shape of the LF at each redshift using this
procedure, we set the normalization by requiring that the total number
of sources predicted on the basis of our LF be equal to the total
number of sources observed over our search fields.  Applying the above
SWML procedure to the observed surface densities of sources in our
different search fields, we determined the maximum-likelihood LFs.

We elected to use a 0.5-mag binning scheme for the LFs at $z\sim4$-8,
consistent with past practice.  To cope with the noise in our SWML LF
determinations that result from deconvolving the transfer function
(implicit in the $V_{i,j}$ term in Eq.~\ref{eq:numcountg}) from the
number counts $n_{observed,i}$, we have adopted a wider binning scheme
at the faint-end of the LF.  This issue also causes the uncertainties
we derive on the bright end of the LF to remain somewhat large at all
redshifts (as uncertainties in the measured flux for individual
sources allow for the possibility that the observed source counts
could arise from ``picket fence''-type LF with the bulk of sources
concentrated in just the odd or even stepwise LF intervals.)

In deriving the LF from such a diverse data set, it is essential to
ensure that our LF determinations across this data set are generally
self consistent.  We therefore derived the $UV$ LFs at $z\sim5$,
$z\sim6$, $z\sim7$, and $z\sim8$ separately from the wide-area
UDS+COSMOS+EGS CANDELS observations, from the CANDELS-DEEP region
within the CANDELS-GN and GS, from the CANDELS-WIDE region within the
CANDELS-GN and GS, and from the BoRG/HIPPIES observations.  As we
demonstrate in Figure~\ref{fig:wdlfcomp} from Appendix E, we find
broad agreement between our LF determinations from all four data sets,
suggesting that the impact of systematics on our LF results is quite
limited in general.

After considering the LF results from each of our fields separately,
we combine our search results from all fields under consideration to
arrive at stepwise LFs at $z\sim4$-8 for our overall sample.  The
results are presented in Figure~\ref{fig:lfall} and in
Table~\ref{tab:swlf}.  Broadly speaking, the LF determinations over
the range $z\sim4$-8 show clear evidence for a steady build-up in the
volume density and luminosity of galaxies with cosmic time.

\subsection{Schechter Function-Fit Results}

We next attempt to represent the $UV$ LFs at $z\sim4$, $z\sim5$,
$z\sim6$, $z\sim7$, $z\sim8$, and $z\sim10$ using a Schechter-like
parameterization ($\phi^* (\ln(10)/2.5)$
$10^{-0.4(M-M^{*})(\alpha+1)}$ $e^{-10^{-0.4(M-M^{*})}}$).  Schechter
functions exhibit a power-law-like slope $\alpha$ at the faint end,
with an exponential cut-off brightward of some characteristic
magnitude $M^*$.  The Schechter parameterization has proven to be
remarkably effective in fitting the luminosity function of galaxies at
both low and high redshifts (e.g., Blanton et al.\ 2003; Reddy \&
Steidel 2009).

The procedure we use to determine the best-fit Schechter parameters is
that of Sandage, Tammann, \& Yahil (1979) and has long been the method
of choice in the literature.  Like the SWML procedure of Efstathiou et
al.\ (1988), this approach determines the LF shape that would most
likely reproduces the observed surface density of galaxies in our many
search fields.  The approach is therefore highly robust against
large-scale structure variations across the survey fields.  As with
the SWML approach, one must normalize the LF derived using this method
in some way, and for this we require that the total number of sources
observed across our search fields match the expected numbers.

We can make use of essentially the same procedure to derive the
maximum likelihood Schechter parameters as we used for the stepwise LF
in the previous section, after we convert model Schechter parameters
to the equivalent stepwise LF.  For this calculation, we adopt a
0.1-mag binning scheme in comparing the stepwise LF to the surface
density of sources in our search fields.  A 0.1-mag binning scheme is
sufficiently high resolution that it will yield essentially the same
results as estimates made without binning the observations at all
(e.g., Su et al.\ 2011).

Our maximum likelihood results for the Schechter fits at $z\sim4$,
$z\sim5$, $z\sim6$, $z\sim7$, and $z\sim8$ are presented in
Figure~\ref{fig:mlcontours}.  Meanwhile, our best-fit Schechter
parameters are presented in Table~\ref{tab:lfparm} using the
XDF+HUDF09-Ps+ERS+CANDELS-GN+CANDELS-GS data set alone and using the
full data set considered here.  These Schechter parameters are also
provided for the $z\sim7$ and $z\sim8$ LFs based on the full data set
but excluding the CANDELS-EGS field, to indicate what the results
would be excluding the large number of bright ($\sim-21.7$ mag)
galaxies found in that CANDELS field.  Finally, in
Table~\ref{tab:southvsnorth} from Appendix E, we also present
determinations of the Schechter parameters from the
CANDELS-GN+XDF+HUDF09-Ps fields and CANDELS-GS+ERS+XDF+HUDF09-Ps
fields separately.

\begin{figure*}
\epsscale{1.1}
\plotone{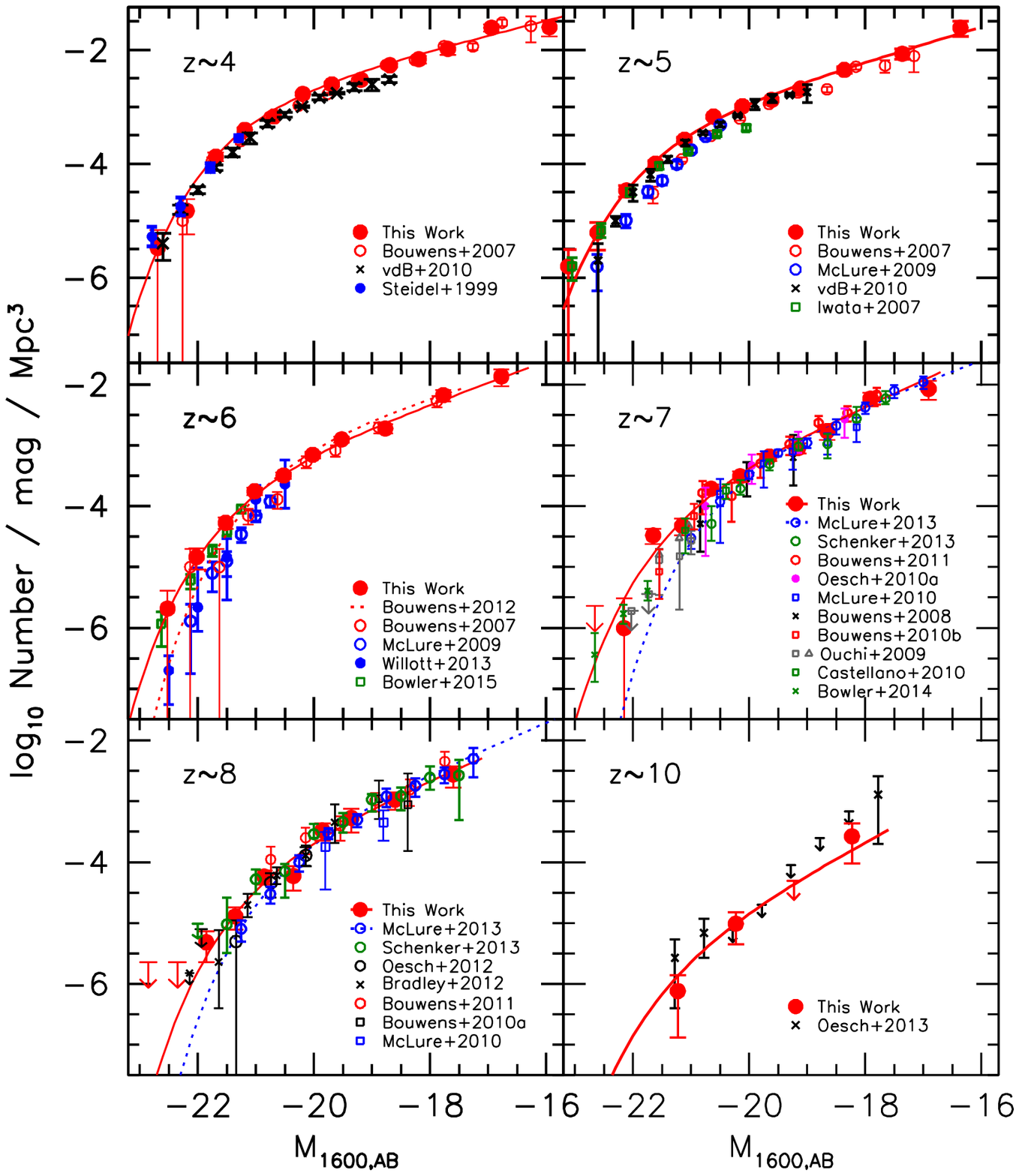}
\caption{Comparisons between the present SWML (\textit{red solid
    circles and $1\sigma$ upper limits}) and STY (\textit{red solid
    lines}) LF determinations at $z\sim4$, $z\sim5$, $z\sim6$,
  $z\sim7$, $z\sim8$, and $z\sim10$ and previous determinations of the
  $UV$ LF at these redshifts in the literature (see Appendices
  F.1-F.5).  For comparison with the present $z\sim4$-10 results, we
  also include the results of Steidel et al.\ (1999: \textit{solid
    blue circles}) at $z\sim4$, Bouwens et al.\ (2007: \textit{open
    red circles}) at $z\sim4$-6, McLure et al.\ (2009: \textit{open
    blue circles}) at $z\sim5$-6, van der Burg et al.\ (2010:
  \textit{black crosses}) at $z\sim4$-5, Iwata et al.\ (2007:
  \textit{open green squares}) at $z\sim5$, Bouwens et al.\ (2012a:
  \textit{dotted red line}) at $z\sim6$, Willott et al.\ (2013:
  \textit{solid blue circles}) at $z\sim6$, Bowler et al.\ (2015:
  \textit{open green squares}) at $z\sim6$, Bouwens et al.\ (2008:
  \textit{black crosses}) at $z\sim7$, McLure et al.\ (2010:
  \textit{blue squares}) at $z\sim7$-8, Oesch et al.\ (2010:
  \textit{solid magenta circles}) at $z\sim7$, Castellano et
  al.\ (2010: \textit{green squares}) at $z\sim7$, Ouchi et
  al.\ (2009: \textit{gray squares and limits} [best estimates] and
  \textit{gray open triangles} [before contamination correction]) at
  $z\sim7$, and Bouwens et al.\ (2010b: \textit{open red squares}) at
  $z\sim7$, Bowler et al.\ (2014: \textit{green cross}) at $z\sim7$,
  Bouwens et al.\ (2011: \textit{open red circles}) at $z\sim7$-8,
  Schenker et al.\ (2013: \textit{open green circles and upper
    limits}) at $z\sim7$-8, and McLure et al.\ (2013: \textit{open
    blue circles}) at $z\sim7$-8, Oesch et al.\ (2012b: \textit{open
    black circles and limits}) at $z\sim8$, and Bradley et al.\ (2012:
  \textit{black crosses}) at $z\sim8$, and Oesch et al.\ (2014:
  \textit{black crosses and limits}) at $z\sim10$.  All limits are
  $1\sigma$.  The brightest point in the $z\sim6$ LF by Willott et
  al.\ (2013) has also been replaced by the Bowler et al.\ (2014)
  re-estimate.  Overall, the present LFs are in broad agreement with
  previous determinations, except at the bright end of the $z\sim6$-7
  LFs.  New results from Bowler et al.\ (2015), however, are in better
  agreement with our $z\sim6$ LF.\label{fig:comp410}}
\end{figure*}

These results suggest that a good fraction of the evolution in the
$UV$ LF at $z>4$ may involve an evolution in both the normalization of
the LF $\phi^*$ and the faint-end slope $\alpha$.  Evolution in
$\phi^*$ would be expected, if galaxies in arbitrarily massive halos
in the early universe were capable of reaching the same maximum
luminosity at essentially all epochs, independent of redshift.
Evolution in the faint-end slope $\alpha$ is also expected due to the
steepening of the halo mass function towards early times (e.g., Trenti
et al.\ 2010: see \S5.5).

These general conclusions are not significantly impacted by possible
systematic errors in our analysis technique.  Even if we make
factor-of-2 changes in the contamination rate across all of our search
fields, we only find $\Delta M \lesssim 0.1$ changes in the
characteristic magnitude $M^*$ at $z\sim4$-7 and $\Delta \log_{10}
\phi^* \lesssim 0.1$ changes in the normalization $\phi^*$.  While the
impact on our faint-end slope $\alpha$ estimates are larger, i.e.,
$\Delta\alpha$ changes of 0.01, 0.03, 0.02, 0.10 at $z\sim4$,
$z\sim5$, $z\sim6$, and $z\sim7$, respectively, these uncertainties
are small relative to the overall evolution apparent from $z\sim7$ to
$z\sim4$.  Larger changes in the characteristic magnitude $M^*$ are
potentially possible (i.e., $\Delta M^* \gtrsim 0.1$), but for this to
occur, contaminating sources must be systematically undercorrected so
as to leave a $20\%$ excess in the number of bright galaxies.  Small
($\sim$10\%) systematic errors in the selection volumes (per 0.5-mag
interval) also likely have a small impact on the best-fit Schechter
parameters.

Some earlier studies have argued that a simple $\phi^*$ evolutionary
model may allow for a better representation of the evolution of the LF
than an evolution in $M^*$ (Beckwith et al.\ 2006; van der Burg et
al.\ 2010).  At slightly higher redshifts ($z\gtrsim6$), McLure et
al.\ (2010), Bouwens et al.\ (2011), McLure et al.\ (2013), and Oesch
et al.\ (2014) all indicated that $\phi^*$ evolution may provide a
slightly better description of the evolution of the $UV$ LF.  Of
course, even distinguishing evolution in $\phi^*$ from $M^*$ over the
range $z\sim6$ to $z\sim8$ can be challenging (as McLure et al.\ 2013
note explicitly).

While a pure $\phi^*$ evolutionary model seems quite effective at
fitting the evolution at the bright end of the LF to high redshift,
such a model does not capture the considerable steepening the $UV$ LF
experiences over a wide-luminosity baseline.  Fitting this steepening
requires either evolution in $\alpha$ or evolution in $M^*$ as had
been preferred by Bouwens et al.\ (2007).  Yan \& Windhorst (2004)
effectively captured both aspects of the approximate evolution with
their best-fit LF at $z\sim6$ (though they offer no clear
justification in their analysis for their decision to fix $M^*$ to the
$z\sim3$ value and to exclusively use the faint-end slope $\alpha$ to
model possible shape changes in the $UV$ LF).

\begin{figure}
\epsscale{1.13}
\plotone{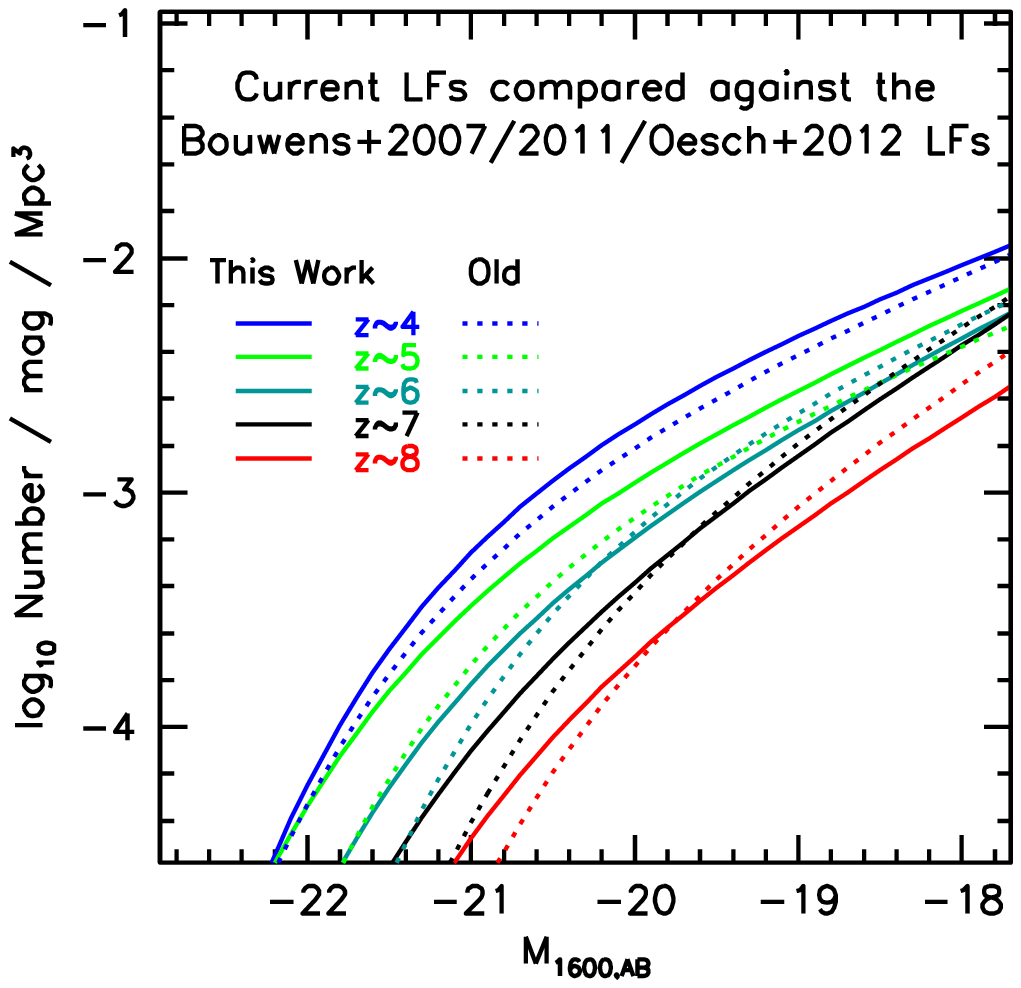}
\caption{Comparisons of the $z\sim4$ (\textit{solid blue line}),
  $z\sim5$ (\textit{solid green line}), $z\sim6$ (\textit{solid cyan
    line}), $z\sim7$ (\textit{solid black line}), and $z\sim8$
  (\textit{solid red line}) $UV$ LFs derived here with those presented
  in Bouwens et al.\ (2007), Bouwens et al.\ (2011), and Oesch et
  al.\ (2012b: \textit{dotted lines}: see \S4.3).  The boundaries on
  this figure (both horizontal and vertical) have been intentionally
  chosen to correspond to the faint-end limit of the HUDF09 $z\sim6$-8
  search and to the minimum volume density that could be probed in the
  $\sim$50 arcmin$^2$ survey area available to Bouwens et al.\ (2011).  In calculating the minimum volume density,
  we assume that a survey must contain at least two sources for the LF
  to be well constrained (given the large Poissonian uncertainties on
  the LF for single-source samples).  The present LF results are in
  excellent agreement with the Bouwens et al.\ (2011) results over the
  range in parameter space both studies probe well.  Our current
  constraints on the bright end of the $z\sim6$ and $z\sim7$ LFs are
  higher than what we previously found.  The present determinations
  provide much superior constraints at the bright end, benefitting
  from a much larger search volume, the availability of near-IR
  coverage to do a proper two-color selection of $z\sim6$ galaxies,
  and consistent coverage with HST to minimize the impact of
  systematics on our results (Appendix F.2 and F.3).  See also
  discussion in Appendix F.6.\label{fig:modelcomp}}
\end{figure}

The present evolutionary scenario in $\phi^*$ and $\alpha$ would
appear to be quite different in form from the evolutionary scenario
proposed by Bouwens et al.\ (2007), McLure et al.\ (2009), and Bouwens
et al.\ (2011), which preferred evolution in the characteristic
luminosity (particularly over the redshift range $z\sim4$-6), with
some evolution in $\phi^*$ and $\alpha$ at $z>6$ (Bouwens et
al.\ 2011; McLure et al.\ 2013).  However, in detail, an
$\phi^*$+$\alpha$ evolutionary scenario is not as different from $M^*$
evolution as one might think given their different parameterizations.
Changes in the characteristic magnitude $M^*$ produce a similar
steepening of the $UV$ LF, as one can accomplish through changes in
the faint-end slope $\alpha$.

Moreover, as we show in \S5.3, the evolution in the $UV$ luminosity we
find for a galaxy (at a fixed cumulative number density) under the
present $\phi^*$+$\alpha$ evolutionary scenario is essentially
identical to what Bouwens et al.\ (2008) and Bouwens et al.\ (2011)
found previously invoking an evolution in the characteristic magnitude
$M^*$ (Figure~\ref{fig:sfrevol}).  Unless one has very wide-area data
to obtain tight constraints on the bright end of the LF at high
redshift (such as one has with the wide-area CANDELS data set), one
can trade off changes in the characteristic magnitude $M^*$ for
changes in both $\alpha$ and $\phi^*$ (without appreciably affecting
the goodness of fit).  We discuss these issues in more detail in
Appendix F.2, F.3, F.6, and Figure~\ref{fig:oldi}.

An alternate way of looking at the evolution in the $UV$ LF is by
rescaling the volume densities of our derived LFs so that they have
the same normalization at $-21.1$ mag.  We chose to rescale the LFs so
they have the same normalization at this luminosity, which
approximately corresponds to the value of $M^*$ at $z\sim4$-7.  This
allows us to look for systematic changes in the shape of the $UV$ LF
without relying on a specific parameterization of the LF.  The results
are presented in Figure~\ref{fig:shapelf}, and it is clear that the LF
adopts an increasingly steep form at higher redshift.  It is also
clear that the volume density of galaxies at $z\sim4$-7 does not fall
off precipitously until brightward of $-22.5$ mag.

\subsection{Comparison against Previous Results}

Before moving onto a discussion of possible non-Schechter-like
features in the luminosity function, field-to-field variations, or our
LF constraints at $z\sim10$, it is useful to compare the present LF
results with previous results from our own team (e.g., Bouwens et
al.\ 2007; Bouwens et al.\ 2011; Oesch et al.\ 2012b; Oesch et
al.\ 2014) as well as those from other groups.  We include a
comprehensive set of comparisons to previous results in
Figure~\ref{fig:comp410}.

Overall, we find broad agreement with previous LF results over the
full redshift range $z\sim4$-10.  However, there are also some
noteworthy differences, particularly with regard to the volume
densities of the most luminous $z\sim6$-8 galaxies.  In our current
results, we find a higher volume density for luminous
($M_{UV,AB}<-20.5$) $z\sim6$-8 galaxies than reported earlier.

This significant update to our measurement of the volume density of
high luminosity $z\geq6$ galaxies is the direct result of our lacking
sufficiently deep ($H\gtrsim25.5$) near-IR observations over very wide
areas prior to the CANDELS program.  With the new searches, we can now
probe $\sim$15$\times$ more volume than was possible in our earlier
$z\sim7$-8 study (Bouwens et al.\ 2011) and $\sim$3$\times$ more
volume than in our earlier Bouwens et al.\ (2007) $z\sim4$-6 LF
analysis.  Our new LF results agree quite well with our earlier
results, if we only consider the LF constraints over a range which
were well constrainted by previous observations
(Figure~\ref{fig:modelcomp}).

Differences with our previous $z\sim6$ constraints (Bouwens et
al.\ 2007) can be attributed to the large increases in search volume,
the availability of near-IR coverage to do a proper two-color
selection of $z\sim6$ galaxies, and consistent coverage with HST to
minimize the impact of systematics on our results, as one can verify
by using the new WFC3/IR information to $k$-correct previous results
(Appendix F.2).  The explanation for the observed differences with
previous ground-based results (McLure et al.\ 2009; Willott et
al.\ 2013) is less clear, but can at least partially be explained by
uncertainties in deriving total luminosities from the $z$-band fluxes
(both from the IGM correction and $k$-correcting the results to
1600\AA: another $\sim$0.13-mag correction) and also possibly large
field-to-field variance (Bowler et al.\ 2015).  In any case, it is
encouraging that our $z\sim6$ catalog and the Willott et al.\ (2013)
catalogs agree quite well over the search fields where there is
overlap (the brightest $z\sim6$ galaxies we find over the CANDELS
COSMOS and EGS fields are exactly the same $z\sim6$ candidates as
found by Willott et al.\ 2013 and we only miss one of the Willott et
al.\ 2013 candidates over that field).  It is also encouraging that
new wide-area search results from Bowler et al.\ (2015) utilizing both
the UltraVISTA and UDS fields are consistent with our determinations.

At $z\sim7$, our LF results also indicate a much higher volume density
of bright sources than indicated previously in Bouwens et al.\ (2011).
However, this was largely due to our reliance on LF constraints
available from the ground (e.g., from Ouchi et al.\ 2009; Castellano
et al.\ 2010).  If there was an overcorrection for contamination in
those studies or the total magnitudes derived for sources were
systematically fainter ($\sim$0.1-0.2 mag) than those found here, it
could explain the observed differences.  We remark that our present
constraints on $M^*$ and $\alpha$ are in excellent agreement with the
Bouwens et al.\ (2011b) constraints on those quantities if only the
HST search results from that study are considered (Figure 8 from
Bouwens et al.\ 2011 and Figure~\ref{fig:agree7} from Appendix F.3).

Our new LF constraints also show a higher volume density of sources at
$\sim-21.5$ mag than was previously found in either McLure et
al.\ (2013), Schenker et al.\ (2013), or Bowler et al.\ (2014)
studies.  Differences with the McLure et al.\ (2013) and Schenker et
al.\ (2013) results appear to occur due to $\sim0.2$-mag bias in the
total magnitudes measured by McLure et al.\ (2013) and Schenker et
al.\ (2013) for the brightest sources (see Figure~\ref{fig:compm13}
from Appendix H).  Differences relative to Bowler et al.\ (2014)
determinations over the UltraVISTA and UDS fields can potentially be
explained, if Bowler et al.\ (2014) overestimated their completeness
at the faint end of their probe, but we note that our constraint at
$-22.2$ mag is consistent with the Bowler et al.\ (2014)
determinations.  Moreover, it is encouraging that we identify exactly
the same two $z\sim7$ galaxies over the CANDELS COSMOS area we probe
as Bowler et al.\ (2014) find as part of their wide-area search
($z\sim7$ candidates 268576 and 270128 from Bowler et al.\ 2014 lie
over that region of the CANDELS COSMOS field that lacks deep optical
ACS data and hence is not included in our search).

At $z\sim8$, our new LF results are generally in excellent agreement
with all previous HST studies.  However, we do note a slight excess at
the bright end of the $z\sim8$ LF relative to previous studies.  This
excess derives from three particularly bright ($H_{160,AB}\sim25$ mag)
$z>7$ candidate galaxies found over the CANDELS EGS program.  Each of
these candidates appears very likely to be at $z>7$, as they each have
$[3.6]-[4.5]$ colors of $\sim$0.8 mag, very similar to that found by
Ono et al.\ (2012), Finkelstein et al.\ (2013), and Laporte et
al.\ (2014).

For a more extensive set of comparisons with previous work, we refer
the reader to Appendix F.

\begin{figure}
\epsscale{1.14}
\plotone{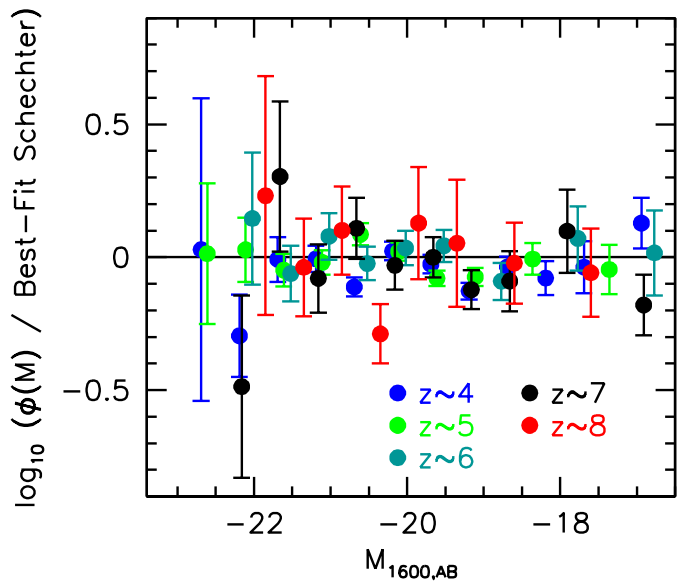}
\caption{Differences (in dex) between the best-fit Schechter LFs at
  $z\sim4$ (\textit{blue circles}), $z\sim5$ (\textit{green circles}),
  $z\sim6$ (\textit{cyan circles}), $z\sim7$ (\textit{black circles}),
  and $z\sim8$ (\textit{red circles}) and the stepwise equivalents.
  Only those bins in the $z\sim4$-8 LFs with uncertainties of $<1$ dex
  are shown.  No significant deviations are found in the stepwise LFs
  relative to the best-fit Schechter functions in this comparison.
  Figure~\ref{fig:powlaw} presents an alternate method for assessing
  the functional form of the $UV$ LFs at
  $z\sim4$-8.\label{fig:spowlaw}}
\end{figure}

\begin{figure*}
\epsscale{0.8}
\plotone{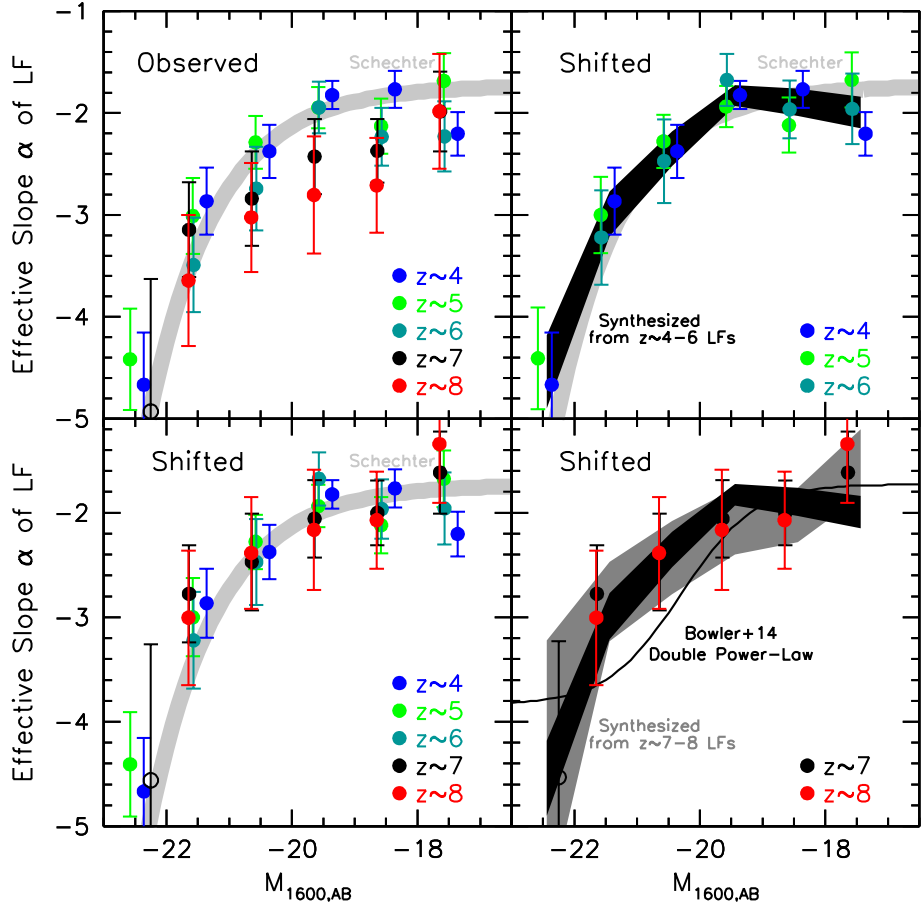}
\caption{(\textit{upper left}) Maximum likelihood determinations of
  the effective slope $d \log_{10} \phi / d\log_{10} L$ of the
  luminosity function at $z\sim4$ (\textit{blue circles}), $z\sim5$
  (\textit{green circles}), $z\sim6$ (\textit{cyan circles}), $z\sim7$
  (\textit{black circles}), and $z\sim8$ (\textit{red circles}) versus
  absolute magnitude $M_{UV}$ (\S4.4).  Each slope determination makes
  use of sources over a 1.5 mag baseline, utilizing the Sandage et
  al.\ (1979) technique.  Uncertainties are $1\sigma$.  Also included
  here is the effective slope of the LF at $z\sim7$ by combining the
  brightest LF bin from Bowler et al.\ (2014) with the second
  brightest LF bin we derive from CANDELS (\textit{open black
    circle}).  The gray curve shows the expected magnitude-dependence
  of the slope for a simple Schechter parameterization with $M^* =
  -21.07$ and $\alpha = -1.73$, while the black line shows this
  dependence for the Bowler et al.\ (2014) double power-law LF.
  (\textit{lower left}) Same determinations of the slope as in the
  upper left panel, but with the mean offset at each redshift removed
  to allow for more direct intercomparisons.  The gray curve is the
  same as shown in the above panel.  (\textit{upper right}) The same
  corrected determinations of the effective slope as in the lower-left
  panels but for the $z=4$-6 LF determinations.  Also shown are
  inverse-variance-weighted constraints on the slope for the average
  $z\sim4$-6 LF vs. luminosity (\textit{dark grey shaded region}).
  Overall, the constraints on the slope of the $UV$ LF all appear
  remarkably similar, after the mean offset is removed, with a form
  very similar to that of a Schechter function.  (\textit{lower
    right}) The same corrected determinations of the effective slope
  as in the lower-left panels but for the $z=7$-8 LF determinations.
  Also shown are the inverse-variance-weighted constraints on the
  slope of the $z=7$-8 LFs (\textit{light grey shaded region}) and
  similar constraints on the LFs at $z=4$-6 (\textit{dark grey shaded
    region}).  The solid black line show the expected dependence for
  the double power-law model preferred by Bowler et al.\ (2014).
  While our current LF constraints are not sufficient to set strong
  constraints on the functional form of the $UV$ LF at $z=7$-8, our
  results seem broadly consistent with the $z=7$-8 LF having a
  Schechter-like form.\label{fig:powlaw}}
\end{figure*}

\begin{figure}
\epsscale{1.14}
\plotone{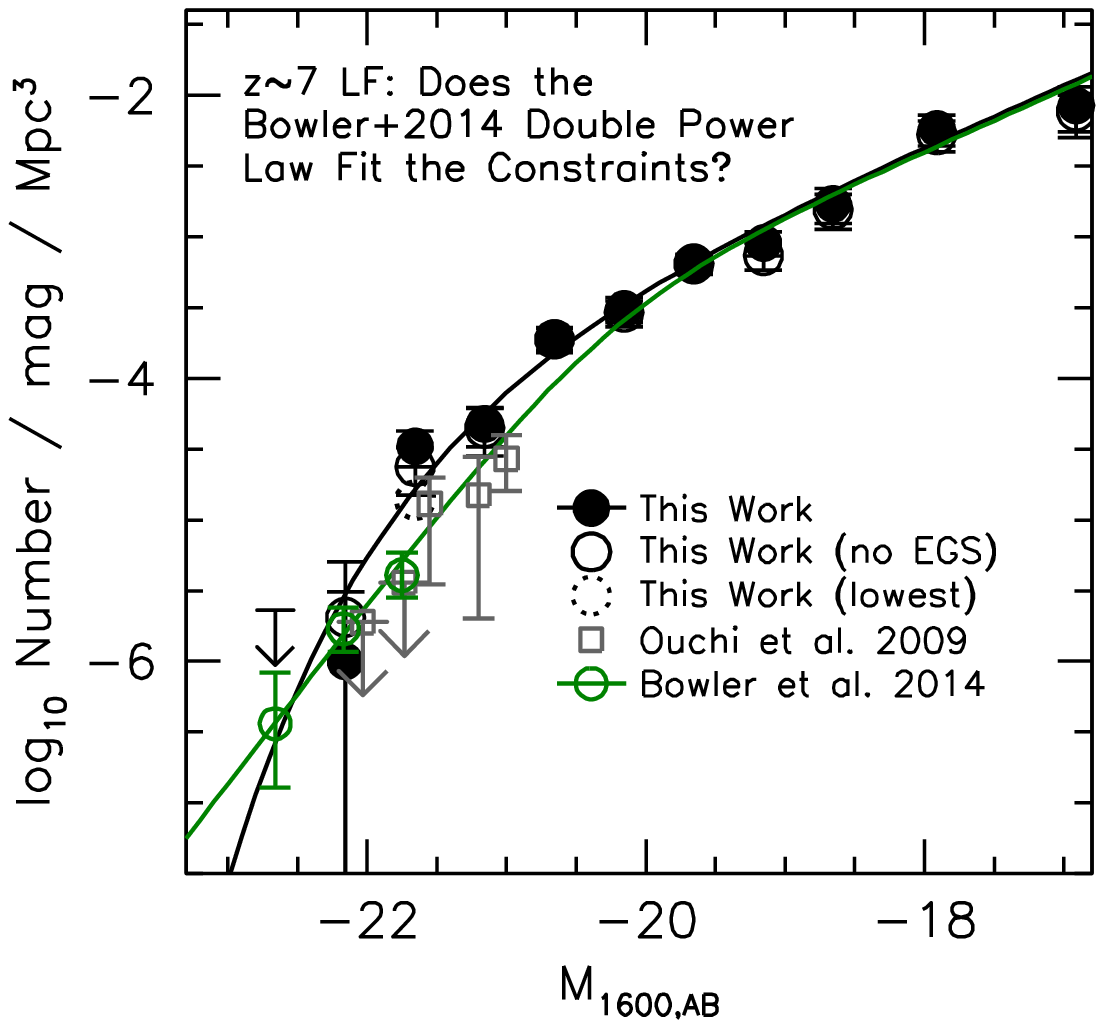}
\caption{Comparison of the present Schechter fit to the $z\sim7$ LF
  (\textit{black line}) with the double power-law fit advocated by
  Bowler et al.\ (2014: \textit{green line}: see \S4.4).  Our
  constraints from the CANDELS+ERS+HUDF09+XDF data set are indicated
  by the solid black circles, while the quoted constraints from the
  ground-based searches of Ouchi et al.\ (2009b) and Bowler et
  al.\ (2014) are shown with the grey open squares (upper limits) and
  open green circles.  Our constraints from the full data set but
  excluding CANDELS-EGS (where a large number of bright $z\sim7$
  galaxies is observed) is shown with the black open circles.  The
  dotted black circle indicates the position of our LF constraint at
  $\sim$$-$21.7 mag, if in addition to excluding the CANDELS-EGS
  field, we suppose (as a worst-case scenario) we have overestimated
  the total magnitude of sources by 0.1 mag and underestimated the
  completeness (and it is 100\%).  While our HST results are generally
  in excellent agreement with the double power-law fit of Bowler et
  al.\ (2014), they disagree with this fit over the range $\sim-22$ to
  $-21$ mag.\label{fig:bowler14}}
\end{figure}

\subsection{Non-Schechter-like Shape of the LF at $z>4$?}

Several previous studies (Bowler et al.\ 2012; Bowler et al.\ 2014)
have presented evidence that the $UV$ LF at $z\sim7$ is not well
represented by a Schechter function, but is rather better represented
by a double power law:
\begin{displaymath}
\phi(M) = \frac{\phi^*}{10^{0.4(\alpha+1)(M-M^*)} + 10^{0.4(\beta+1)(M-M^*)}}
\end{displaymath}
Bowler et al.\ (2014) derive their constraints on the bright end of
the $z\sim7$ LF from the UltraVISTA+UDS fields.  If true, the Bowler
et al.\ (2014) claim would be interesting, as it would imply that the
$UV$ LF at $z\sim7$ does not cut-off abruptly at a specific luminosity
(as it would if the luminosity function were exponential), perhaps
indicating that mass quenching or dust extinction were not as
important early in the history of the universe as they were at later
times.

The depth, area, and \textit{redshift range} provided by our present
samples put us in an unprecedented position to examine the general
shape of the $z\sim4$-8 $UV$ LF and to see whether the $UV$ LF is
better represented by a Schechter-like function, a power law, a double
power-law, or some other functional form.\footnote{This is
  particularly true, given that we also partially make use of search
  constraints available from the $\sim1.7$ deg$^2$ $z\sim7$ LF
  analysis from Bowler et al.\ (2014).}  More precise constraints will
eventually be possible, of course, integrating current HST constraints
with even wider-area probes of the LF.\footnote{Of course, to
  successfully make use of the very wide-area observations for such
  purposes, one must ensure that the total magnitude measurements and
  volume density measurements are made much more consistently than has
  generally been the case in the literature (with $\sim$0.1 mag
  systematic differences in the measured magnitudes being quite
  common: see Appendix F.2 and Skelton et al.\ 2014).  Even 0.05 mag
  differences can result in 0.08 dex (20\%) systematics in the LF
  assuming an effective slope of the LF of $-5$, which is typical at
  the bright end $\sim-22$ mag (see Figure~\ref{fig:powlaw}).}

There are at least two different facets to this endeavor.  The first
regards the shape of the $UV$ LF at the bright end.  Does the $UV$ LF
show an exponential-like cut-off at the bright end or is the bright
end of the LF better represented by a steep power law?  This was the
question Bowler et al.\ (2014) attempted to answer.  The second
regards the shape of the $UV$ LF at the faint end.  Does the effective
slope of the $UV$ LF asymptote to a constant power-law slope (after
modulation by an exponential), or does the effective slope of the $UV$
LF show some dependence on luminosity even at very faint magnitudes?
This second question was considered by Mu{\~n}oz \& Loeb (2011), as
Figure 3 from their paper illustrates quite well.

Perhaps the easiest way to look for deviations from a Schechter-like
form of the $UV$ LF is by comparing the stepwise maximum-likelihood LF
to the Schechter LF determined using the STY technique and computing
the residuals as a function of luminosity.  The result is shown in
Figure~\ref{fig:spowlaw}.  The lack of a significant trend relative to
the best-fit Schechter functions suggest that the $UV$ LFs at
$z\sim4$-8 can be described reasonably well with a Schechter function.

We can look at the overall functional form of the $UV$ LF more
directly by computing the effective slope of the LFs as a function of
luminosity.  This will allow us to assess whether other functional
forms, i.e., a double power law, a rolling power law, or simple power
law, also provide a reasonable representation of the $UV$ LF.  As with
determinations of the $UV$ LF itself, the effective slope of the LF
can be derived over limited range in luminosity using the same
maximum-likelihood technique as we used on the $UV$ LF itself (i.e.,
by Sandage et al.\ 1979).  For simplicity, we only attempt to derive
these slopes at six distinct luminosities along the LF, i.e., $-22.5$,
$-21.5$, $-20.5$, $-19.5$, $-18.5$, and $-17.5$.

In deriving the effective slope of the $UV$ LF at these luminosities,
we only consider sources 0.75 mag brighter and fainter than these
luminosities, providing us with a total luminosity baseline for these
slope measurements of 1.5 mag.  Since this luminosity baseline is
slightly longer than the separation between our slope measurements, we
caution that the slope measurements we derive will not be entirely
independent of each other.  The longer luminosity baseline for each
slope determination is quite useful, though, given the reductions in
uncertainty on each slope measurement.

The result is shown in the panel of Figure~\ref{fig:powlaw} for the
$z\sim4$, $z\sim5$, $z\sim6$, $z\sim7$, and $z\sim8$ LFs.  Also shown
on this figure (\textit{open black circle}) is the effective slope of
the $z\sim7$ LF at $-$22 mag by comparing our $-21.6$ mag $z=7$ LF
constraint with the volume density of $-22.5$ mag $z\sim7$ galaxies
obtained by Bowler et al.\ (2014).  It is clear from these results
that the effective slope of the LF at $z\sim6$, $z\sim7$, and $z\sim8$
is generally steeper than it is at $z\sim4$-5.  We already saw this in
the fit results from the previous section.

To obtain a more precise constraint on the general shape of the LFs at
$z\sim4$-8, we can try to combine the constraints from the LFs at
$z\sim4$, $z\sim5$, $z\sim6$, $z\sim7$, and $z\sim8$ considered
individually.  Motivated by the results from \S4.2, perhaps the best
way of accomplishing is to assume that the effective slope results at
$z=4$-8 LFs are all the same, modulo a change in the zero point (e.g.,
$\Delta\alpha_{z=5}$, $\Delta\alpha_{z=6}$, $\Delta\alpha_{z=7}$,
$\Delta\alpha_{z=8}$).  If we do so and find the offsets that minimize
the overall differences in the inverse-variance-weighted mean
corrected offset at each redshift (specifically minimizing $\Sigma_{M}
\Sigma_z (\alpha_{M,z} + \Delta\alpha_z - \overline{\alpha_M +
  \Delta\alpha})^2/\sigma(\alpha_{M,z})^2$ where $\alpha_{M,z}$
indicates the effective slope measurements at a given absolute
magnitude $M$ and redshift $z$), we find the following offsets in
slope $\Delta\alpha_z$ for the $z\sim5$, $z\sim6$, $z\sim7$, and
$z\sim8$ LFs relative to the $z\sim4$ LF: 0.01, 0.27, 0.37, and 0.64.

We then apply these offsets to the slopes of the $z\sim4$, $z\sim5$,
$z\sim6$, $z\sim7$, and $z\sim8$ LFs shown in the top panel and show
the result in the lower left panel of Figure~\ref{fig:powlaw}.  For
context, we also show in Figure~\ref{fig:powlaw} the luminosity
dependence we would expect adopting the typical Schechter function
results derived in the previous section (\textit{shaded curve}), with
$M^* = -21.07$ and $\alpha=-1.73$.  Overall, the constraints we have
on the slope of the $UV$ LF as a function of luminosity all appear to
be remarkably similar to each other (after one removes the general
offset in slope).

It is interesting to try to combine the constraints we have available
on the $z\sim4$-8 LFs to examine the overall form of the $UV$ LF at
$z\gtrsim4$.  We examine the $z=4$-6 case and the $z=7$-8 cases
separately, given possible evolution in both the shape and functional
form of the LF.  In the two cases, we compute the
inverse-variance-weighted mean effective slope and variance as a
function of luminosity (after removing the zero-point offset in
effective slope).  

The estimated 68\% confidence intervals on the effective slope of the
$z=4$-6 and $z=7$-8 LFs are indicated in the upper right and lower
right panels of Figure~\ref{fig:powlaw} with the dark gray and light
gray regions, respectively.  In general, we find that our
luminosity-dependent slope results are in broad agreement with the
expectations of a Schechter function.  At the low-luminosity end, we
see no evidence for the effective slope of the LF being especially
steeper at $-19.5$ mag than at $-17.5$ mag.  This argues against the
effective slope of the $UV$ LF being strongly luminosity dependent, as
one might expect if there is curvature in the halo mass function or if
galaxy formation were less efficient at lower masses (e.g., Mu{\~n}oz
\& Loeb 2011).

At high luminosities, the $z=4$-6 $UV$ LF show evidence for a similar
exponential-like cut-off at bright magnitudes as that present in a
Schechter function (compare the dark grey region in the upper left
panel of Figure~\ref{fig:powlaw} with the light grey region).  Not
surprisingly, at $z=7$-8, our overall constraints on the shape of the
$UV$ LF at high luminosities are much weaker and clearly not
sufficient to constrain the functional form of the LF.  However, our
results do seem consistent with that observed at $z=4$-6 and also
adopting a Schechter function (compare the light grey region in the
lower left panel with the dark grey region).  For context, we also
show the effective slope results implied from the Bowler et
al.\ (2014) double power-law fit (shown in the lower right panel of
Figure~\ref{fig:powlaw} as the solid black line), i.e., with
$\alpha=-2.1$, $\beta=-4.2$, $M^*=-20.3$, and $\phi^* =
3.9\times10^{-4}$ Mpc$^{-3}$ mag$^{-1}$.  While it is reasonable to
imagine that the $UV$ LF may exhibit a slightly non-Schechter shape at
early enough times, we find no strong evidence for such a behavior
here.

It is interesting to ask why our conclusions appear to differ from
those of Bowler et al.\ (2014).  For the purpose of this discussion,
we compare our LF constraints with the double-power-law fit they find
for their $z\sim7$ LF in Figure~\ref{fig:bowler14}.  While we find
good agreement between the Bowler et al.\ (2014) power-law fit and our
results at both the bright and faint ends, our LF is in excess of
their double power-law fit at moderately high luminosities ($-21.7$
mag), suggesting this is the origin of our different conclusions.  

How reliable are our $z\sim7$ LF constraints at $\sim-21.7$ mag?  In
the luminosity interval $-21.91$ mag to $-21.41$ mag, we find 16
galaxy candidates in total (3, 1, 1, 4, and 7 from the CANDELS-GS, GN,
UDS, COSMOS, and EGS fields, respectively), so the uncertainties from
shot noise (0.12 dex) are relatively limited.  In addition, all 16
appear to be relatively robust $z\sim6.3$-7.3 galaxy candidates, as
inferred from the tests we run in Appendix G and \S3.2.2-\S3.2.3 (for
distinguishing stars and galaxies).  Nevertheless, there are other
issues which could have an impact.  If the large number of bright
sources in the CANDELS-EGS field represent a rare overdensity and we
exclude that field, if our total magnitude estimates are too bright by
0.1 mag, or if the completeness is underestimated (and it is instead
100\%), then our $\sim-21.7$-mag point in the $z\sim7$ LF would be
lower by 0.15 dex, 0.09 dex, and 0.14 dex, respectively.  Even if we
assume all 3 issues to be the case (as a worst-case scenario), our LF
estimates (\textit{open dotted circle} in Figure~\ref{fig:bowler14})
would only be lower by 0.38 dex and still be in tension with the
Bowler et al.\ (2014) $\sim-21.7$-mag point by $\sim$0.4 dex.  Such a
constraint, however, would be in excellent agreement with a Schechter
fit to the LF at $z\sim7$ (but we note that fits to other functional
forms would also be possible given the uncertainties).

\begin{figure}
\epsscale{1.14}
\plotone{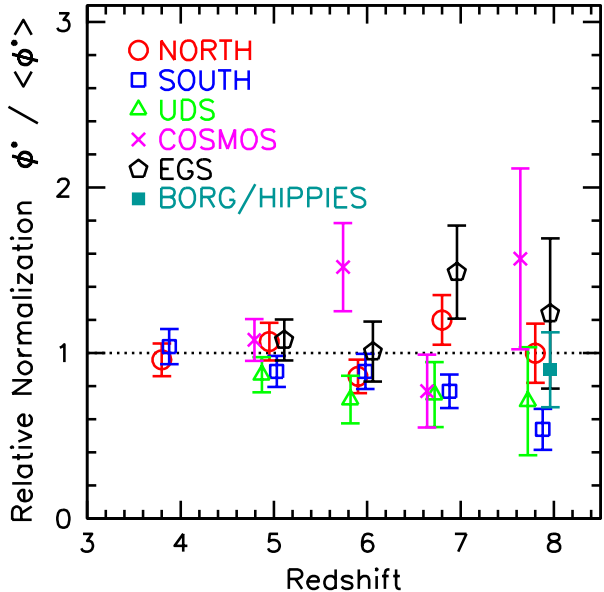}
\caption{The relative normalization $\phi^*$ of the $UV$ LF at various
  redshifts based on sources from the CANDELS-GN (\textit{open red
    circles}), CANDELS-GS (\textit{open blue squares}), CANDELS-UDS
  (\textit{open green triangles}), CANDELS-COSMOS (\textit{magenta
    crosses}), CANDELS-EGS (\textit{open black pentagons}), and
  BoRG/HIPPIES (\textit{solid cyan square}) fields versus redshift
  (\S4.6).  In deriving the relative normalization $\phi^*$ of the LF
  from the individual CANDELS fields, we fix the characteristic
  magnitude $M^*$ and faint-end slope $\alpha$ to the value derived
  based on our entire search area and fit for $\phi^*$.  The plotted
  $1\sigma$ uncertainty estimates are calculated assuming Poissonian
  uncertainties based on the number of sources in each field and
  allowing for small ($\sim$10\%) systematic errors in the calculated
  selection volumes field-to-field.  Specific search fields show a
  significantly higher surface density of candidate galaxies at
  specific redshifts than other search fields (e.g., the CANDELS-EGS
  and CANDELS-GN fields show a higher surface density of $z\sim7$
  candidates than the CANDELS-GS or CANDELS-UDS
  fields).\label{fig:f2f}}
\end{figure}

\subsection{Field-to-Field Variations}

One generic concern for the determination of any luminosity function
is the presence of large-scale structure.  As a result of such
structure, the volume density of sources seen in one's survey fields
can lie significantly above or below that of the cosmic average --
resulting in sizeable field-to-field variations.  While normally these
field-to-field variations introduce considerable uncertainties in our
LF determinations, the availability of deep HST + ground-based
observations over five independent survey fields allows us to largely
overcome this issue.  We estimate that the overall uncertainty on our
LF results to be just 10\%, by using the Trenti \& Stiavelli (2008)
cosmic variance calculator and accounting for the fact that we have
observational constraints over 5 independent $\sim$150 arcmin$^2$
search fields.

\begin{figure}
\epsscale{1.15}
\plotone{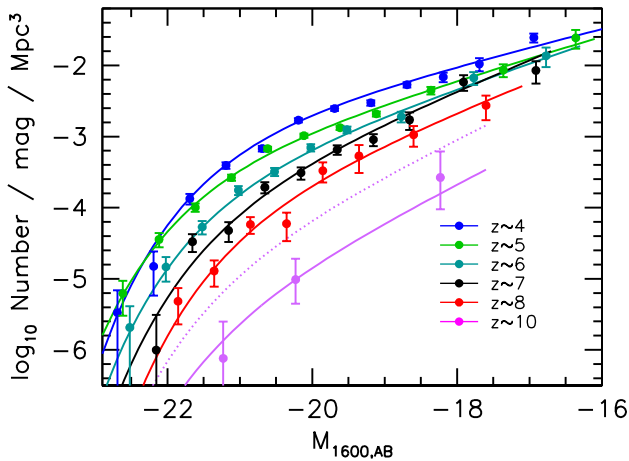}
\caption{SWML determinations of the $UV$ LFs at $z\sim10$
  (\textit{magenta points and $1\sigma$ upper limits}) compared to
  those at lower redshifts (see caption to Figure~\ref{fig:lfall}).
  Also shown are our Schechter fits to the $z\sim10$ LF
  (\textit{magenta line}: see \S4.6).  The dotted magenta line shows
  the LF we would expect extrapolating the $z\sim4$-8 LF results to
  $z\sim10$ using the fitting formula we derive in \S5.1.  We note a
  deficit of fainter ($M_{UV,AB}\gtrsim-19.5$) $z\sim10$ candidates
  relative to the predictions from the fitting formula we present in
  \S5.1, in agreement with the earlier findings of Oesch et
  al.\ (2012a) and Oesch et al.\ (2013a).\label{fig:lfall10}}
\end{figure}

Due to the large number of independent search fields, we can perform a
different test.  Instead of our results on the $UV$ LF being
significantly limited by the impact of cosmic variance, we can use the
current samples to set interesting constraints on the amplitude of the
field-to-field variations themselves.  For simplicity, we assume that
we can capture all variations in the LF through a change in its
normalization $\phi^*$, keeping the characteristic magnitude $M^*$ and
faint-end slope $\alpha$ for galaxies at a given redshift fixed.  The
best-fit values for $\phi^*$ we derive for sources in each field
relative to that found for all fields is shown in Figure~\ref{fig:f2f}
for sources in all five samples considered here.  Bouwens et
al.\ (2007) previously attempted to quantify the differences in
surface densities of $z\sim4$, $z\sim5$, and $z\sim6$ sources over
GOODS North and GOODS South (see also Bouwens et al.\ 2006 and Oesch
et al.\ 2007).  Uncertainties on the value of $\phi^*$ in a field
relative to the average of all search fields is calculated based on
the number of sources in each field assuming Poissonian uncertainties,
allowing for small ($\sim$10\%) systematic errors in the calculated
selection volumes field-to-field.

While the volume density of high-redshift candidates in most wide-area
fields does not differ greatly (typically varying $\lesssim$20\%
field-to-field), there are still sizeable differences present for
select samples field-to-field.  One of the largest deviations from the
cosmic average occurs for $z\sim7$ galaxies over the EGS field where
the volume density appears to be almost double what it is over the
CANDELS-GS, COSMOS, or UDS fields, for example.  The CANDELS-GN also
shows a similar excess at $z\sim7$ relative to these other fields (see
also Finkelstein et al.\ 2013).  The relative surface density of
$z\sim4$, $z\sim5$, and $z\sim6$ candidates over the CANDELS-GN and GS
fields are similar to what Bouwens et al.\ (2007) found previously
(see Table B1 from that work), with the GS field showing a slight
excess in $z\sim4$ and $z\sim6$ candidates relative to GN and the GN
field showing an excess of $z\sim5$ candidates.

Generally however, the observed field-to-field variations are well
within the expected $\sim$20\% variations in volume densities for the
large volumes probed in the present high-redshift samples.

\subsection{$z\sim10$ LF Results}

We also took advantage of our large search areas to set constraints on
the $UV$ LF at $z\sim10$.  Only a small number of $z\sim10$ candidates
were found, but they still provide, along with the upper limits, a
valuable addition to the $z\sim4$-8.  In doing so, we slightly update
the recent LF results of Oesch et al.\ (2014) to consider the
additional search area provided by the CANDELS-UDS, CANDELS-COSMOS,
and CANDELS-EGS fields.

Due to the fact that the majority of our search fields contain zero
$z\sim10$ candidates, we cannot use the bulk of the present fields to
constrain the shape of the LF, making the SWML and STY fitting
techniques less appropriate.  In such cases, it can be useful to
simply derive the $UV$ LF assuming that the source counts are
Poissonian-distributed (given that field-to-field variations will be
smaller than the very large Poissonian uncertainties).  One then
maximizes the likelihood of both the stepwise and model LFs by
comparing the observed surface density of $z\sim10$ candidates with
the expected surface density of $z\sim10$ galaxies in the same way as
we have done before (e.g., Bouwens et al.\ 2008).

Figure~\ref{fig:lfall10} shows the constraints we derive on the
stepwise LF at $z\sim10$ based on the present searches (the $z\sim10$
results are also provided in Table~\ref{tab:swlf}).  A 1-mag binning
scheme is used, given the very small number of $z\sim10$ candidates in
the present search.  Also included on Figure~\ref{fig:lfall10} is our
best-fit Schechter function at $z\sim10$.  For the latter fit, we fix
the characteristic magnitude $M^*$ equal to $-20.92$ and the faint-end
slope $\alpha$ to $-2.27$, consistent with the approximate
characteristic magnitude $M^*$ and faint-end slope $\alpha$ we
estimate based on the LF fitting formula we present in \S5.1.

The best-fit $\phi^*$ we estimate using our $z\sim10$ search over all
of our search fields is $0.000008_{-0.000003}^{+0.000004}$ Mpc$^{-3}$.
We tabulate this value of $\phi^*$ in Table~\ref{tab:lfparm}.  As we
will discuss in Appendix F.5, the best-fit parameters we derive here
are consistent with what Oesch et al.\ (2014) derived previously from
a search over the CANDELS-GN+GS+XDF+HUDF09-Ps fields.  These
parameters are also consistent with the $10\times$ evolution in volume
density that Oesch et al.\ (2013b, 2014) find from $z\sim10$ to
$z\sim8$.

\section{Discussion}

\begin{figure}
\epsscale{1.05}
\plotone{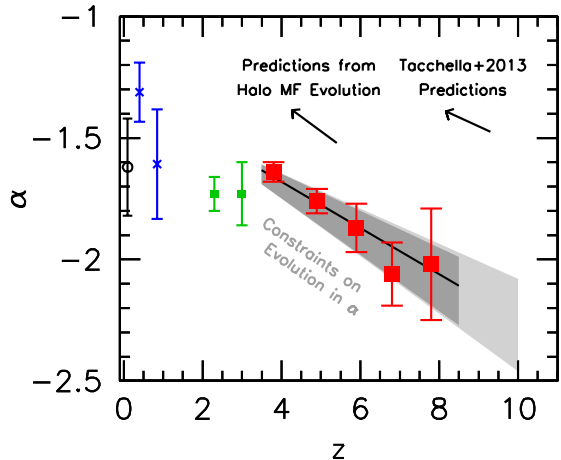}
\caption{Current determinations of the faint-end slope to the $UV$ LF
  (\textit{solid red squares}) versus redshift.  Also shown are the
  faint-end slope determinations from Treyer et al.\ (1998:
  \textit{black open circle}) at $z\sim0$, from Arnouts et al.\ (2005)
  at $z\sim0$-2 (\textit{blue crosses}), and from Reddy et al.\ (2009)
  at $z\sim2$-3 (\textit{green squares}). The solid line is a fit of
  the $z\sim4$-8 faint-end slope determinations to a line, with the
  1$\sigma$ errors (gray area: calculated by marginalizing over the
  likelihood for all slopes and intercepts).  The light gray region
  gives the range of expected faint-end slopes at $z>8.5$ assuming a
  linear dependence of $\alpha$ on redshift.  The best-fit trend with
  redshift is $d\alpha/dz=-0.10\pm0.03$ (\S5.1).  If we keep $M^*$
  fixed, the trend is an even steeper $d\alpha/dz=-0.10\pm0.02$
  (\S5.1).  The overplotted arrows indicate the predicted change in
  the slope of the LF per unit redshift, $d\alpha/dz$, from the
  evolution of the halo mass function based on the conditional LF
  model from \S5.5 and from the Tacchella et al.\ (2013) model (see
  \S5.5.1).  We observe strong evidence for a steepening of the $UV$
  LF from $z\sim8$ to $z\sim4$ (\S5.1).\label{fig:slopeevol}}
\end{figure}

\begin{figure}
\epsscale{1.0}
\plotone{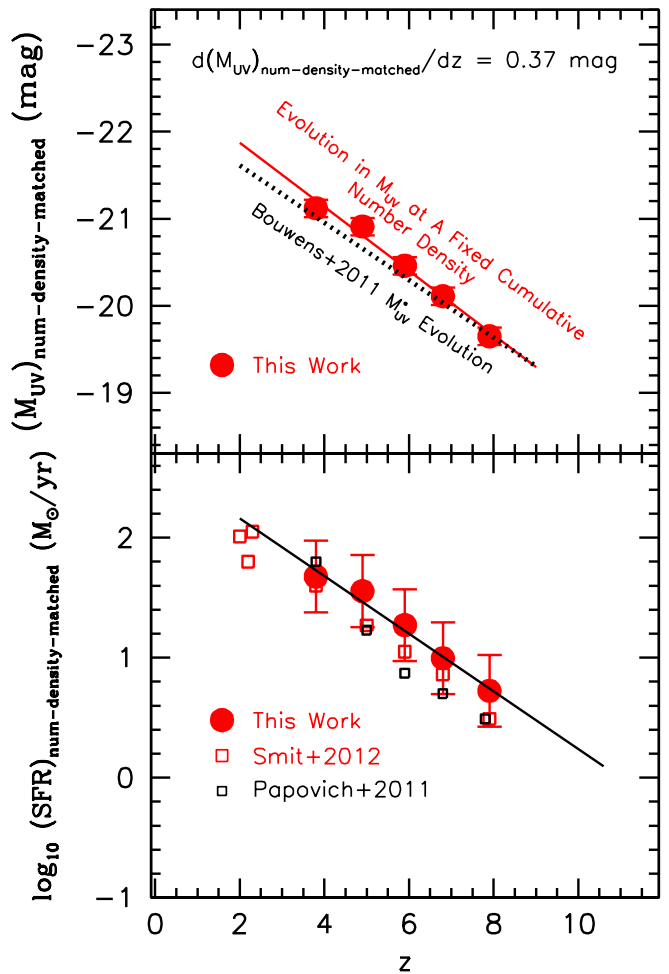}
\caption{(\textit{upper}) The $UV$ luminosities we estimate for
  galaxies from our derived LFs taking galaxies at a fixed cumulative
  number density, i.e., $n(>L_{UV}) = 2\times10^{-4}$ Mpc$^{-3}$
  (identical to the criterion employed by Papovich et al.\ 2011 and
  Smit et al.\ 2012: \S5.3).  Interestingly enough, the best-fit
  evolution in $UV$ luminosity we estimate at a fixed cumulative
  number density (\textit{solid red line}) is quite similar to what
  Bouwens et al.\ (2011) estimated for the evolution in the
  characteristic magnitude $M^*$ (\textit{dotted black line}), before
  strong constraints were available on the bright end of the $UV$ LF
  at $z\gtrsim6$.  (\textit{lower}) The star formation rate we
  estimate for galaxies from our derived LFs to the same cumulative
  number density as in the upper panel.  Results from the literature
  are corrected to assume the same Salpeter IMF assumed for our own
  determinations.  The $z\sim2$ results are based on the mid-IR and
  H$\alpha$ LF results (Reddy et al.\ 2008; Magnelli et al.\ 2011;
  Sobral et al.\ 2013).  The best fit $\textrm{SFR}$ versus redshift
  relation is shown with the black line and can be described as
  follows $(15.8 M_{\odot}\textrm{/yr})10^{-0.24(z-6)}$.  By selecting
  galaxies that lie at a fixed cumulative number density at many
  distinct points in cosmic time, we can plausibly trace the evolution
  in the SFRs of individual galaxies with cosmic
  time.\label{fig:sfrevol}}
\end{figure}

\subsection{Empirical Fitting Formula for Interpolating and Extrapolating our 
LF Results to $z>8$}

As in previous work (e.g., Bouwens et al.\ 2008), it is useful to take
the present constraints on the $UV$ LF and condense them into a
fitting formula for describing the evolution of the $UV$ LF with
cosmic time.  This enterprise has utility not only for extrapolating
the present results to $z>8$, but also for interpolating between the
present LF determinations at $z\sim4$, $z\sim5$, $z\sim6$, $z\sim7$,
and $z\sim8$ when making use of a semi-empirical model.  We will
assume that each of the three Schechter parameter ($M^*$, $\alpha$,
$\log_{10} \phi^*$) depends linearly on redshift when deriving this
formula.  The resultant fitting formula is as follows:
\begin{eqnarray*}
M_{UV} ^{*} =& (-20.95\pm0.10) + (0.01\pm0.06) (z - 6)\\
\phi^* =& (0.47_{-0.10}^{+0.11}) 10^{(-0.27\pm0.05)(z-6)}10^{-3} \textrm{Mpc}^{-3}\\
\alpha =& (-1.87\pm0.05) + (-0.10\pm0.03)(z-6)
\label{eq:empfit}
\end{eqnarray*}
Constraints from Reddy \& Steidel (2009) on the faint-end slope of the
LF at $z\sim3$ were included in deriving the above best-fit relations.
As is evident from these relations, the evolution in the faint-end
slope $\alpha$ is significant at $3.4\sigma$.  The evolution in the
normalization $\phi^*$ of the LF is significant at $5.4\sigma$.  We
find no significant evolution in the value of $M^*$.

Given the considerable degeneracies that exist between the Schechter
parameters, it is also useful to derive the best-fit model if we fix
the characteristic magnitude $M^*$ to some constant value and assume
that all of the evolution in the effective shape of the $UV$ LF is due
to evolution in the faint-end slope $\alpha$.  For these assumptions,
the resultant fitting formula is as follows:
\begin{eqnarray*}
M_{UV} ^{*} =& (-20.97\pm0.06) ~~~~ \textrm{(fixed)} \\
\phi^* =& (0.44\pm0.06) 10^{(-0.28\pm0.02)(z-6)}10^{-3} \textrm{Mpc}^{-3}\\
\alpha =& (-1.87\pm0.04) + (-0.100\pm0.018)(z-6)
\label{eq:empfit2}
\end{eqnarray*}
From this fitting formula, we can see that the steepening in the
effective shape of the $UV$ LF (as seen in Figure~\ref{fig:shapelf})
appears to be significant at 5.7$\sigma$.

The apparent evolution in the faint-end slope $\alpha$ is quite
significant.  Even if we allow for large factor-of-2 errors in the
contamination rate or sizeable ($\sim10$\%) uncertainties in the
selection volume (as we consider in \S4.2), the formal evolution is
still significant at $2.9\sigma$, while the apparent steepening of the
$UV$ LF presented in Figure~\ref{fig:shapelf} remains significant at
$5\sigma$ (instead of $5.7\sigma$).

\subsection{Faint-End Slope Evolution}

The best-fit faint-end slopes $\alpha$ we find in the present analysis
are presented in Figure~\ref{fig:slopeevol}.  The faint-end slope
$\alpha$ we determine is equal to $-1.87\pm0.10$, $-2.06\pm0.13$, and
$-2.02\pm0.23$ at $z\sim6$, $z\sim7$, and $z\sim8$, respectively.
Faint-end slopes $\alpha$ of $\sim -2$ are very steep, and the
integral flux from low luminosity sources can be very large since the
luminosity density in this case is formally divergent.  While clearly
the $UV$ LF must cut off at some luminosity, the $UV$ light from
galaxies fainter than $-$16 should dominate the overall luminosity
density (Bouwens et al.\ 2012a).

In combination with the results at somewhat lower redshifts, the
present results strongly argue for increasingly steep faint-end slopes
$\alpha$ at higher redshifts.  Results from \S5.1 suggest that this
evolution is significant at $3.1\sigma$ if we consider just the formal
evolution in the faint-end slope $\alpha$ itself.  The evolution is
significant at $5.7\sigma$ if we consider the evolution in the shape of
the $UV$ LF (Figure~\ref{fig:shapelf}).

\begin{deluxetable*}{ccccc}
\tablewidth{13cm}
\tabletypesize{\footnotesize}
\tablecaption{$UV$ Luminosity Densities and Star Formation Rate Densities to $-17.0$ AB mag (0.03 $L_{z=3} ^{*}$: see \S5.4).\tablenotemark{a}\label{tab:sfrdens}}
\tablehead{
\colhead{} & \colhead{} & \colhead{$\textrm{log}_{10} \mathcal{L}$} & \multicolumn{2}{c}{$\textrm{log}_{10}$ SFR density} \\
\colhead{Dropout} & \colhead{} & \colhead{(ergs s$^{-1}$} & \multicolumn{2}{c}{($M_{\odot}$ Mpc$^{-3}$ yr$^{-1}$)} \\
\colhead{Sample} & \colhead{$<z>$} & \colhead{Hz$^{-1}$ Mpc$^{-3}$)} & \colhead{Dust Uncorrected} & \colhead{Dust Corrected}}
\startdata
$B$ & 3.8 & 26.52$\pm$0.06 & $-1.38\pm$0.06 & $-1.00\pm0.06$ \\
$V$ & 4.9 & 26.30$\pm$0.06 & $-1.60\pm$0.06 & $-1.26\pm0.06$ \\
$i$ & 5.9 & 26.10$\pm$0.06 & $-1.80\pm$0.06 & $-1.55\pm0.06$ \\
$z$ & 6.8 & 25.98$\pm$0.06 & $-1.92\pm$0.06 & $-1.69\pm0.06$ \\
$Y$ & 7.9 & 25.67$\pm$0.06 & $-2.23\pm$0.07 & $-2.08\pm0.07$ \\
$J$ & 10.4 & 24.62$_{-0.45}^{+0.36}$ & $-3.28$$_{-0.45}^{+0.36}$ & $-3.13$$_{-0.45}^{+0.36}$
\enddata
\tablenotetext{a}{Integrated down to 0.05 $L_{z=3}^{*}$.  Based upon
  LF parameters in Table 2 of Bouwens et al.\ (2011b: see also Bouwens
  et al.\ 2007) (see \S5.4).  The SFR density estimates assume
  $\gtrsim100$ Myr constant SFR and a Salpeter IMF (e.g., Madau et
  al.\ 1998).  Conversion to a Chabrier (2003) IMF would result in a
  factor of $\sim$1.8 (0.25 dex) decrease in the SFR density estimates
  given here.}
\end{deluxetable*}

\begin{figure*}
\epsscale{1.05}
\plotone{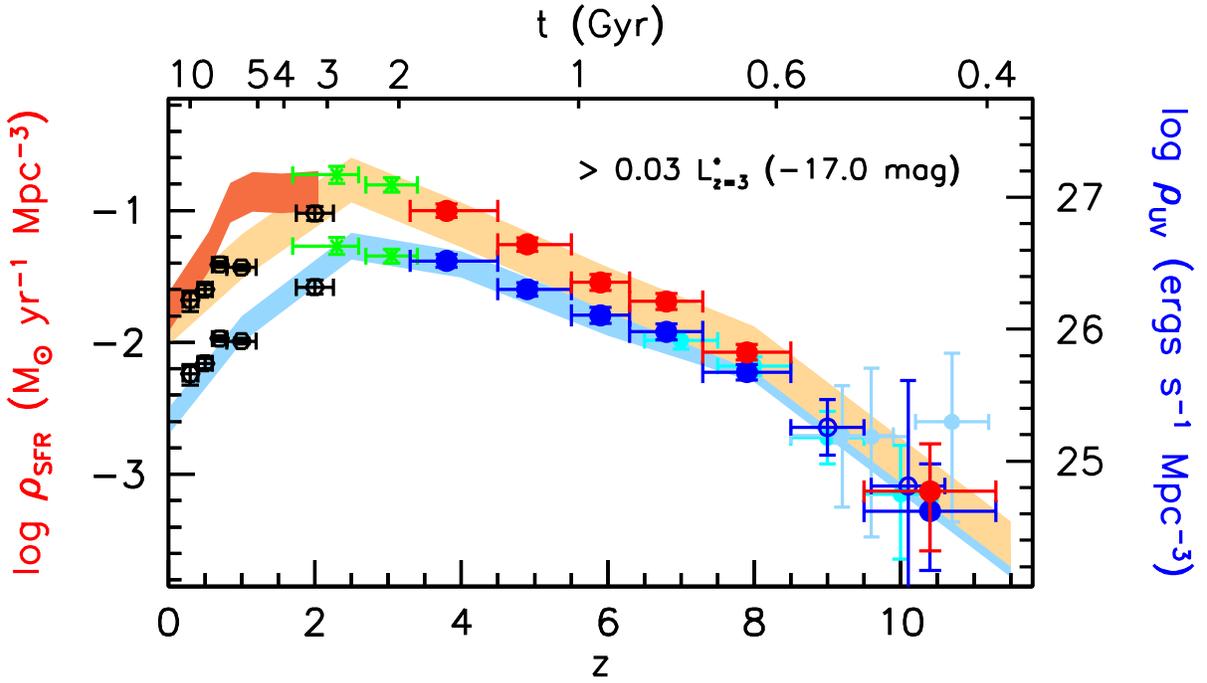}
\caption{Updated determinations of the derived SFR (\textit{left
    axis}) and $UV$ luminosity (\textit{right axis}) densities versus
  redshift (\S5.4).  The left axis gives the SFR densities we would
  infer from the measured luminosity densities, assuming the Madau et
  al.\ (1998) conversion factor relevant for star-forming galaxies
  with ages of $\gtrsim10^8$ yr (see also Kennicutt 1998).  The right
  axis gives the $UV$ luminosities we infer integrating the present
  and published LFs to a faint-end limit of $-17$ mag (0.03
  $L_{z=3}^{*}$) -- which is the approximate limit we can probe to
  $z\sim8$ in our deepest data set.  The upper and lower set of points
  (\textit{red and blue circles, respectively}) and shaded regions
  show the SFR and $UV$ luminosity densities corrected and uncorrected
  for the effects of dust extinction using the observed $UV$ slopes
  $\beta$ (from Bouwens et al.\ 2014a) and the IRX-$\beta$
  relationship (Meurer et al.\ 1999).  Also shown are the SFR
  densities at $z\sim2-3$ from Reddy et al.\ (2009: \textit{green
    crosses}), at $z\sim0$-2 from Schiminovich et al.\ (2005:
  \textit{black hexagons}), at $z\sim7$-8 from McLure et al.\ (2013:
  \textit{cyan circles}), and at $z\sim9$-10 from Ellis et al.\ (2013:
  \textit{cyan circles}), from CLASH (Zheng et al.\ 2012; Coe et
  al.\ 2013; Bouwens et al.\ 2014b: \textit{light blue circles}), and
  Oesch et al.\ (2013b, 2014: \textit{blue open circles}), as well as
  the likely contribution from IR bright sources at $z\sim0.5$-2
  (Magnelli et al.\ 2009, 2011; Daddi et al.\ 2009: \textit{dark red
    shaded region}).  The $z\sim9$-11 constraints on the $UV$
  luminosity density have been adjusted upwards to a limiting
  magnitude of $-17.0$ mag assuming a faint-end slope $\alpha$ of
  $-2.0$ (consistent with our constraints on $\alpha$ at both $z\sim7$
  and at $z\sim8$).\label{fig:sfz}}
\end{figure*}

While consistent with previous results, the present results suggest
slightly steeper faint-end slopes $\alpha$ than reported in Bouwens et
al.\ (2011), McLure et al.\ (2013), and Schenker et al.\ (2013) at
$z\sim7$.  These steeper faint-end slope are a direct consequence of
the somewhat brighter values for $M^*$ that we find in the current
study and the trade-off between fainter values for $M^*$ and steeper
faint-end slopes $\alpha$.  These results only serve to strengthen
earlier findings suggesting that the faint-end slope $\alpha$ is
steeper at $z\sim7$ (and likely $z\sim8$) than it is at $z\sim3$.
Similar conclusions have been drawn from follow-up work on gamma-ray
hosts (Robertson et al.\ 2012; Trenti et al.\ 2012b; Tanvir et
al.\ 2012; Trenti et al.\ 2013).

\subsection{SFR Evolution in Individual Galaxies}

Given the apparent evolution of the $UV$ LF, one might ask how rapidly
the $UV$ luminosity or SFR of an individual galaxy likely increases
with cosmic time.  Fortunately, we can make progress on this question
using a number density-matching procedure,\footnote{Cumulative
  number-density matching can be a powerful way for following the
  evolution of individual galaxies with cosmic time.  This is due to
  the fact that galaxies within a given volume of the universe largely
  grow in a self-similar fashion, so that $n$th brightest or most
  massive galaxy at some point in cosmic time generally maintains its
  ranking in terms of brightness or mass at some later point in cosmic
  time (van Dokkum et al.\ 2010; Papovich et al.\ 2011; Lundgren et
  al.\ 2014).} by ordering galaxies in terms of their observed $UV$
luminosities and following the evolution of those sources with a fixed
cumulative number density.

For convenience, we adopt the same integrated number density
$2\times10^{-4}$ Mpc$^{-3}$ (the approximate cumulative number density
for $L^*$ galaxies) for this question as Papovich et al.\ (2011: see
also Lundgren et al.\ 2014) had previously considered in quantifying
the growth in the SFR of an individual galaxy with cosmic time.  Dust
corrections are performed using the measured $\beta$'s for galaxies at
$z\sim4$-8 (Bouwens et al.\ 2014a) and the well-known IRX-$\beta$
relationship from Meurer et al.\ (1999).

The results are presented in Figure~\ref{fig:sfrevol}.  The $UV$
luminosity at a fixed cumulative number density evolves as $M_{UV}(z)
= -20.40 + 0.37(z-6)$.  Interestingly enough, the evolution in the
$UV$ luminosity we infer for galaxies at some fixed cumulative number
density is almost identical to what Bouwens et al.\ (2011) had
previously inferred for the evolution in the characteristic magnitude
$M^*$ with redshift (i.e., $-20.29 + 0.33(z-6)$: \textit{dotted black
  line}).

Upon reflection, it is clear why this must be so.  For pure luminosity
evolution, one would expect both the characteristic magnitude $M^*$ of
the $UV$ LF and the $UV$ luminosity of individual galaxies to evolve
in exactly the same manner.  Even though we now see that such a
scenario does not work for the brightest, rarest galaxies, one can
nevertheless roughly parameterize the evolution of fainter galaxies
assuming pure luminosity evolution.  For these galaxies, the Bouwens
et al.\ (2008, 2011) fitting formula for $M^*$ evolution works
remarkably well in describing their steadily-increasing $UV$
luminosities.  In this way, the modeling of the evolution of the LF
using $M^*$ evolution by Bouwens et al.\ (2008, 2011) -- a treatment
built on by Stark et al.\ (2009) -- effectively foreshadowed later
work using a sophisticated cumulative number density-matching
formalism to trace the star-formation history of individual systems at
$z>2$ (Papovich et al.\ 2011; Lundgren et al.\ 2014).

The SFR for a galaxy in this number density-matched scenario evolves
as $\textrm{SFR} = (16.2 M_{\odot}\textrm{/yr})10^{-0.24(z-6)}$.  The
evolution in the SFR is remarkably similar to the relations found by
Papovich et al.\ (2011) and Smit et al.\ (2012).  Not surprisingly,
the best-fit trends for galaxies with $L^*$-like volume densities
(i.e., at $\sim$$2\times10^{-4}$ Mpc$^{-3}$) show little dependence on
the parameterization of the Schechter function and whether one fits
the evolution through a change in $M^*$ or a change in $\phi^*$ and
$\alpha$.

\subsection{Luminosity and Star Formation Rate Densities}

We will take advantage of our new LF determinations at $z\sim4$-10 to
provide updated measurements of the $UV$ luminosity density at
$z\sim4$-10.  As in previous work (Bouwens et al.\ 2007, 2008, 2011;
Oesch et al.\ 2012), we only derive the $UV$ luminosity density to the
limiting luminosity probed by the current study at $z\sim8$, i.e.,
$-17$ mag (0.03 $L_{z=3}^{*}$), to keep these determinations as
empirical as possible.  Since this is slightly fainter than what one
can probe in searches for galaxies at $z\sim10$, we make a slight
correction to our $z\sim9$ and $z\sim 10$ results.  The best-fit
faint-end slope $\alpha=-2$ we find at $z\sim8$ is assumed in this
correction.  The use of even steeper faint-end slopes (i.e., $-2.3$)
as implied by our LF fitting formula in \S5.1 would yield similar
results, only increasing the luminosity density by $\sim$0.015 dex.

\begin{figure*}
\epsscale{0.8}
\plotone{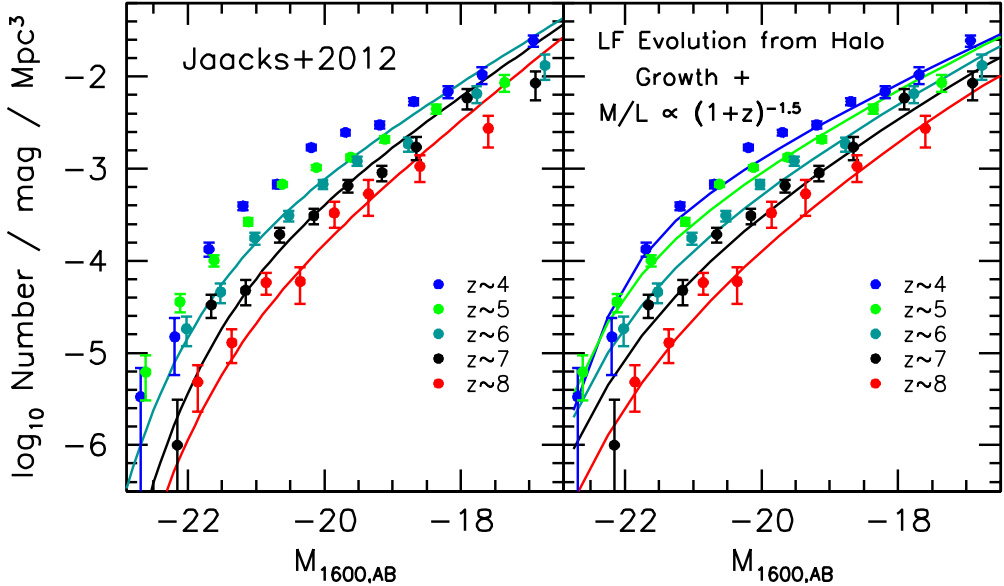}
\caption{Comparison of the observed $UV$ LFs with the simulation
  results from Jaacks et al.\ (2012: \textit{left panel}) and the
  predictions of a simple conditional luminosity function (CLF) model
  based on halo growth (Bouwens et al.\ 2008: \textit{right panel}).
  The Jaacks et al.\ (2012) curves are for $z\sim8$, $z\sim7$, and
  $z\sim6$.  As described in \S5.5, the Jaacks et al.\ (2012) results
  show the predictions of a sophisticated cosmological hydrodynamical
  simulation for the LF, while the CLF model shows the predicted
  evolution based on the expected evolution of the halo mass function
  and a mass-to-light ratio that evolves as $(1+z)^{-1.5}$ (see
  Appendix I).  While the Jaacks et al.\ (2012) model overpredicts the
  observed steepening of the $UV$ LF towards high redshift
  ($d\alpha/dz\sim-0.17$ vs. $d\alpha/dz=-0.10 \pm 0.02$), the simple
  conditional LF model considered here predicts the observed
  steepening quite well ($d\alpha/dz\sim-0.12$
  vs. $d\alpha/dz=-0.10 \pm 0.02$).  The luminosity per unit halo mass
  for lower-mass galaxies may increase more rapidly towards high
  redshift than for higher-mass galaxies.  Our CLF model predicts a
  cut-off in the $UV$ LF at $z>6$ brightward of $-23$ mag, in apparent
  agreement with the observations.\label{fig:theorylf}}
\end{figure*}

In combination with our estimates of the luminosity density, we also
take this opportunity to provide updated measurements of the star
formation rate density at $z\sim4$-10.  In making these estimates of
the SFR density at $z\sim4$-10, we correct for dust extinction using
the well-known IRX-$\beta$ relationship (Meurer et al.\ 1999) combined
with the latest measurements of $\beta$ from Bouwens et al.\ (2014a).
As before, we assume that the extinction $A_{UV}$ at rest-frame $UV$
wavelengths is $4.43+1.99\beta$, with an intrinsic scatter of 0.35 in
the $\beta$ distribution.  This is consistent with what has been found
for bright galaxies at $z\sim4$-5 (Bouwens et al.\ 2012b; Castellano
et al.\ 2012).  The new $\beta$ determinations from Bouwens et
al.\ (2014a) utilize large $>$4000-source samples constructed from the
XDF, HUDF09-1, HUDF09-2, ERS, CANDELS-GN, and CANDELS-GS data sets and
were constructed to provide much more accurate and robust measurements
of the $\beta$ distribution than has been provided in the past.  The
mean dust extinction we estimate based on the Meurer et al.\ (1999)
law for the observed $\beta$ distribution is 2.4, 2.2, 1.8, 1.66, 1.4,
and 1.4 (in units of $L_{IR}/L_{UV}+1$ where $L_{IR}$ and $L_{UV}$ are
the bolometric and UV luminosities of a galaxy, respectively) for the
observed galaxies at $z\sim4$, $z\sim5$, $z\sim6$, $z\sim7$, $z\sim8$,
and $z\sim10$, respectively.

The dust-corrected $UV$ luminosity densities are then converted into
SFR densities using the canonical Madau et al.\ (1998) and Kennicutt
et al.\ (1998) relation:
\begin{equation}
L_{UV} = \left( \frac{\textrm{SFR}}{M_{\odot} \textrm{yr}^{-1}} \right) 8.0 \times 10^{27} \textrm{ergs}\, \textrm{s}^{-1}\, \textrm{Hz}^{-1}\label{eq:mad}
\end{equation}
where a $0.1$-$125\,M_{\odot}$ Salpeter IMF and a constant star
formation rate for ages of $\gtrsim100$ Myr are assumed.  In light of
the very high EWs of the $H\alpha$ and [OIII] emission lines in
$z\sim4$-8 galaxies (Schaerer \& de Barros 2009; Shim et al.\ 2011;
Stark et al.\ 2013; Schenker et al.\ 2013; Gonz{\'a}lez et al.\ 2012;
Labb{\'e} et al.\ 2013; Smit et al.\ 2014; Gonz{\'a}lez et al.\ 2014),
it is probable that the adopted conversion factors underestimate the
actual SFRs (perhaps by as much as a factor of 2: Castellano et
al.\ 2014).

Our updated results on both the luminosity density and star-formation
rate density are presented in Table~\ref{tab:sfrdens} and
Figure~\ref{fig:sfz}.  As before, we have included select results from
the literature (Schiminovich et al.\ 2005; Reddy \& Steidel 2009) to
show the trends at $z<4$, as well as presenting recent determinations
of the star formation rate density at $z\sim0.5$-2.0 from IR bright
sources (Daddi et al.\ 2009; Magnelli et al.\ 2009, 2011).  We also
include select $z\geq6$ results from the literature for comparison
with previous results (Zheng et al.\ 2012; Coe et al.\ 2013; McLure et
al.\ 2013; Ellis et al.\ 2013; Bouwens et al.\ 2014b).

We observe very good agreement with previous results over the full
range in redshift $z\sim4$-10.  The most noteworthy changes occur at
$z\sim5$ where the volume density we find is higher than estimated
previously (Bouwens et al.\ 2007) and better matches the evolutionary
trend connecting the $z\sim4$ and $z\sim6$ results.  The improved
robustness of the present $z\sim5$ results is likely a direct
consequence of the significantly broader wavelength baseline available
to select $z\sim5$ galaxies over the $z\sim4.5$-5.5 volume than was
available in the earlier purely optical/ACS data set (e.g., see
discussion in Duncan et al.\ 2014).

\subsection{Comparison with Theoretical Models}

It is interesting to compare the current observational results with
what is found from large hydrodynamical simulations and also from
simple theoretical models.  Such comparisons are useful for
interpreting the present results and also for ascertaining whether any
of our observational results are unexpected or challenge the current
paradigm in any way.  We first describe the models and then in the
following subsections we discuss comparisons with our new LF results.

The first set of cosmological hydrodynamical simulations we consider
are those from Jaacks et al.\ (2012).  These results provide a very
detailed investigation as to how the shape of the $UV$ LF might evolve
with cosmic time.  Jaacks et al.\ (2012) make use of some large
simulations done on a modified version of the GADGET-3 code (Springel
et al.\ 2005) that includes cooling by H+He+metal line cooling,
heating by a modified Haardt \& Madau (1996) spectrum (Katz et
al.\ 1996), an Eisenstein \& Hu (1999) initial power spectrum,
``Pressure model'' star formation (Schaye \& Dalla Vecchia 2008),
supernovae feedback, and multiple-component variable velocity wind
model (Choi \& Nagamine 2011).  Simulations are done with a range of
box sizes from 10 $h^{-1}$ Mpc to 100 $h^{-1}$ Mpc ($2\times 600^3$ or
$3\times 400^3$ particles).

As an alternative to the results from large hydrodynamical
simulations, we make use of a much more simple-minded theoretical
model using a conditional luminosity function (CLF: Yang et al.\ 2003;
Cooray \& Milosavljevi{\'c} 2005) formalism where one derives the LF
from the halo mass function using some mass-to-light kernel.  We adopt
the same CLF model as Bouwens et al.\ (2008) had previously used in
their analysis of the $UV$ LF, but have modified the model to include
a faster evolution in the M/L of halos, i.e., $\propto (1+z)^{-1.5}$.
This evolution better reproduces changes in the observed $UV$ LF from
$z\sim8$ to $z\sim4$.  The $(1+z)^{-1.5}$ factor also matches the
expected evolution of the dynamical time scale.  A detailed
description of this model is provided in Appendix I.  The advantage of
this approach is that it can give us insight into the extent to which
the evolution in the $UV$ LF is driven by the growth of dark matter
halos themselves and to what extent the evolution arises from changes
in the mass-to-light ratio of those halos and hence gas dynamical
processes (e.g., gas cooling or SFR time scales).

Finally, we consider the predictions by Tacchella et al.\ (2013),
which are based on a minimal model that also links the evolution of
the UV galaxy luminosity function to that of the dark-matter halo mass
function.  The model is constructed by assuming that a halo of mass
$M_h$ at redshift $z$ has a stellar mass $M_*=\epsilon(M_h)*M_h$, of
which a small fraction (10\%) is formed at the halo assembly time
$z_a$, while the remaining is formed at a constant rate from $z_a$ to
$z$.  Since halos have shorter assembly times as redshift increases,
the UV light to halo mass ratio increases with redshift.
$\epsilon(M_h)$ describes the efficiency of the accretion in forming
stars; Tacchella et al.\ (2013) calibrate it at $z=4$ via abundance
matching.

Before conducting detailed comparisons of the observational results
with the above theoretical models, we first present a comparison of
the binned LF results with the first two theoretical models to
illustrate the broad overall agreeement between the two sets of
results (Figure~\ref{fig:theorylf}).

\subsubsection{Expected Evolution of the Faint-End Slope}

The present observational results provide compelling evidence for
significant evolution in the effective slope of the $UV$ LF
(Figure~\ref{fig:shapelf}).  While some of the evolution in the
effective slope of the $UV$ LF may be due to a change in the
characteristic magnitude $M^*$, most of the evolution appears to
result from an evolving faint-end slope $\alpha$.

In comparing the present observational results with theory, let us
assume that we can effectively parameterize the entire shape evolution
of the LF using the faint-end slope $\alpha$ (and because we do not
finding convincing evidence for evolution in $M^*$).  This assumption
is useful, since it distills the shape information present in the
moderately-degenerate $M^*$+$\alpha$ combination into a single
parameter, resulting in a smaller formal error on the evolution.  As
shown in \S5.1, we derive $d\alpha/dz=-0.10 \pm 0.02$ from the
observations, if we force $M^*$ to be constant in our fits.

Remarkably enough, our simple-minded conditional LF model (Appendix I)
is in remarkable agreement with our observational results, predicting
that the faint-end slope $\alpha$ of the LF evolves as $d\alpha/dz
\sim -0.12$.  This compares with $d\alpha/dz \sim -0.17$ predicted
from the Jaacks et al.\ (2012) simulation results.  Finally, the
Tacchella et al.\ (2013) model predict an evolutionary trend
$d\alpha/dz$ of $-0.08$.  Each of these predictions is very similar to
the observed evolution (see Figure~\ref{fig:slopeevol}) of $d\alpha/dz
= -0.10 \pm 0.02$.

\subsubsection{Expected Evolution in the Characteristic Luminosity?}

Our discovery of modest numbers of highly luminous galaxies in each of
our high redshift samples, even at $z\sim10$, provides strong evidence
against a rapid evolution in the luminosity where the $UV$ LF cuts
off.  Over the redshift range $z\sim4$ to $z\sim7$, we find no
significant evolution in $M^*$ (see Table~\ref{tab:lfparm}).  Over the
slightly wider redshift range $z\sim4$ to $z\sim8$, our best-fit
estimate for the evolution in the characteristic magnitude $M^*$ is
just $dM^*/dz\sim 0.01\pm0.06$ (see the fitting formula in \S5.1) or
just $dM^*\sim0.25 \pm0.37$ from $z\sim8$ to $z\sim4$.  Given the
observed luminosity of the brightest $z\sim10$ candidates found over
the CANDELS fields (Oesch et al.\ 2014), i.e., $-21.4$ mag, it seems
unlikely that the bright-end cut-off $M^*$ is especially fainter than
$M^*\sim-20$ (limiting the evolution in $M^*$ to $\lesssim\,$1 mag
over the redshift range $z\sim4$-10).

\begin{figure}
\epsscale{0.8}
\plotone{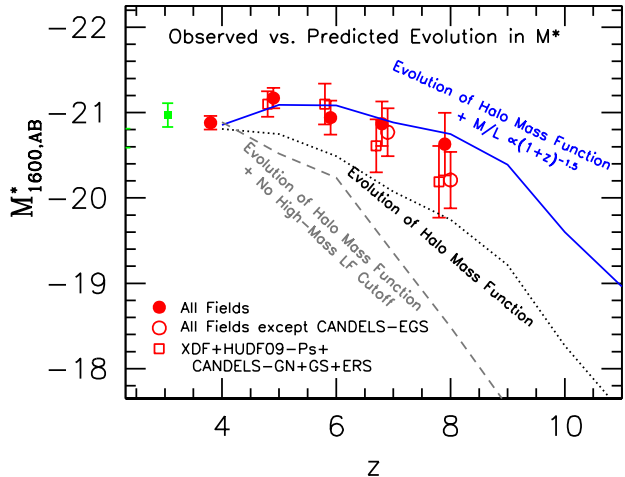}
\caption{Comparison of the observed evolution in the characteristic
  magnitude $M^*$ with that expected from a simple CLF model based on
  the growth in the halo mass function (Bouwens et al.\ 2008: Appendix
  I).  Shown separately (and horizontally offset for clarity) are our
  characteristic magnitudes $M^*$ determinations
  (Table~\ref{tab:lfparm}) for all of the fields in our analysis
  (\textit{solid red circles}), all of the fields in our analysis but
  CANDELS-EGS (\textit{open red circles}), and the
  XDF+HUDF09-Ps+ERS+CANDELS-GN+GS fields (\textit{open red squares}).
  The green cross is the characteristic magnitude determination at
  $z\sim3$ from Reddy \& Steidel (2009).  The gray dashed line shows
  the expected evolution in $M^*$ for simple-minded CLF models that do
  not include a cut-off at the bright end of the $UV$ LF
  (renormalizing the mass-to-light ratio to match $M^*$ at $z\sim4$).
  The black dotted and blue solid lines show the expected evolution in
  $M^*$ for CLF models where the mass-to-light ratio of halos is
  constant in time or evolves as the dynamical time scale, i.e., as
  $(1+z)^{-3/2}$ (\textit{blue line}).  At sufficiently high redshift,
  it seems clear that we would expect the characteristic magnitude
  $M^*$ to be fainter due to evolution in the halo mass function.  In
  practice, the evolution in the characteristic magnitude $M^*$ may be
  more limited (1) if the bright-end cut-off to the $UV$ LF (above
  some mass threshold) is instead set by a physical process (e.g.,
  dust obscuration or quenching) and (2) if halos at higher redshifts
  have systematically lower mass-to-light ratios.\label{fig:theoryms}}
\end{figure}

This implies that whatever physical mechanism imposes a cut-off at the
bright end of $z\gtrsim 4$ $UV$ LFs, this cut-off luminosity does not
vary dramatically with redshift, at least out to $z\sim7$.  Indeed,
for the three mechanisms discussed by Bouwens et al.\ (2008) to impose
a cut-off at the bright end of the $UV$ LF, i.e., heating from an AGN
(Croton et al.\ 2005), the inefficiency of gas cooling for high-mass
halos (e.g., Binney 1977; Rees \& Ostriker 1977; Silk 1977), and the
increasing importance of dust attenuation for the most luminous and
likely most massive galaxies (Bouwens et al.\ 2009; Pannella et
al.\ 2009; Reddy et al.\ 2010; Bouwens et al.\ 2012; Finkelstein et
al.\ 2012; Bouwens et al.\ 2014a), there is no obvious reason any of
these mechanisms should depend significantly on redshift or cosmic
time.

Indeed, the results of the simulations or theoretical models bear out
these expectations.  The best-fit characteristic magnitudes $M^*$
derived from the Jaacks et al.\ (2012) simulations show very little
evolution with cosmic time.  Jaacks et al.\ (2012) derive $-21.15$,
$-20.85$, and $-21.0$ for their $z\sim6$, $z\sim7$, and $z\sim8$ LFs,
respectively.

Simple fits to our CLF results also show only limited evolution in the
characteristic magnitude $M^*$ with redshift, even out to $z\sim10$.
The characteristic magnitudes we derive from fitting the model LFs at
$z\sim4$-10 (minimizing the square of the logarithmic residuals) are
presented in Figure~\ref{fig:theoryms} for comparison with our
observational determinations of this same quantity
(Table~\ref{tab:lfparm}).  Both a model assuming fixed mass-to-light
ratios for the halos (\textit{black line}) and a model with
mass-to-light ratios evolving as the dynamical time ($(1+z)^{-3/2}$:
\textit{blue line}) are considered.

It is useful to contrast these results with a CLF model where no
cut-off is imposed at the bright-end of the $UV$ LF and where there is
no evolution in the mass-to-light ratio of halos.  For the model
described in Appendix I, this could be achieved by replacing the
$(1+(M/m_c))$ term in Eq.~\ref{eq:kernel} by unity and renormalizing
the mass-to-light ratio so that $M^*$ for the model LF is equal to
$-21$ at $z\sim4$.  The evolution in the characteristic magnitude
$M^*$ we would predict for this model is shown with the dashed gray
line in Figure~\ref{fig:theoryms}.

At sufficiently high redshift, it seems clear from the gray line that
we would expect the characteristic magnitude $M^*$ to be fainter due
to evolution in the halo mass function.  In practice, however, the
evolution in the characteristic magnitude $M^*$ may be more limited if
the bright-end cut-off to the $UV$ LF is instead set by a physical
process that becomes dominant at some mass threshold (e.g., dust
obscuration or quenching), as the dotted black line in
Figure~\ref{fig:theoryms} illustrates.  Even less evolution would be
expected in the characteristic magnitude $M^*$ with cosmic time if
halos at higher redshifts had systematically lower mass-to-light
ratios, as illustrated by the blue line in this same figure.

In reality, of course, we should emphasize that almost all LFs
predicted by simulations or CLF models can only be approximately
modelled using a Schechter-function-like parameterization, and
therefore there can be considerable ambiguity in actually extracting
the Schechter parameters from the model results and hence representing
their evolution with cosmic time.

\section{Summary}

The HUDF/XDF, HUDF09-1, HUDF09-2, ERS, and the five CANDELS fields
contain a great wealth of deep, wide-area multiwavelength observations
from the Hubble Space Telescopes and other facilities like Spitzer.
Observations over these fields reach as deep as 30 mag ($5\sigma$),
cover a total area of $\sim$750 arcmin$^2$, and include deep coverage
in at least six passbands from HST and Spitzer, from
$\sim$0.6-4.5$\mu$m.  $\sim$1000 arcmin$^2$ area is leveraged in total
including the BoRG/HIPPIES program.  These exceptional depths, area,
and quality make these fields a great resource for identifying
galaxies over a wide range in both luminosity and redshift, from
$z\sim4$ to $z\sim10$.

Making use of this significant data set and a more efficient selection
methodology we have developed, we have identified $\sim$10400
star-forming galaxies over the redshift range $z\sim4$-10, including
more than 698 probable galaxies at $z\sim7$-8, and 6 candidate
galaxies at $z\sim10$.  This is the largest such sample of galaxies
assembled to date.  The color criteria we introduce here for the
selection of galaxies in the redshift range $z\sim4$-10 now makes full
use of the wavelength leverage available from the near-IR observations
and has been optimized to be essentially complete, with no gaps in
redshift between adjacent samples (Figure~\ref{fig:zdist}).  This
methodology produces comparably-sized samples and redshift segregation
to what one can achieve segregating samples by their best-fit
photometric redshifts, but retains the essential simplicity,
reproducibility, and robustness against contamination that color
criteria can particularly provide (\S3.2).

We make use of these unprecedented samples to derive the $UV$ LF in
six distinct redshift intervals, at $z\sim4$, $z\sim5$, $z\sim6$,
$z\sim7$, $z\sim8$, and $z\sim10$.  We utilize essentially the same
procedures as we previously utilized in Bouwens et al.\ (2007) and
Bouwens et al.\ (2011).  The selection volumes and selection
efficiency for these samples are calculated by pixel-by-pixel
redshifting actual $z\sim4$ galaxies from the HUDF to higher redshift
according to the observed size-redshift $(1+z)^{-1.2}$ relationship
(Oesch et al.\ 2010; Ono et al.\ 2013), inserting these sources into
the actual observations, and then attempting to reselect these sources
and measure their properties using the same procedure that we use on
the real observations.  We explicitly verified that the size of the
average sources in our simulations was well matched to the size of
sources in the observations, as a function of both redshift and
luminosity (Figure~\ref{fig:sizecalib}: Appendix D).

Five different types of contamination are considered for our samples,
i.e., contamination from photometric scatter, contamination from
stars, contamination from extreme emission line galaxies,
contamination from supernovae, and contamination from spurious
sources.  We estimate a contamination level of 2\%, 3\%, 6\%, 7\%, and
10\% for all but the faintest sources in our $z\sim4$, $z\sim5$,
$z\sim6$, $z\sim7$, and $z\sim8$ samples, respectively.  As in most of
our previous studies, the only significant source of contamination is
from the impact of noise on the photometry of individual sources
(``photometric scatter'').

The low contamination rate is the result of great care being taken
throughout the selection process to minimize the impact of potential
contamination on our high-redshift samples.  We validated our
selection volume estimates in our wide-area fields, with a
sophisticated set of degradation experiments, through the repeated
addition of noise to our deepest data sets to match that found in our
shallower data (\S3.5.5).  Similar use of these degradation
experiments was made to determine the impact of higher noise levels on
the total magnitudes measured for sources in our fields.

Extensive comparisons were made between the present LF results and
some of the more noteworthy LF determinations from the literature
(\S4.3 and Appendix F).  This is to provide us with the most
comprehensive possible perspective from which to identify systematics
in current and previous studies of the LF.  In cases of differences,
substantial effort was made to understand those differences, so as to
make our final LF results as accurate as possible.

Our use of all five CANDELS fields to derive our high-redshift
luminosity functions makes our results quite robust against the impact
of cosmic variance, given that each CANDELS field provides us with an
entirely independent sightline on the high redshift universe.  The
availability of different sightlines puts us in position to quantify
the variation in the $UV$ LF from field to field and therefore set
accurate empirical constraints on the large-scale structure
uncertainties (\S4.5, Figure~\ref{fig:f2f}, and Appendix G).

Our conclusions are as follows:
\begin{itemize}
\item{Taking advantage of the widest-area systematic search for
  galaxies in the redshift range $z\sim4$-10, we show that galaxies
  remain moderately prevalent ($\gtrsim$5$\times$10$^{-6}$ Mpc$^{-3}$)
  to $UV$ luminosities of $-22$ mag over the entire redshift range
  $z\sim4$ to $z\sim8$ (\S4.1).  The volume density of galaxies only
  begins to fall off rapidly brightward of this magnitude.  Sharp
  cut-offs in the $UV$ LF were previously only found brightward of
  $-22.5$ by van der Burg et al.\ (2010) for $z\sim4$-5 samples and
  brightward of $-22$ mag for $z\sim6$ samples by Willott et
  al.\ (2013) using the CFHT deep legacy survey fields.}
\item{While our $z\sim4$-7 LFs are still in excellent agreement with
  our previous results over the range in luminosity and volume density
  well probed by our previous studies (Bouwens et al.\ 2007, 2008,
  2011: Figure~\ref{fig:modelcomp}), the relatively robust constraints
  we have on the volume density of bright ($M_{UV,AB}<-21$) galaxies
  at $z\sim4$-8 from the wide-area CANDELS program allow for at most
  modest evolution in the characteristic magnitude $M^*$ with cosmic
  time (assuming a Schechter form for th LF).  This suggests that
  whatever physical mechanism is responsible for imposing a cut-off in
  the $UV$ LF at high luminosities (i.e., AGN feedback, inefficient
  gas cooling, high dust extinction) does not evolve dramatically with
  cosmic time (\S5.5.2).  The limited evolution in $M^*$ we observe is
  also consistent with the observational results of van der Burg et
  al.\ (2010) and simulation results of Jaacks et al.\ (2012).}

\item{We find significant evidence ($3.4\sigma$) for a steepening of
  the faint-end slope $\alpha$ from $\alpha=-1.64\pm0.04$ at $z\sim4$
  to $\alpha= -2.06\pm0.13$ at $z\sim7$ and $\alpha=-2.02\pm 0.23$ at
  $z\sim8$.  Previously, some evidence for a steepening of the $UV$ LF
  was presented by Bouwens et al.\ (2011), Su et al.\ (2011), Bradley
  et al.\ (2012), Schenker et al.\ (2013), McLure et al.\ (2013), and
  Calvi et al.\ (2013).  The present study considerably strengthens
  the conclusions from these earlier studies, given the much tighter
  constraints we now have on the faint-end slope $\alpha$ of the $UV$
  LF at $z\sim5$-6 and self-consistent approach we have used to treat
  the $UV$ LFs over the range $z\sim4$ to $z\sim 8$.  The observed
  evolution appears to be in excellent agreement with that predicted
  from the steepening of the halo mass function (see \S5.5.2), e.g.,
  as seen in the results of Jaacks et al.\ (2012) and Tacchella et
  al.\ (2013).}
\item{Due to the strong limits we can set on the evolution in the
  characteristic magnitude $M^*$ from the current samples and the
  significant evolution in the $UV$ LF itself with cosmic time, some
  evolution in the normalization $\phi^*$ of the LF appears to be
  required.  From $z\sim7$ to $z\sim4$, $\phi^*$ increases by nearly
  $6\times$ from 0.000008 Mpc$^{-3}$ to 0.00197 Mpc$^{-3}$ (see
  Figure~\ref{fig:mlcontours} and Table~\ref{tab:lfparm}: \S4.2).
  While such a scenario might seem similar to that preferred by van
  der Burg et al.\ (2010) and Beckwith et al.\ (2006), a good fit to
  the overall evolution of the $UV$ LF also requires considerable
  evolution in the steepness of the $UV$ LF with cosmic time, as one
  can accomplish through a change in the faint-end slope $\alpha$ (or
  also somewhat through changes in the characteristic magnitude
  $M^*$).}
\item{The best-fit characteristic magnitude $M^*\sim-21$ we find for
  the $UV$ LF at $z\sim6$ and $z\sim7$ is brighter than has been found
  in many previous studies (Bouwens et al.\ 2006; Bouwens et
  al.\ 2007; McLure et al.\ 2009; Su et al.\ 2011; Bouwens et
  al.\ 2011; Lorenzoni et al.\ 2011; Grazian et al.\ 2012; Willott et
  al.\ 2012; Schenker et al.\ 2013; McLure et al.\ 2013).  The
  improved constraints at the bright end and larger numbers of sources
  show that the evolution in $M^*$ that has been widely accepted as
  the dominant change in the LF with time should be revised.
  Evolution in $\phi^*$ appears to be dominating the change in the LF
  with time.  Interestingly, the evolution seen in $\alpha$, when
  combined with that found in $\phi^*$, can be mimicked by an
  evolution in $M^*$ in noisier data, helping to clarify why the
  earlier, more limited datasets may have led to the the conclusion
  that $M^*$ was evolving (Appendix F.6).}
\item{Despite changes in the form of the evolution at the bright end
  of the LF, the best-fit evolution in $M^*$ preferred by Bouwens et
  al.\ (2008) and Bouwens et al.\ (2011) is in remarkably good
  agreement with the evolution in luminosity for the typical
  $UV$-bright galaxy (at a fixed cumulative number density: see \S5.3
  and Figure~\ref{fig:sfrevol}).  The $UV$ luminosity for such a
  number density-matched galaxy increases by $\sim$0.37 mag per unit
  redshift, which is almost identical to what Bouwens et al.\ (2008)
  and Bouwens et al.\ (2011) had inferred for the evolution in the
  characteristic magnitude $M^*$ of the $UV$ LF over a similar
  redshift range to what we consider here.  In this way, the schematic
  $M^*$-evolutionary model of Bouwens et al.\ (2008) effectively
  foreshadowed later work using a cumulative number density-matched
  formalism to trace the steadily-increasing $UV$ luminosities and
  star formation rates of individual galaxies (Papovich et al.\ 2011;
  Smit et al.\ 2012; Lundgren et al.\ 2014).}
\item{Our LF results appear to be perfectly consistent with the LF
  having a Schechter-like form over the entire redshift range
  $z\sim4$-8 (\S4.4).  The consistency of our results with the
  Schechter form can be seen in Figure~\ref{fig:spowlaw} where we
  present the differences between stepwise and Schechter
  representations of the LFs.  We draw a similar conclusion looking at
  the effective slope of the $z\sim4$-8 LFs, as a function of
  luminosity (Figure~\ref{fig:powlaw}).  We observe this both at high
  and low luminosities.  At high luminosities, the $UV$ LF exhibits a
  very similar exponential-like cut-off to that present in a Schechter
  function.  At lower luminosities, the effective slope of the LF
  shows no significant change from $-19.5$ to $-17.5$, consistent with
  this slope asymptoting to some fixed value.  While our LF results
  are completely consistent with having a Schechter form (at both
  $z\sim4-6$ and $z\sim7$-8), we cannot
  exclude the LF having an alternate functional form at $z>6$ (such as
  a double power-law shape: despite the clear tension between our
  $z\sim7$ LF results and those from Bowler et al.\ 2014).  Although
  it is reasonable to imagine that the $UV$ LF would exhibit a
  slightly non-Schechter shape at early enough times or at low enough
  luminosities (e.g., Mu{\~n}oz \& Loeb 2011), we find no strong
  evidence for such a behavior here.}
\item{The deep, wide-area search data over five independent sightlines
  in the high-redshift universe have made it possible for us to
  quantify the importance of field-to-field variations on the bright
  ends of the $z\sim4$, $z\sim5$, $z\sim6$, $z\sim7$, and $z\sim8$ LFs
  (\S4.5).  While most of our search fields show only modest
  differences ($\lesssim$20\%) in the volume density for sources at
  different redshifts, we find larger field-to-field variations in the
  volume density of galaxies in our samples at $z\sim7$ and $z\sim8$,
  with the CANDELS-GN and EGS fields showing almost double the surface
  density of $z\sim7$ galaxies as the CANDELS-GS and UDS fields.  The
  relative surface density of $z\sim4$, $z\sim5$, and $z\sim6$
  galaxies we find over the CANDELS-GN and GS are in excellent
  agreement with the relative surface densities found previously by
  Bouwens et al.\ (2007).}
\item{We have taken advantage of our new LF constraints to derive a
  fitting formula to match the evolution seen in our sample over the
  redshift range $z\sim8$ to $z\sim4$ (\S5.1).  Our best fit relation
  is $M_{UV} ^{*} = (-20.95\pm0.10) + (0.01\pm0.06) (z - 6)$, $\phi^*
  = (0.47_{-0.10}^{+0.11}) 10^{(-0.27\pm0.05)(z-6)}10^{-3}
  \textrm{Mpc}^{-3}$, and $\alpha = (-1.87\pm0.05) +
  (-0.10\pm0.03)(z-6)$.  From this fitting formula, we find strong
  evidence for significant evolution in the volume density $\phi^*$
  and $\alpha$.  Evolution in the characteristic magnitude $M^*$ may
  be present, but it is less significant than found previously
  (Bouwens et al.\ 2008, 2011), as we noted above.  Results from this
  fitting formula are in excellent agreement with our previous fitting
  formula (which preferred a more significant $M^*$ evolution) over
  the more limited range of luminosities and volume densities that was
  well probed by previous studies.}
\item{We find we can approximately match the evolution of the $UV$ LF
  from $z\sim8$ to $z\sim4$ with a simple conditional luminosity
  function (CLF) model based on halo growth and a modest evolution in
  the mass-to-light ratio $(\propto (1+z)^{-1.5})$ of the halos
  (\S5.5).  This CLF is successfully at reproducing the approximate
  evolution in all three Schechter parameters (see
  Figure~\ref{fig:slopeevol}, Figure~\ref{fig:theoryms}, \S5.5.1, and
  \S5.5.2).  The CLF model we present here is identical to the model
  we previously developed in Bouwens et al.\ (2008) except for the
  assumed evolution in the mass-to-light ratio of the halos.}
\end{itemize}
The extraordinary depth, area, and wavelength baseline of the CANDELS,
HUDF09, and HUDF12 data sets have provided us with substantial
leverage to study the evolution of the $UV$ LF with cosmic time.  The
most remarkable results of this study has been to demonstrate the
progressive steepening of the $UV$ LF to high redshift.  As
illustrated in Figure~\ref{fig:shapelf}, the $UV$ LF results at
$z\sim7$ and $z\sim8$ are clearly much steeper than at $z\sim3$ and
$z\sim4$.  Meanwhile, our use of $\sim$1000 arcmin$^2$ search area
along 5 independent sightlines (and numerous independent sightlines
from the BoRG/HIPPIES pure parallel programs) has allowed us to
demonstrate the existence of modest numbers of highly luminous
($\lesssim-21$ mag) galaxies in the early universe at redshifts as
high as $z\sim10$ (see also Oesch et al.\ 2014).  The existence of
such luminous galaxies at early times clearly demonstrates that the
characteristic magnitude $M^*$ can only experience limited evolution
with cosmic time.

In the future, we can expect stronger constraints on the evolution of
the UV LF at $z\sim4$-10 using data from the new Frontier Field
Initiative, which will obtain 140 orbits of optical + near-IR
observations over 4-6 cluster and parallel fields.  These fields
should be particularly effective in ensuring that current LF results
are robust, since combining these new fields together with the 3
existing deep fields (XDF + two HUDF09-parallel) we will have 11-15
fields from which to map out the shape of the $UV$ LF.  The new
Frontier Fields will also allow us to assess whether the results we
have derived here based on the XDF and the HUDF09 parallel fields are
representative and will add especially useful new constraints at
$z\sim9$-11.

\acknowledgements

We thank Peter Capak, Mark Dickinson, Jim Dunlop, Richard Ellis, Steve
Finkelstein, Kristian Finlator, Jason Jaacks, Ross McLure, Ken
Nagamine, Masami Ouchi, Rogier Windhorst, and Haojing Yan for valuable
conversations.  Feedback from our referee substantially improved this
paper.  We acknowledge the support of NASA grant NAG5-7697, NASA grant
HST-GO-11563, ERC grant HIGHZ \#227749, and a NWO vrij competitie
grant 600.065.140.11N211.  This research has benefitted from the SpeX
Prism Spectral Libraries, maintained by Adam Burgasser at
http://pono.ucsd.edu/$\sim$adam/browndwarfs/spexprism.

\appendix

\section{A.  Other Detailed Information on the Observational Data Sets that we 
utilize}

\subsection{A.1  Important Ground-Based Observations over the CANDELS 
UDS, COSMOS, and EGS Fields}

The $\sim$450 arcmin$^2$ region provided by the CANDELS UDS, EGS, and
COSMOS fields provides valuable constraints on the volume density of
the brightest, rarest sources at high redshift and as an additional
control on the impact of field-to-field variations (``cosmic
variance'') on the high-redshift luminosity functions.

To ensure that sources in our high-redshift selections were as free of
lower redshift contamination as possible, we also made use of the very
deep, optical ground-based data available over the three wide-area
CANDELS fields.  Deep observations at optical wavelengths are
important for ensuring that high-redshift candidates exhibit a robust
Lyman break and therefore are not likely at lower redshifts.  For each
of our fields, the ground-based imaging data reach as deep or deeper
than the HST observations, particularly for extended sources (as most
lower redshift contaminants typically are).  Over both the CANDELS
COSMOS and CANDELS EGS fields, we made use of the CFHT legacy survey
deep observations in the $u$, $g$, $r$, $i$ (``$i_1$''), $y$
(``$i_2$''), and $z$
bands.\footnote{http://www.cfht.hawaii.edu/Science/CFHTLS} Over the
COSMOS field, we also made use of the very deep Subaru observations
made available by Capak et al.\ (2007) in the $B$, $g$, $V$, $r$, $i$,
and $z$ bands.  Finally, over the CANDELS UDS field, we made use of
the very deep ($\sim$ 28 mag depths at $5\sigma$: $2''$-diameter
apertures) Subaru observations taken as part of the Subaru XMM-Newton
Deep Field (SXDF) program in the $B$, $V$, $R$, $i$, and $z$ bands
(Furusawa et al. 2008).  

Moderately deep $YK_s$-band and $YJHK_s$ observations are available
over the CANDELS-UDS and CANDELS-COSMOS fields with HAWK-I and VISTA,
respectively, from the HUGS (Fontana et al.\ 2014) and UltraVISTA
(McCracken et al.\ 2012) programs.  The $Y$-band observations are of
value for determining which $z\sim7$-8 candidates from the
CANDELS-UDS/COSMOS fields are more likely at $z\sim7$ and which are
more likely at $z\sim8$.  The $JHK_s$ observations also provide us
with useful information on the overall magnitude and spectral slope of
candidates redward of the putative Lyman breaks.  Our reduction of the
HUGS observations is described in L. Spitler et al.\ (2014, in prep).
Meanwhile, for a reduction of the three-year UltraVISTA observations,
we use the official ESO release
(http://www.eso.org/sci/observing/phase3/data\_releases/uvista\_dr2.pdf).

Intermediate-depth $K_s$-band observations are available over the
CANDELS EGS field from the WIRCam Deep Survey (McCracken et al.\ 2010;
Bielby et al.\ 2012).  While these observations only reach to 24.1 mag
(1.2$''$-diameter apertures) for the typical source over the CANDELS
EGS field (Skelton et al.\ 2014), they do provide a probe of the
spectral slope of galaxies redward of the CANDELS near-IR observations
and therefore have some value in ascertaining the nature of the
brightest sources over the CANDELS EGS field.  We use these
observations in deriving the best-fit redshifts for individual sources
with EAZY.

\subsection{A.2  BoRG/HIPPIES Fields}

To obtain the most accurate constraints on the volume density of the
rarest, brightest galaxies at $z\sim8$, we also made use of the
wide-area BoRG/HIPPIES pure-parallel programs (Trenti et al.\ 2011;
Yan et al.\ 2011) and similar parallel data from the COS GTO team
(Trenti et al.\ 2011).  The BoRG/HIPPIES program features moderately
deep observations ($\sim$0.5 orbit to $\sim$3 orbit) in at least four
different bands, i.e., $V_{606}$/$V_{600}$, $Y_{098}$/$Y_{105}$,
$J_{125}$, and $H_{160}$ bands, over a wide variety of different
positions in the sky outside the galactic plane.

To ensure that the candidates we select from the BoRG/HIPPIES data set
are robust, we only made use of the highest quality BoRG/HIPPIES
fields, excluding those search fields with average exposure times in
the $J_{125}+H_{160}$ bands of less than 1200 seconds or search fields
where the exposure time in the optical $V_{606}$ or $V_{600}$ bands is
less than the average exposure time in $J_{125}$ and $H_{160}$
observations.

The total search area in BoRG+HIPPIES and similar programs that
satisfy both of these requirements was 218 arcmin$^2$.

Where reductions are of the BoRG/HIPPIES search fields were already
publicly available from Bradley et al.\ (2012: 0.08$''$-pixel scale),
we made use of those reductions.  For the remaining search fields, the
reductions were made using our \textsc{wfc3red.py} pipeline (Magee et
al.\ 2011).  We did not include the cycle-18 HIPPIES program (GO
12286: PI Yan) in our analysis due to the lack of the $Y_{098}$ data
and the challenge in selecting contamination-free $z\sim8$ galaxies
over a similar redshift range as our other samples using the
$Y_{105}$-band data from that program.

Though not formally part of the BoRG/HIPPIES program, we also
incorporated the 28 orbits of parallel WFC3/IR observations over Abell
1689 (GO 11710: Alamo-Mart{\'{\i}}nez et al.\ 2013) and the 18-orbit
GO-12905 program (PI: Trenti) over the purported BoRG protocluster of
$z\sim8$ galaxies (Trenti et al.\ 2012a; Schmidt et al.\ 2014) into
the BoRG/HIPPIES data set, due to the similar filter choices available
over these fields.  The Abell 1689 parallel field has thus far not
been used in searches for $z\sim8$ galaxies.

\section{B.  Initial Photometric Set of $z\sim5$-8 Candidates from the CANDELS-UDS, CANDELS-COSMOS, and CANDELS-EGS Fields}

To derive our final of $z\sim5$-8 candidates over the CANDELS UDS,
COSMOS, and EGS fields, we first considered a selection of all those
sources which satisfied Lyman-break-like selection criteria at
$z\sim5$, $z\sim6$, and $z\sim7$-8.

\begin{figure*}
\epsscale{0.89}
\plotone{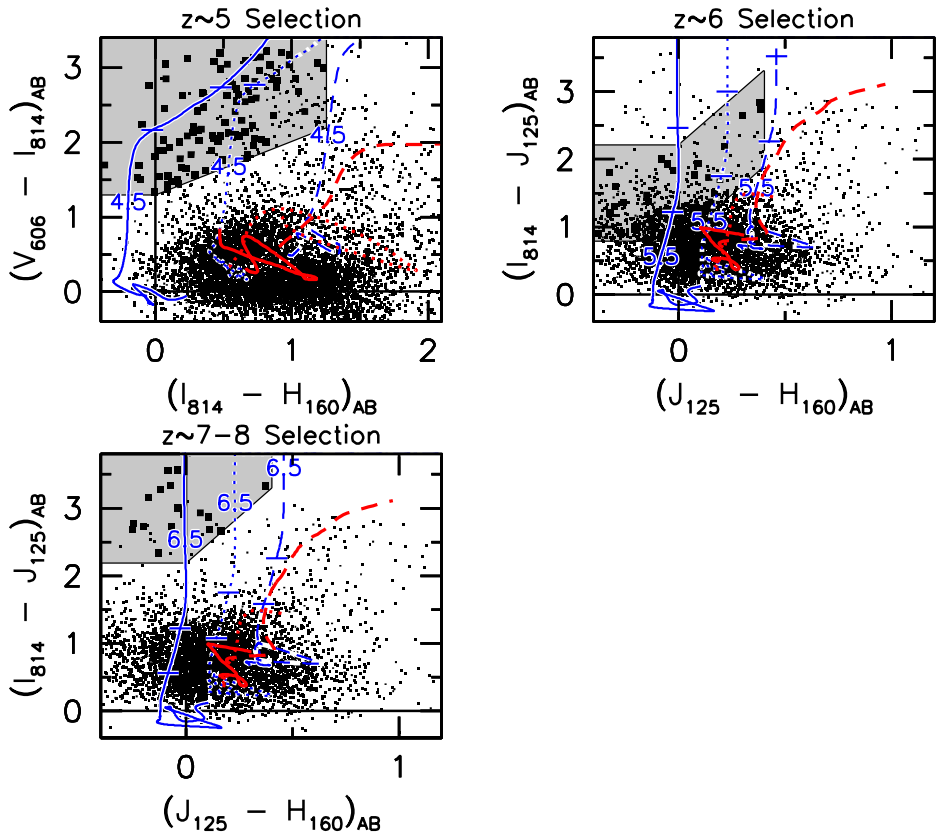}
\caption{Color-color criteria used to provide an initial selection of
  candidate $z\sim5$, $z\sim6$, and $z\sim7$-8 galaxies over the
  CANDELS UDS, CANDELS COSMOS, and CANDELS EGS wide fields (Appendix
  B.1).  Lines and symbols are as in Figure~\ref{fig:sel1}.  The small
  black dots represent sources from the EGS data set, while the large
  black squares indicate sources identified as part of each
  high-redshift selection.  Candidate $z\sim10$ galaxies are selected
  over these fields using a similar strategy as for the XDF,
  CANDELS-GN, and CANDELS-GS.  The lack of observations in certain
  bands (i.e., $i_{775}$, $z_{850}$, or $Y_{105}$ bands) necessitate
  that we utilize different selection criteria to select star-forming
  galaxies at $z\sim5$, $z\sim6$, $z\sim7$ and $z\sim8$ than we do
  over the CANDELS-GN and GS.  Sources identified as part of these
  $z\sim5$-8 samples are redistributed across these samples based on
  the photometric redshifts we derive from their HST + ground-based +
  Spitzer/IRAC photometry (\S3.2.3).  Faint galaxies at $z\sim6$,
  $z\sim7$, and $z\sim8$ are only selected to a bright limit of 26.7
  mag to ensure good redshift separation given the limited depth of
  both the $I_{814}$-band observations and ground-based
  observations.\label{fig:sel3}}
\end{figure*}

We selected these sources using the following color criteria
\begin{eqnarray*}
(V_{606}-I_{814}>1.3)\wedge (I_{814}-H_{160}<1.25) \wedge (V_{606}-I_{814} > 0.72(I_{814}-H_{160})+1.3)
\end{eqnarray*}
for our initial $z\sim5$ selection,
\begin{eqnarray*}
(I_{814}-J_{125}>0.8)\wedge (J_{125}-H_{160}<0.4) \wedge (I_{814}-J_{125} > 2(J_{125}-H_{160})+0.8) 
\end{eqnarray*}
for our initial $z\sim6$ selection, and
\begin{eqnarray*}
(I_{814}-J_{125}>2.2)\wedge (J_{125}-H_{160}<0.4) \wedge (I_{814}-J_{125} > 2(J_{125}-H_{160})+2.2)
\end{eqnarray*}
for our initial $z\sim7$-8 selection (similar to the color criteria
adopted by Grazian et al.\ 2012).  These color criteria were
constructed in an analogous manner to the criteria we describe in
\S3.2.2 of this paper, such that sources entered the two color
selection window at approximately the same redshift independent of the
$UV$-continuum slope of the source.  The color criteria for our
initial $z\sim5$, $z\sim6$, and $z\sim7$-8 selections are illustrated
in Figure~\ref{fig:sel3}.

These criteria are used to identify the initial set of candidate
$z\sim5$-8 galaxies, to which we add deep ground-based optical+near-IR
and Spitzer/IRAC photometry and measure photometric redshifts to
derive our final $z\sim5$, $z\sim6$, $z\sim7$, and $z\sim8$ samples
(\S3.2.3).

\begin{deluxetable*}{cccccc}
\tablecaption{Observed Surface densities of $z\sim4$, $z\sim5$, $z\sim6$, $z\sim7$, $z\sim8$, and $z\sim10$ galaxy candidates from all fields.\tablenotemark{*}\label{tab:surfdens}}
\tablehead{
\colhead{} & \colhead{Surface Density\tablenotemark{b}} &
\colhead{} & \colhead{Surface Density\tablenotemark{b}} &
\colhead{} & \colhead{Surface Density\tablenotemark{b}}\\
\colhead{Magnitude} & \colhead{(arcmin$^{-2}$)} &
\colhead{Magnitude} & \colhead{(arcmin$^{-2}$)} &
\colhead{Magnitude} & \colhead{(arcmin$^{-2}$)}}\\
\startdata
$22.50<i_{775}<23.00$ & $<$ 0.0039\tablenotemark{a} & $22.40<Y_{105}<22.90$ & $<$ 0.0015\tablenotemark{a} & $22.50<H_{160}<23.00$ & $<$ 0.0015\tablenotemark{a}\\
$23.00<i_{775}<23.50$ & 0.0106 $\pm$ 0.0061 & $22.90<Y_{105}<23.40$ & $<$ 0.0015\tablenotemark{a} & $23.00<H_{160}<23.50$ & $<$ 0.0015\tablenotemark{a}\\
$23.50<i_{775}<24.00$ & 0.0354 $\pm$ 0.0112 & $23.40<Y_{105}<23.90$ & $<$ 0.0015\tablenotemark{a} & $23.50<H_{160}<24.00$ & $<$ 0.0015\tablenotemark{a}\\
$24.00<i_{775}<24.50$ & 0.2376 $\pm$ 0.0290 & $23.90<Y_{105}<24.40$ & 0.0014 $\pm$ 0.0014 & $24.00<H_{160}<24.50$ & $<$ 0.0015\tablenotemark{a}\\
$24.50<i_{775}<25.00$ & 0.6494 $\pm$ 0.0480 & $24.40<Y_{105}<24.90$ & 0.0081 $\pm$ 0.0033 & $24.50<H_{160}<25.00$ & $<$ 0.0015\tablenotemark{a}\\
$25.00<i_{775}<25.50$ & 1.4575 $\pm$ 0.0718 & $24.90<Y_{105}<25.40$ & 0.0350 $\pm$ 0.0069 & $25.00<H_{160}<25.50$ & 0.0041 $\pm$ 0.0023\\
$25.50<i_{775}<26.00$ & 2.4695 $\pm$ 0.0935 & $25.40<Y_{105}<25.90$ & 0.0981 $\pm$ 0.0115 & $25.50<H_{160}<26.00$ & 0.0081 $\pm$ 0.0033\\
$26.00<i_{775}<26.50$ & 3.7300 $\pm$ 0.1627 & $25.90<Y_{105}<26.40$ & 0.2584 $\pm$ 0.0419 & $26.00<H_{160}<26.50$ & 0.0527 $\pm$ 0.0190\\
$26.50<i_{775}<27.00$ & 4.7275 $\pm$ 0.7609 & $26.40<Y_{105}<26.90$ & 0.3806 $\pm$ 0.1638 & $26.50<H_{160}<27.00$ & 0.1441 $\pm$ 0.1019\\
$27.00<i_{775}<27.50$ & 6.6043 $\pm$ 0.8993 & $26.90<Y_{105}<27.40$ & 1.0717 $\pm$ 0.2749 & $27.00<H_{160}<27.50$ & 0.4270 $\pm$ 0.1754\\
$27.50<i_{775}<28.00$ & 6.5582 $\pm$ 0.8962 & $27.40<Y_{105}<27.90$ & 1.2049 $\pm$ 0.2915 & $27.50<H_{160}<28.00$ & 0.4992 $\pm$ 0.1897\\
$28.00<i_{775}<28.50$ & 8.3582 $\pm$ 1.0117 & $27.90<Y_{105}<28.40$ & 1.8070 $\pm$ 0.3570 & $28.00<H_{160}<28.50$ & 0.6403 $\pm$ 0.2148\\
$28.50<i_{775}<29.00$ & 10.4910 $\pm$ 1.4912 & $28.40<Y_{105}<28.90$ & 2.9412 $\pm$ 0.7896 & $28.50<H_{160}<29.00$ & 1.0643 $\pm$ 0.4908\\
$29.00<i_{775}<29.50$ & 16.8280 $\pm$ 1.8886 & $28.90<Y_{105}<29.40$ & 5.8268 $\pm$ 1.1113 & $29.00<H_{160}<29.50$ & 1.3466 $\pm$ 0.5520\\
$29.50<i_{775}<30.00$ & 10.7412 $\pm$ 1.5089 & $29.40<Y_{105}<29.90$ & 4.5725 $\pm$ 0.9845 & $29.50<H_{160}<30.00$ & 1.6986 $\pm$ 0.6200\\
\multicolumn{2}{c}{$z\sim5$} & $29.90<Y_{105}<30.40$ & 2.0457 $\pm$ 0.6585 & \multicolumn{2}{c}{$z\sim10$}\\
$22.50<Y_{105}<23.00$ & $<$ 0.0015\tablenotemark{a} & \multicolumn{2}{c}{$z\sim7$} & $22.20<H_{160}<23.20$ & $<$ 0.0014\tablenotemark{a}\\
$23.00<Y_{105}<23.50$ & 0.0014 $\pm$ 0.0014 & $22.95<J_{125}<23.45$ & $<$ 0.0015\tablenotemark{a} & $22.70<H_{160}<23.70$ & $<$ 0.0014\tablenotemark{a}\\
$23.50<Y_{105}<24.00$ & 0.0041 $\pm$ 0.0023 & $23.45<J_{125}<23.95$ & $<$ 0.0015\tablenotemark{a} & $23.70<H_{160}<24.70$ & $<$ 0.0014\tablenotemark{a}\\
$24.00<Y_{105}<24.50$ & 0.0231 $\pm$ 0.0055 & $23.95<J_{125}<24.45$ & $<$ 0.0015\tablenotemark{a} & $24.70<H_{160}<25.70$ & $<$ 0.0014\tablenotemark{a}\\
$24.50<Y_{105}<25.00$ & 0.0893 $\pm$ 0.0110 & $24.45<J_{125}<24.95$ & 0.0014 $\pm$ 0.0014 & $25.70<H_{160}<26.70$ & 0.0070 $\pm$ 0.0070\\
$25.00<Y_{105}<25.50$ & 0.2771 $\pm$ 0.0194 & $24.95<J_{125}<25.45$ & 0.0215 $\pm$ 0.0054 & $26.70<H_{160}<27.70$ & $<$ 0.0792\tablenotemark{a}\\
$25.50<Y_{105}<26.00$ & 0.5549 $\pm$ 0.0274 & $25.45<J_{125}<25.95$ & 0.0333 $\pm$ 0.0067 & $27.70<H_{160}<28.70$ & $<$ 0.2488\tablenotemark{a}\\
$26.00<Y_{105}<26.50$ & 1.1366 $\pm$ 0.0884 & $25.95<J_{125}<26.45$ & 0.1569 $\pm$ 0.0327 & $28.70<H_{160}<29.70$ & 0.4523 $\pm$ 0.3198\\
$26.50<Y_{105}<27.00$ & 1.9991 $\pm$ 0.3950 & $26.45<J_{125}<26.95$ & 0.2821 $\pm$ 0.1411 &  & \\
$27.00<Y_{105}<27.50$ & 2.2056 $\pm$ 0.4149 & $26.95<J_{125}<27.45$ & 0.3527 $\pm$ 0.1577 &  & \\
$27.50<Y_{105}<28.00$ & 3.1493 $\pm$ 0.4958 & $27.45<J_{125}<27.95$ & 0.8306 $\pm$ 0.2420 &  & \\
$28.00<Y_{105}<28.50$ & 4.3133 $\pm$ 0.5802 & $27.95<J_{125}<28.45$ & 1.2726 $\pm$ 0.2996 &  & \\
$28.50<Y_{105}<29.00$ & 4.6413 $\pm$ 0.9919 & $28.45<J_{125}<28.95$ & 1.2638 $\pm$ 0.5176 &  & \\
$29.00<Y_{105}<29.50$ & 6.3116 $\pm$ 1.1566 & $28.95<J_{125}<29.45$ & 4.2857 $\pm$ 0.9531 &  & \\
$29.50<Y_{105}<30.00$ & 5.2184 $\pm$ 1.0517 & $29.45<J_{125}<29.95$ & 3.4843 $\pm$ 0.8594 &  & 
\enddata
\tablenotetext{*}{See Figure~\ref{fig:surfdens} for a presentation of these surface
  densities in graphical form.}
\tablenotetext{a}{$1\sigma$ upper limits}
\tablenotetext{b}{The surface densities of galaxies in a given
  magnitude interval are only estimated from fields that are largely
  complete in that magnitude interval.}
\end{deluxetable*}

\section{C.  Surface Density of $z\sim4$-10 Galaxies}

For convenience, we have calculated the surface density of $z\sim4$-10
galaxy candidates found across all of our search fields and tabulated
these surface densities in Table~\ref{tab:surfdens}.  In calculating
the average surface density of sources over a given magnitude range,
we have only included those regions from our multi-field probe where
our simulations (\S4.1) indicated we should be at least 80\% complete
relative to our completeness level at brighter magnitudes
(i.e.,$\sim$25 mag).  This included our search results to $\sim$26 mag
from all fields, results from our CANDELS-DEEP search fields to
$\sim$27.0 mag, results from the HUDF09-1 and HUDF09-2 fields to
$\sim$28.0 mag, and results from the XDF to $\sim$30.0 mag.  While we
would not expect the XDF results to be complete at $\sim$30 mag (it is
expected to be similarly complete to $\sim$29), we quote the recovered
surface density of sources in this field, since it represents our only
probe of the surface density of galaxies to this magnitude level.

\begin{figure*}
\epsscale{1.1}
\plotone{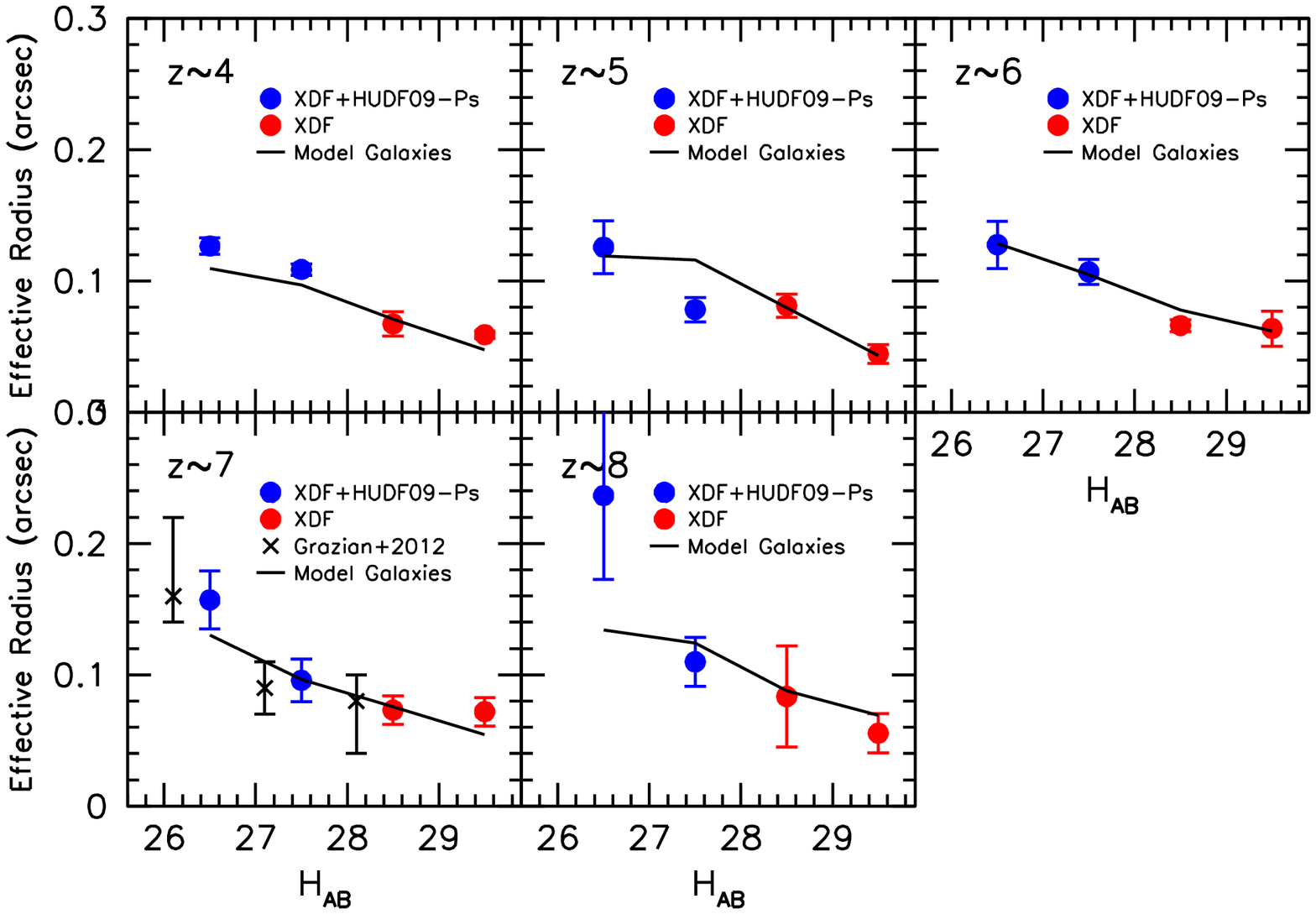}
\caption{Intrinsic half-light radii of the average candidate galaxy
  selected as part of our $z\sim4$ (\textit{upper-left panel}),
  $z\sim5$ (\textit{upper-middle panel}), $z\sim6$
  (\textit{upper-right panel}), $z\sim7$ (\textit{lower-left panel}),
  and $z\sim8$ (\textit{lower-middle panel}) samples versus their
  $H_{160}$-band magnitude.  The red circles are for galaxies found in
  the XDF+HUDF09-Ps data set, while the blue circles are for galaxies
  found in the XDF data set.  Uncertainties on these sizes are
  computed by bootstrap resampling.  The black crosses are the median
  sizes of $z\sim7$ galaxies, as derived by Grazian et al.\ (2012) and
  plotted at the equivalent $H_{160}$-band magnitude based on the
  Bouwens et al.\ (2014a) $\beta$-$M_{UV}$ relation.  The black solid
  line in each panel shows the average size of sources selected to be
  part of these samples in the simulations we use to derive the
  selection volumes.  Sizes and surface brightnesses of galaxies in
  the simulations appear to be very well matched to the observations
  (see Appendix D).
  \label{fig:sizecalib}}
\end{figure*}

\begin{figure*}
\epsscale{0.85}
\plotone{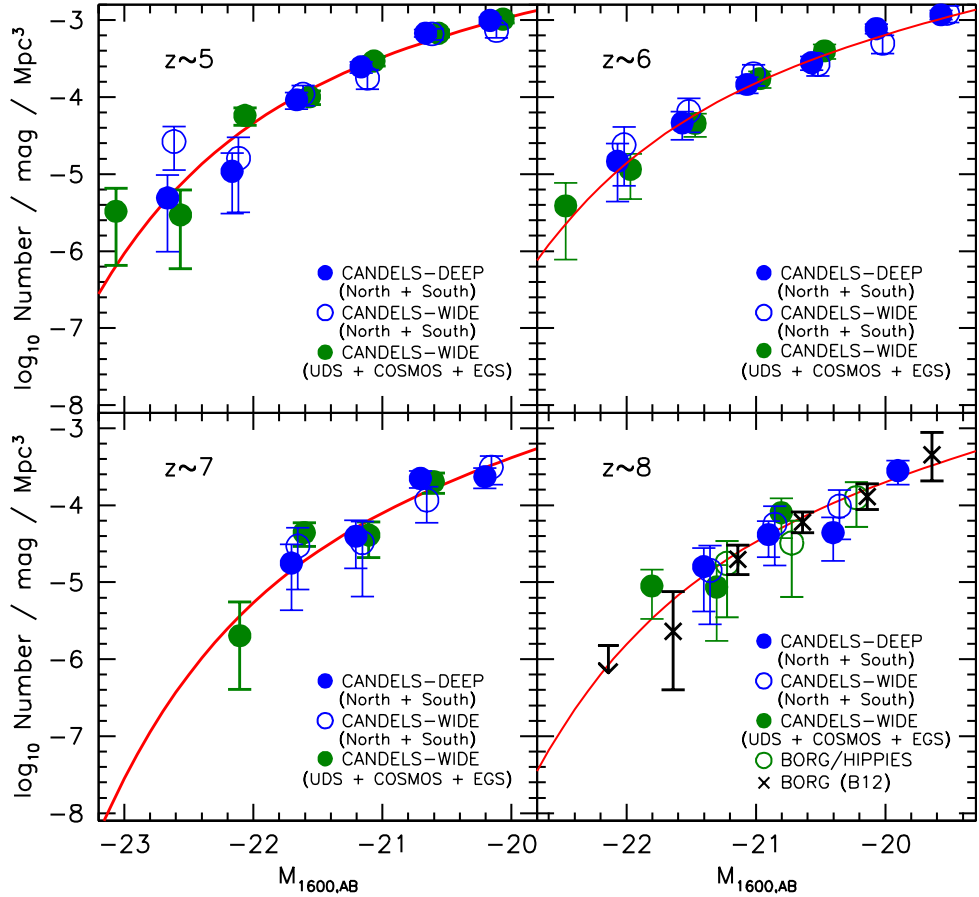}
\caption{SWML determinations of the $UV$ LFs at $z\sim5$, $z\sim6$,
  $z\sim7$, and $z\sim8$ using our high-redshift samples from the
  CANDELS-DEEP-GN and GS fields (\textit{solid red circles}),
  CANDELS-WIDE-GN and GS fields (\textit{open red circles}),
  CANDELS-WIDE UDS+COSMOS+EGS fields (\textit{solid green circles}),
  and the BoRG/HIPPIES fields (\textit{open green circles}).  The SWML
  determinations are offset slight from each other (by $\pm$0.05 mag)
  for clarity.  Also shown are earlier determinations of the $z\sim8$
  LF from the BoRG data set (\textit{black crosses}: Bradley et
  al.\ 2012).  For comparison, we also overplotted our best-fit
  Schechter function results from \S4.2 (\textit{red lines}).  By
  subdividing our search fields according to depth and ancillary depth
  and deriving our LF results from each subset independently, we can
  ensure that our LF determination procedure is largely free of
  systematics specific to a data set.  Overall, we observe broad
  consistency between our LF results using data sets with a variety of
  depths and supporting data -- particularly faintward of $L^*$.  This
  strongly suggests that systematic errors in our LF determinations
  are small and our LF results are robust.\label{fig:wdlfcomp}}
\end{figure*}

\begin{deluxetable*}{cccccccc}
\tablewidth{0pt}
\tabletypesize{\footnotesize}

\tablecaption{Comparisons of the Schechter Parameters for the $UV$ LFs
  derived using constraints from the XDF+HUDF09-Ps and alternatively from one of the two GOODS fields
  (CANDELS-GN or CANDELS-GS+ERS).\label{tab:southvsnorth}}
\tablehead{
\colhead{} & \colhead{} & \multicolumn{3}{c}{XDF+HUDF09-Ps+CANDELS-GS+ERS} &
\multicolumn{3}{c}{XDF+HUDF09-Ps+CANDELS-GN}\\
\colhead{Dropout} & \colhead{} & \colhead{} & \colhead{$\phi^*$
$(10^{-3}$} & \colhead{} & \colhead{} & \colhead{$\phi^*$ $(10^{-3}$} & \colhead{} \\
\colhead{Sample} & \colhead{$<z>$} &
\colhead{$M_{UV} ^{*}$\tablenotemark{a}} & \colhead{Mpc$^{-3}$)} &
\colhead{$\alpha$} & \colhead{$M_{UV} ^{*}$\tablenotemark{a}} & 
\colhead{Mpc$^{-3}$)} & \colhead{$\alpha$}}
\startdata
$B$ & 3.8 & $-20.99\pm0.11$ & $1.65_{-0.29}^{+0.36}$ & $-1.67\pm0.05$ & $-21.02\pm0.12$ & $1.62_{-0.31}^{+0.38}$ & $-1.68\pm0.05$\\ 
$V$ & 4.9 & $-20.86\pm0.15$ & $1.13_{-0.26}^{+0.34}$ & $-1.69\pm0.07$ & $-21.14\pm0.18$ & $0.90_{-0.22}^{+0.28}$ & $-1.69\pm0.07$ \\
$i$ & 5.9 & $-21.06\pm0.27$ & $0.43_{-0.16}^{+0.24}$ & $-1.88\pm0.11$ & $-21.62\pm0.24$ & $0.15_{-0.06}^{+0.10}$ & $-2.15\pm0.11$\\
$z$ & 6.8 & $-20.63\pm0.31$ & $0.48_{-0.21}^{+0.37}$ & $-1.98\pm0.15$ & $-20.73\pm0.34$ & $0.55_{-0.25}^{+0.45}$ & $-1.88\pm0.15$\\
$Y$ & 7.9 & $-20.09\pm0.52$ & $0.51_{-0.32}^{+0.81}$ & $-1.76\pm0.29$ & $-20.31\pm0.47$ & $0.43_{-0.25}^{+0.58}$ & $-1.81\pm0.27$
\enddata
\end{deluxetable*}

\section{D.  Ensuring the Model Size Distribution of Galaxies Matches the
Observed Size Distribution}

It is essential that we have an accurate measurement of the selection
volume to obtain reliable estimates of the luminosity function at high
redshift.  The most important input for determining the selection
volume for a high-redshift sample is the size or surface brightness
distribution of the high-redshift star-forming galaxies from which the
luminosity function is derived.  Adopting sizes that are too large for
model galaxies in the simulations will result in an underestimate of
the selection volume, while adopting sizes that are too small for the
model galaxies will result in an overestimate of the selection volume.

While this issue had already been considered in many studies of the
$UV$ LF (e.g., Bouwens et al.\ 2006; Oesch et al.\ 2007), Grazian et
al.\ (2011) demonstrated the sizeable impact this issue could have
determinations of the faint-end slope at $z\sim7$-8, if not treated
properly.  Fortunately, care was taken in Bouwens et al.\ (2011),
Schenker et al.\ (2013), and McLure et al.\ (2013) to ensure that the
model galaxies in the simulations had a similar size distribution to
what was used in the real observations (though the use of point-source
profiles in deriving selection volumes by McLure et al.\ 2013 may
result in a slight overestimate of the selection volume for bright
galaxies).

To ensure an accurate match between the size distribution of galaxies
in our simulations and that found in the observations, we subdivided
galaxies in our $z\sim4$-8 samples from the XDF+HUDF09-1+HUDF09-2
fields by their apparent $H_{160}$-band magnitude and stacked the
sources (after re-pixelating them to the same centroid position).  We
then measured their sizes using \textit{galfit} (Peng et al.\ 2002).
This process was then repeated using sources that we selected from the
selection volume simulations described in \S4.1.  The two results are
compared in Figure~\ref{fig:sizecalib} as a function of the
$H_{160}$-band magnitude, for all of our high-redshift samples except
our $z\sim10$ samples (where the small sample size precludes detailed
comparisons).  We experimented with the size scale of the $z\sim4$
HUDF galaxy we were using in the simulations until good agreement was
obtained.  The initial agreement was quite good, with the best match
being obtained for sizes slightly ($\sim$10\%) smaller than expected
from a $(1+z)^{-1.2}$ scaling for fixed-luminosity sources.

\begin{deluxetable*}{lclclc}
\tablewidth{0pt}
\tabletypesize{\footnotesize}
\tablecaption{Stepwise Determination of the rest-frame $UV$ LF at $z\sim7$ and $z\sim8$ using the SWML method (\S4.1) using all of our search fields except CANDELS EGS.\tablenotemark{a}\label{tab:swlfegs}}
\tablehead{
\colhead{$M_{1600,AB}$\tablenotemark{a}} & \colhead{$\phi_k$ (Mpc$^{-3}$ mag$^{-1}$)} & \colhead{$M_{1600,AB}$\tablenotemark{b}} & \colhead{$\phi_k$ (Mpc$^{-3}$ mag$^{-1}$)}}
\startdata
\multicolumn{2}{c}{$z\sim7$ galaxies} & \multicolumn{2}{c}{$z\sim8$ galaxies}\\
$-$22.66 & $<$0.000003\tablenotemark{c} & $-$22.85 & $<$0.000003\tablenotemark{c}\\
$-$22.16 & 0.000002$\pm$0.000003 & $-$22.35 & $<$0.000003\tablenotemark{c}\\
$-$21.66 & 0.000024$\pm$0.000009 & $-$21.85 & $<$0.000003\tablenotemark{c}\\
$-$21.16 & 0.000045$\pm$0.000017 & $-$21.35 & 0.000019$\pm$0.000007\\
$-$20.66 & 0.000189$\pm$0.000037 & $-$20.85 & 0.000054$\pm$0.000016\\
$-$20.16 & 0.000293$\pm$0.000060 & $-$20.35 & 0.000060$\pm$0.000026\\
$-$19.66 & 0.000645$\pm$0.000099 & $-$19.85 & 0.000320$\pm$0.000100\\
$-$19.16 & 0.000740$\pm$0.000158 & $-$19.35 & 0.000497$\pm$0.000212\\
$-$18.66 & 0.001566$\pm$0.000431 & $-$18.60 & 0.001020$\pm$0.000340\\
$-$17.91 & 0.005300$\pm$0.001320 & $-$17.60 & 0.002620$\pm$0.001000\\
$-$16.91 & 0.007720$\pm$0.002680 &  & 
\enddata
\tablenotetext{a}{The results in this table are derived in exactly the
  same way as the results in Table~\ref{tab:swlf}, but exclude the
  $z\sim7$ + $z\sim8$ search results over the CANDELS EGS field.  While
  our simulation results (Figure~\ref{fig:zdistsel}) suggest that it
  is possible to identify $z\sim7$ and $z\sim8$ galaxies using the
  available observations over the CANDELS EGS field (albeit with some
  intercontamination between $z\sim7$ and $z\sim8$ samples), the lack
  of deep $Y$-band observations over this search field make the
  results less robust than over the other CANDELS fields.}
\tablenotetext{b}{Derived at a rest-frame wavelength of 1600\AA.}
\tablenotetext{c}{Upper limits are $1\sigma$.}
\end{deluxetable*}

\section{E.  Testing our $z\sim4$-8 LF results for Internal Consistency}

Given the large numbers of $z\sim4$-8 galaxies that have been
identified at $z\geq4$, the entire enterprise of quantifying the LF at
high redshift has increasingly become about minimizing the impact of
systematic errors on one's determination of the LF at high redshift.

To ensure that systematic errors in our high-redshift LFs are as small
as possible, we have performed a considerable number of tests to
ensure that our results are accurate and robust.  

\subsection{E.1  LF Results for Data Sets with Different Depths or 
Wavelength Coverage}

One of the most important tests we performed was to divide our data
set according to the depth, filter sets, and quality of data, to
derive the $UV$ LF on each data set independently, and then to compare
the results to test for an overall consistency of the results.

We provide such a comparison in Figure~\ref{fig:wdlfcomp} for our
wide-area data sets, considering separately the $\sim$130 arcmin$^2$
CANDELS-DEEP region over GOODS North and GOODS South, the $\sim$100
arcmin$^2$ CANDELS-WIDE region over GOODS North and GOODS South, and
the $\sim$450 arcmin$^2$ CANDELS-WIDE region over the CANDELS-UDS,
CANDELS-COSMOS, and CANDELS-EGS fields.  Overall, our LF results show
excellent consistency overall, particularly at the faint end and at
$z\sim6$, which is encouraging given significant differences in the
depths and nature of the data sets used to derive the LFs.

However, at the bright end ($M_{UV}<-21$), our stepwise determinations
show larger differences.  The most significant differences between our
determinations appear to be at $z\sim5$ and $z\sim7$.  At $z\sim5$,
these differences appear to be partially the result of shot noise and
large field-to-field variations in the volume densities of the
brightest galaxies.  In particular, we find $>$2$\times$ the surface
density of bright ($H_{160,AB}<24.3$) galaxies over the CANDELS-UDS,
EGS, and COSMOS fields, as we find over the CANDELS-GN+GS+ERS fields.
Slight differences in the $k$-corrections we apply in deriving the
absolute magnitude of galaxies at $1600\AA$ may contribute at a low
level as well to the observed differences.  For $z\sim5$ galaxies from
the CANDELS-GS+GN+ERS fields, these magnitudes are derived based on
the $Y_{105}$-band fluxes; however, for $z\sim5$ galaxies from the
CANDELS-UDS+COSMOS+EGS fields, these magnitudes are derived from the
$J_{125}$-band fluxes.  

At $z\sim7$, we also observe noteworthy differences between our
different determinations plotted in Figure~\ref{fig:wdlfcomp}.  As was
the situation at $z\sim5$, these differences appear to arise from
substantial differences in the surface density of bright galaxies,
from field to field.  The surface density of particularly bright
$z\sim7$ galaxies is $\sim2\times$ higher over the
CANDELS-UDS/COSMOS/EGS fields as what it is over CANDELS-GN+GS+ERS
fields.

In summary, our LF determinations generally show excellent consistency
across the data sets considered in this analysis particularly at the
faint end.  While we do observe modest differences between our derived
LFs at the bright end, these differences appear consistent with
arising from large-scale structure variations.

\subsection{E.2  Dependence of the LF Results on the GOODS Field 
Used for the Bright Constraints}

A second test we performed was to compare the best-fit Schechter
parameters for the $UV$ LFs at $z\sim4$, $z\sim5$, $z\sim6$, $z\sim7$,
and $z\sim8$ we derived for a variety of different search field
combinations.  The results are presented in
Table~\ref{tab:southvsnorth}.  One of the comparisons we consider is
to contrast the results from the XDF+HUDF09-Ps+CANDELS-GS+ERS data set
with the XDF+HUDF09-Ps+CANDELS-GN data set.  The best-fit Schechter
parameters we derive from the two data sets are generally consistent
with each other at $<1\sigma$.

However, the parameters do differ at $\sim$2$\sigma$ for the $z\sim5$
and $z\sim6$ LF determinations.  The differences appear to be the
result of the CANDELS-GN field showing a $2.5\times$ higher surface
density of bright ($H_{160,AB}<24.5$) $z\sim5$ galaxies as the
CANDELS-GS+ERS field shows.  Meanwhile, differences at $z\sim6$ appear
to be explainable due to the $5\times$ higher surface densities of
bright ($H<25$) $z\sim6$ galaxies in the CANDELS-GS+ERS field relative
to the CANDELS-GN field.

\subsection{E.3  LF Results for Our Entire Data Set Excluding the 
CANDELS-EGS field}

Of all of our $z\sim6$-8 samples over CANDELS, we find the most
prominent excess of luminous galaxies at $z\sim7$ over the CANDELS-EGS
field.  It is possible that such an excess could act to skew our
overall LF results and cause them to be less representative.  A second
concern is the lack of deep $Y$-band observations over the CANDELS-EGS
field.  While one can compensate for this by utilizing the
Spitzer/IRAC [3.6]$-$[4.5] colors to distinguish $z\gtrsim7$ galaxies
from $z\sim8$ galaxies, this schema will not work for all galaxies
(see Figure~\ref{fig:zdistsel}), and therefore we might expected some
intercontamination between the bright $z\sim7$ samples over the
CANDELS-EGS field and bright $z\sim8$ samples.

For these reasons, it is reasonable to quantify the LFs at $z\sim7$
and $z\sim8$ without including the CANDELS-EGS field.  The stepwise
results are provided in Table~\ref{tab:swlfegs}.  Meanwhile, the
best-fit Schechter LF results are placed in Table~\ref{tab:lfparm}.
While we find slight differences between these determinations and our
primary determinations (Tables~\ref{tab:swlfegs} and
\ref{tab:lfparm}), the two results are fully consistent within the
$1\sigma$ uncertainties.  This is not surprising, as the two
determinations are not independent.

\section{F.  Comparisons against Previous $z\sim4$-10 LF Determinations}

Here we compare the present results with a few of the most noteworthy
LF results at these redshifts from the literature in an attempt to
understand the differences.  For a comprehensive comparison with older
LF results at $z\sim4$-6 and $z\sim7$-8, we refer the reader to
Bouwens et al.\ (2007) and Bouwens et al.\ (2011).

Not only are the comparisons provided in this section useful for
improving our confidence in the latest results, but they are also
helpful for identifying biases that have existed in past work (most of
which have occurred due to limitations in various data sets) to
improve future determinations of the LF.  Our new LFs differ from our
previous LFs primarily because of the much larger number of bright
objects from the wide-area CANDELS dataset which provide substantially
more robust constraints at the bright end.

\subsection{F.1  $z\sim4$-5 Results}

We compare the present LF determinations at $z\sim4$-5 to select
previous determinations in Figure~\ref{fig:comp410}.  Included in
the comparisons are the $z\sim4$-5 LF results of Bouwens et
al.\ (2007) using the GOODS+HUDF+HUDF-Parallel fields (Bouwens et
al.\ 2004a), the $z\sim4$ LF results of Steidel et al.\ (1999) who
make use of $z\sim4$ searches over 0.23 square degree, and the
$z\sim4$-5 LF results from van der Burg et al.\ (2010), who analyze
the deep, wide-area (4 square degree) CFHT legacy survey deep field
observations.

Our LF results at $z\sim4$ are in excellent agreement with the
previous results from Bouwens et al.\ (2007) and also the results in
Steidel et al.\ (1999) and van der Burg et al.\ (2010: though our
best-fit Schechter function would appear to be $\sim$0.1 mag
brightward of the van der Burg et al.\ 2010 stepwise constraints).
Similar results were also obtained by Ouchi et al.\ (2004), Giavalisco
(2005), and Yoshida et al.\ (2006) in the past, with the LFs of Ouchi
et al.\ (2004) and Yoshida et al.\ (2006) only showing a modest excess
in their faintest (and most uncertain) bin.  Overall the results from
these surveys are consistent in implying a value for $M^*$ around
$-21$ mag.

The present $z\sim5$ LF shows excellent agreement overall with the
wide-area determination by van der Burg et al.\ (2010), except at
$-22$ mag where our LF determination is $0.5$ dex high (but consistent
within the quoted $1\sigma$ errors), and with the determination by
Iwata et al.\ (2007), except at the faint end of their search (where
completeness and contamination are the most difficult to accurately
model).  At the faint end, our $z\sim5$ LF is generally 0.1 dex higher
than the $z\sim5$ LF from Bouwens et al.\ (2007), but otherwise in
reasonable agreement.  The 0.1 dex difference likely resulted from
Bouwens et al.\ (2007) underestimating the fraction of $z\sim5$
galaxies which would scatter outside their two color selection windows
(see Duncan et al.\ 2014 for a discussion of the challenges) and hence
overestimating the selection volume.  The selection of $z\sim5$
galaxies using the full ACS+WFC3/IR photometry is much cleaner
overall, making estimates of the selection volume more robust.

Our new $z\sim5$ LF is also in excess of the Bouwens et al.\ (2007) LF
determination at the bright end.  Such differences might again be
attributed to Bouwens et al.\ (2007) overestimating their selection
volumes.  It is also possible that large-scale structure effects
contributed (while here we efficiently probe the full redshift
interval $z=4.5$-5.5, most of Bouwens et al.\ 2007 $z\sim5$ selection
volume derives from the redshift interval $z=4.5$-5.0).

Our new LF results are also in excess of the McLure et al.\ (2009)
determination at $z\sim5$.  We remark that differences with McLure et
al.\ (2009) could be resolved if the magnitudes in the McLure et
al.\ (2009) determination were systematically too faint (by $\sim$0.2
mag) or if the UDS field were substantially ($\sim2\times$) underdense
in bright $z\sim5$ galaxies.  While such an underdensity over such a
wide-area might seem implausible for standard models of large-scale
structure or bias (e.g., Somerville et al.\ 2004; Trenti \& Stiavelli
2008), Bowler et al.\ (2015) report evidence for $\sim$1.8$\times$
variations in the surface density of bright $z\sim6$ galaxies on
square-degree scales, with the UDS being underdense relative to the
UltraVISTA field.  Interestingly enough, of all the CANDELS fields we
consider, the CANDELS-UDS field appears to be among the poorest in
bright $(H_{160,AB}<24.5$) $z\sim5$ galaxies, containing $4\times$
fewer bright $z\sim5$ galaxies than the CANDELS-COSMOS field.

\begin{figure*}
\epsscale{1.2}
\plotone{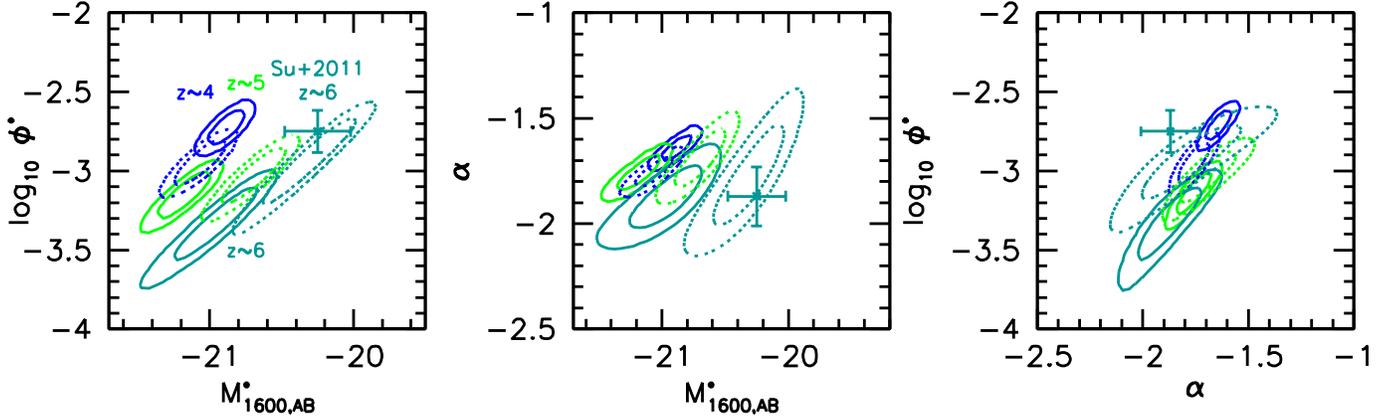}
\caption{Comparison of the 68\% and 95\% confidence intervals on the
  Schechter parameters $M^*$, $\phi^*$, and $\alpha$ we derive for the
  $UV$ LFs at $z\sim4$ (\textit{dark blue contours}), $z\sim5$
  (\textit{green contours}), and $z\sim6$ (\textit{blue contours})
  from the XDF+HUDF09-Ps+ERS+CANDELS-GS+GN fields with
  those found by Bouwens et al.\ (2007: \textit{dotted contours}) who
  considered the optical/ACS data over similar fields.  Also shown are the
  $M^*$, $\phi^*$, and $\alpha$ determinations that Su et al.\ (2011)
  derived for the LF at $z\sim6$ using almost the same data set as
  Bouwens et al.\ (2007).  While our current constraints on the
  Schechter parameters for the LF at $z\sim4$-5 are in reasonable
  agreement with the Bouwens et al.\ (2007) determinations, there is a
  clear disagreement between our current constraints on the $M^*$ and
  $\phi^*$ at $z\sim6$ and the Bouwens et al.\ (2007) and Su et
  al.\ (2011) determinations of these parameters.  Differences between
  the current $z\sim6$ LFs and the Bouwens et al.\ (2007)/Su et
  al.\ (2011) determinations could easily explained as resulting from
  limitations in the data set used by Bouwens et al.\ (2007) and
  uncertainties in the corrections required to cope with
  contamination, IGM absorption, and band-shifting concerns (see
  Figure~\ref{fig:oldi}).\label{fig:contourcomp}}
\end{figure*}

\begin{figure*}
\epsscale{1.15}
\plotone{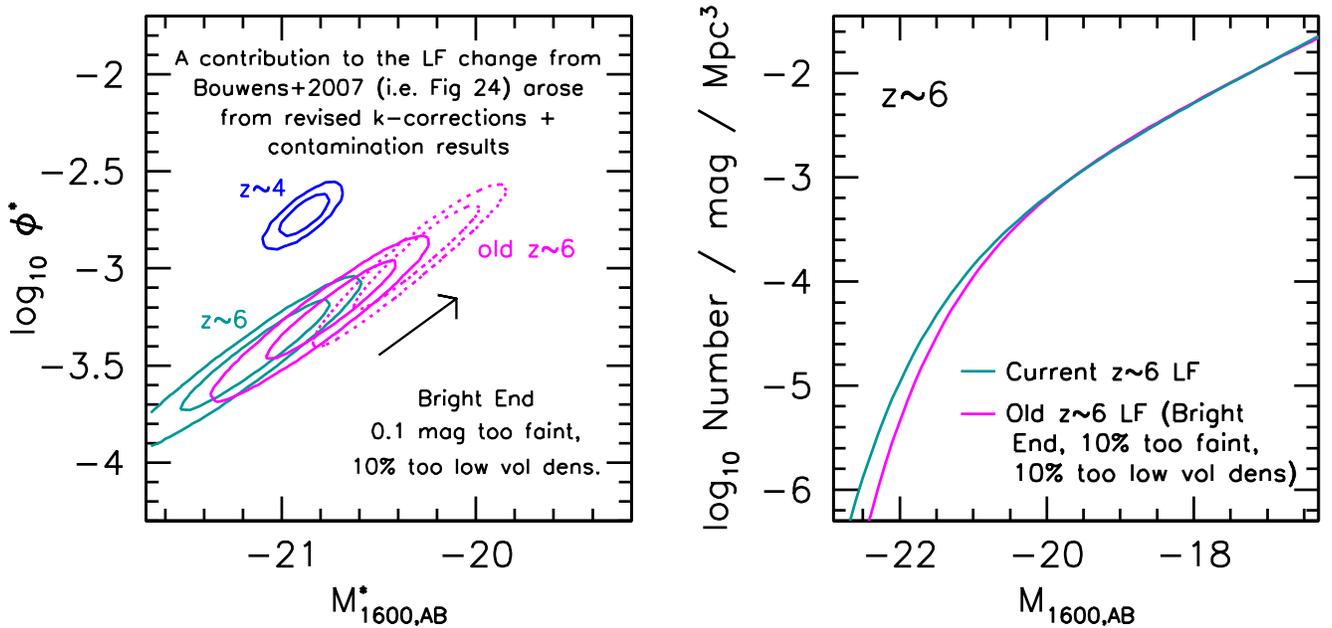}
\caption{(\textit{left}) Possible impact of limitations in the Bouwens
  et al.\ (2007) data set on their $z\sim6$ LF determination (Appendix
  F.2).  The magenta lines give the 68\% and 95\% likelihood contours
  we find for the $z\sim6$ values of $M^*$ and $\phi^*$ based on the
  XDF + HUDF09-Ps + ERS + CANDELS-GN + GS data set.  This figure shows
  the impact on the inferred $M^*$ and $\phi^*$ for the LF that can
  result if the measured magnitudes or volume densities of bright
  sources are estimated in just a slightly different way from the
  faint sources, so that there are small 10\% systematic differences
  between the magnitude measurements and volume density estimates
  between bright and faint sources.  The cyan contours show how the
  present results would change if we suffered from the same
  systematics that affected the Bouwens et al.\ (2007) study, where
  the magnitudes and volume densities of bright sources were likely
  too faint by 10\% (due to limitations in their knowledge of the
  proper $k$-correction) and too low by 10\% (due to limitations in
  their knowledge of the magnitude-dependent contamination rate).  The
  dotted magenta lines show the 68\% and 95\% confidence intervals
  presented by Bouwens et al.\ (2007) on the $z\sim6$ LF.  See also
  Appendix F.2 and F.3.  (\textit{right}) Comparisons of the present
  $z\sim6$ LF determination (\textit{magenta line}) with similar
  determinations of the $z\sim6$ LF modified to include the
  aforementioned biases (\textit{cyan line}).  It is apparent that LFs
  with a brighter characteristic magnitude $M^*$ and lower value for
  $\phi^*$ (steeper faint-end slope $\alpha$) can look very similar
  overall to LFs with a fainter $M^*$ and higher value for $\phi^*$
  (shallower faint-end slope $\alpha$).\label{fig:oldi}}
\end{figure*}

\subsection{F.2  $z\sim6$ Results}

The present LF results are in good agreement with the $z\sim6$ LF
results of Bouwens et al.\ (2007) at the faint end (see
Figure~\ref{fig:comp410}).  At the bright end, however, the $z\sim6$
LF results of Bouwens et al.\ (2007) appear to be slightly lower than
what we find here (albeit of only modest significance for most LF
bins).

It is also useful to compare the present constraints on the Schechter
parameters for the $z\sim6$ LF with the previous constraints on these
parameters from Bouwens et al.\ (2007).  Figure~\ref{fig:contourcomp}
presents both the 68\% and 95\% confidence intervals on these
parameters, as derived by our two studies.  Also included on this
figure are the constraints that Su et al.\ (2011) using a similar
ACS/optical data set, as Bouwens et al.\ (2011) utilize.  There is a
clear disagreement between the current constraints on the $z\sim6$ LF
and previous constraints from Bouwens et al.\ (2007) and Su et
al.\ (2011).

Given that we now have much better datasets at $z\sim 6$ with deep
near-IR coverage and also much larger bright samples, we can assess
how the LFs at $z\sim 6$ from the previous samples came to differ.
This is an opportunity to assess and learn about what issues can arise
with more limited datasets and does not indicate that the approach
used then was inadequate, or that the current results are subject to
significant systematic uncertainties.

After some investigation, we have concluded the differences largely
arose due to Bouwens et al.\ (2007)'s only having ACS/optical data
available to derive the rest-frame $UV$ LF at $z\sim6$ (see also Su et
al.\ 2011).  This necessitated that (1) Bouwens et al.\ (2007)
$k$-correct the measured fluxes of their sources to 1600$\AA$ to
compare with LF results at $z\sim4$-5, (2) Bouwens et al.\ (2007)
correct their measured fluxes for IGM absorption (which is heavily
dependent on the uncertain redshift distribution of $z\sim6$
candidates), and (3) Bouwens et al.\ (2007) correct for contamination
of their selections made on the basis of the optical data alone.  Each
of these steps was uncertain and could have resulted in minor
systematics in the derived Schechter parameters.

Indeed, one contributing factor appears to be the $k$-correction that
Bouwens et al.\ (2007) utilize in calculating the equivalent
luminosity of sources at a rest-frame wavelength of 1600\AA$\,\,$when
the passband in which the sources were observed in the optical
$z_{850}$-band data had an equivalent rest-frame wavelength of
1350\AA.  Bouwens et al.\ (2007) derived the equivalent luminosity at
1600\AA$~$assuming a $UV$-continuum slope of $-2$ based on the
measurements available at that time (Stanway et al.\ 2005; Bouwens et
al.\ 2006).  However, as subsequent research has shown (Wilkins et
al.\ 2011; Bouwens et al.\ 2012, 2014a; Finkelstein et al.\ 2012;
Willott et al.\ 2013; Rogers et al.\ 2014), the most luminous $z\sim6$
galaxies have moderately red $UV$-continuum slopes $\beta\sim-1.5$.
Use of the appropriate $UV$-continuum slopes $\beta$ by Bouwens et
al.\ (2007) would have resulted in 0.1-mag higher estimates of the
luminosity for bright sources than what Bouwens et al.\ (2007) used.

Another likely contributing factor is the correction that Bouwens et
al.\ (2007) applied to account for contamination of their $z\sim6$
$i_{775}-z_{850}>1.3$ selection by $z\sim1$-3 galaxies that were
intrinsically red.  Such corrections were necessary due to the lack of
deep near-IR observations over the entire GOODS North and GOODS South
areas, but could only be estimated from the deep ISAAC $K_s$-band
observations that were available over the GOODS South field.  Bouwens
et al.\ (2006) found that $18_{-9}^{+13}$\% of the sources brighter
than 26 mag were likely contaminants using these near-IR observations
and less than 2\% faintward of 26 mag (see also Stanway et al.\ 2003
who estimated a 25\% contamination rate for such a selection using
sources from the GOODS South).  Bouwens et al.\ (2007) made use of an
almost identical correction.  The availability of the deep WFC3/IR
observations over the GOODS North and South allow us to directly test
the accuracy of this correction.  Using the new WFC3/IR imaging data
to determine the nature of $z\sim6$ candidates in the Bouwens et
al.\ (2007) catalogs, we find that 10.5\% of the candidates brightward
of $z_{850,AB}\sim26.0$ mag have particularly red
$(z_{850}-H_{160})_{AB}\gtrsim2$ colors and appear likely to be lower
redshifts contaminants.

To determine the effect of these systematics on the derived Schechter
parameters at $z\sim6$, we introduced similar systematics into the
surface densities of $z\sim6$ candidates over CANDELS-GN and GS and
rederived the Schechter parameters.  The characteristic luminosity
$M^*$ and $\phi^*$ we recovered is $-20.73\pm0.24$ and $\phi^* =
0.00066_{-0.00022}^{+0.00034}$ Mpc$^{-3}$ (0.4 mag fainter and
$2\times$ higher than for our primary determinations).  Interestingly,
these Schechter parameters are consistent within $2\sigma$ with what
we derived earlier in Bouwens et al.\ (2007: \textit{dotted magenta
  contours in Figure~\ref{fig:oldi}}) for $M^*$, i.e., $M_{UV}^* =
-20.29\pm0.19$, indicating it is possible to fully reconcile our
present and previous results if we consider the above issues.  See
Figure~\ref{fig:oldi} for details.

Bouwens et al.\ (2006) derived an even fainter characteristic
luminosity $M^*$ and higher $\phi^*$ than Bouwens et al.\ (2007)
derived, i.e., $M^*=-20.25\pm0.20$ and $\phi^* =
0.00202_{-0.00076}^{+0.00086}$ Mpc$^{-3}$.  However, that determination
of the Schechter parameters was biased by the procedure that Bouwens
et al.\ (2006) used to correct for field-to-field variations.  Small
systematics in the degradation experiments resulted in the surface
density of sources in the deeper fields being overcorrected upwards
(by 10-15\%) relative to the shallower fields.  Even in the case of
perfect corrections, Trenti \& Stiavelli (2008) showed through
extensive simulations that the procedure Bouwens et al.\ (2006) used
to cope with large-scale structure result in minor biases in the
measured $\phi^*$, $M^*$, and $\alpha$ parameters (with $M^*$ too
faint and $\alpha$ too steep).

The present LF results at $z\sim6$ also imply a higher
($\sim$2-3$\times$) volume density of luminous ($\sim$-21.5 mag)
galaxies than the recent $z\sim6$ results from McLure et al.\ (2009)
and Willott et al.\ (2013).  The McLure et al.\ (2009) probe utilizes
the deep Subaru observations over the Subaru XMM-Newton Deep Field
together the deep near-IR observations from the UKIDSS Ultra-Deep
Survey, while the Willott et al.\ (2012) probe uses the full 4 square
degree probed by the CFHT Legacy Survey deep fields.  Given the very
wide areas probed, it is unlikely that differences between our
$z\sim6$ LF and previous determinations result from large-scale
structure variations or shot noise (see Appendix G and
Table~\ref{tab:brightnumbers}).

At face value, this might suggest that previous ground-based probes of
the LF were 90-99\% incomplete (i.e., missing hundreds of bona-fide
$z\sim6$ galaxies) or that the present probe contains large numbers of
contaminants.  However, explicit comparisons between the brightest
candidates identified in the Willott et al.\ (2013) search over the
CANDELS COSMOS and EGS regions and our own catalogs show very good
agreement, as we describe in \S3.4.  Our $z\sim6$ catalog contains 3
of the 4 bright $z\sim6$ candidates identified by Willott et
al.\ (2013) that fall within the CANDELS fields, i.e., WHM 14, WHM 15,
and WHM 16.  In addition, we find no sources which are brighter than
the candidates common to both of our catalogs.

This suggests there must be some other explanation for the
differences, as it is unlikely to arise from the composition of the
bright samples (either from contamination or incompleteness in
previous ground-based probes).  One significant factor might be
Willott et al.\ (2013) and McLure et al.\ (2009)'s use of deep
$z$-band observations to derive total luminosities for their sources,
due to the impact of IGM absorption on the fluxes (significant in the
$z$-band at $z>5.9$) and $k$-corrections required for comparisons with
LFs derived at 1600\AA.  Both of these factors would tend to make the
total luminosities of $z\sim6$ galaxies measured in Willott et
al.\ (2013) and McLure et al.\ (2009) fainter than derived here, if
not fully corrected.  Use of the median $z_{850}-Y_{105}$ colors of
bright ($H_{160,AB}<26$) $z=5.7$-5.9 galaxies suggests a 0.13-mag
correction from the $k$-correction alone (i.e., from 1350\AA$\,$to
1600\AA).  Absorption by the IGM would also lower the total luminosity
inferred for individual sources (by $\sim$0.15 mag) if not fully
corrected.

While one can speculate on the explanation for such differences in
measurements of the total luminosity, clearly the new WFC3/IR data are
much deeper than what was previously available and should allow for
the best determinations of the total magnitudes.  As a check on our
total magnitude measurements, we have made a comparison with those
from Skelton et al.\ (2014).  For our brightest ($H_{160,AB}<26$)
candidate $z\sim6$-8 galaxies over the CANDELS UDS/COSMOS/EGS fields,
we find excellent overall agreement (with our magnitudes being just
0.04 mag brighter in the median).

Other possible explanations for differences include a slightly too
aggressive removal of possible contaminating sources in previous
studies (e.g., Figure 1 of Steinhardt et al.\ 2014 shows that
photometric redshift techniques could err on the side of
overcorrecting for contamination in $z\sim4$-5 selections if not
calibrated properly) or slight differences in the way total magnitudes
were derived (with systematic differences catalog-to-catalog as large
as $\sim$0.2 mag and more typically $\sim$0.1 mag: e.g., Figures 35-36
from Skelton et al.\ 2014).  Of course, there is no reason to
necessarily expect the total-magnitude measurements in ground-based
probes to be too faint (as the blurring effect of the PSF makes flux
measurements less sensitive to source size).

\subsection{F.3  $z\sim7$ Results}

The $UV$ LF we derive at $z\sim7$ (Figure~\ref{fig:comp410}) is
similar to previous determination of the LF at $z\sim7$ using the ERS
and HUDF09 fields (Bouwens et al.\ 2011) given the uncertainties, but
show a slightly larger volume density of bright sources.  The larger
volume density of bright galaxies is a direct result of the fact that
the CANDELS-GN and EGS fields (Table~\ref{tab:brightnumbers}) show a
larger volume density of bright sources than were found within the
$\sim$50 arcmin$^2$ search area that we previously considered (from
the ERS, HUDF/HUDF09, HUDF09-1, and HUDF09-2 search fields).  The
modest differences we observe at the bright end of the LF are not
especially surprising as we are now probing $\sim$15$\times$ more
volume at the bright end of the LF (and $5\times$ as many sightlines),
as we did in the Bouwens et al.\ (2011) study.

The present LFs are in good agreement with the bright constraints set
by the wide-area searches by Castellano et al.\ (2010) from HAWK-I
(161 arcmin$^2$: \textit{open green squares} on
Figure~\ref{fig:comp410}) and by Bouwens et al.\ (2010b) from NICMOS
(88 arcmin$^2$: \textit{open red squares} on
Figure~\ref{fig:comp410}).  However, the present LF results show a
$\sim1.7$-2$\times$ higher volume density for bright sources than was
found by Ouchi et al.\ (2009b) in their wide-area (1568 arcmin$^2$)
search for $z\sim7$ galaxies over the Subaru Deep Field and GOODS
North to $\sim$26 mag (\textit{grey open squares} on
Figure~\ref{fig:comp410}).

% 25.1, 25.3 vs. 25.19 -- COSZ-0237620370 10:00:23.76 2:20:37.0 (25.35)
% 25.4, 25.3 vs. 25.02 -- COSZ-0301815598 10:00:30.18 2:15:59.8 (25.15)

Given the seeming robustness of the present constraints on the bright
end of the LF (due to the high quality of the present data set and
large areas surveyed: see Appendix G), it would seem more likely that
the issue lies with the Ouchi et al.\ (2009b) determination of the
$z\sim7$ LF.  One particular concern is the large ($\sim$50\%)
contamination correction that Ouchi et al.\ (2009b) apply to their
original sample of 22 $z\sim7$ sources in arriving at their final LF
results.  It is possible that the correction that Ouchi et
al.\ (2009b) apply is too large.  Even though Ouchi et al.\ (2009b)
appear to have taken great care in accurately estimating the number of
low-mass stars, lower-redshift interlopers, and spurious sources that
would contaminate their probe, contamination correction are, by their
very nature, highly uncertain, and Ouchi et al.\ (2009b) explicitly
allow for the possibility that they have significantly overestimated
the contamination rate by also presenting the $z\sim7$ LF without any
contamination correction whatsoever (shown in Figure~\ref{fig:comp410}
as the grey open triangles).  While this does not resolve the slight
tension we observe with the brightest constraints from Ouchi et
al.\ (2009b), where no contamination corrections were applied, such
tensions could be resolved if there were slight differences
($\sim$0.1-0.2 mag) between our measured magnitudes for the brightest
sources and the magnitudes derived by Ouchi et al.\ (2009b: see
Appendix F.2).\footnote{In fact, Ono et al.\ (2012), themselves,
  concede that there is already some tension between the earlier LF
  results from Ouchi et al.\ (2009b) and the total magnitude
  [$JH_{140,AB}=25.17\pm0.07$] they measure for a $z=7.2$ galaxy
  (GN-108036) found in the same search and for which they have
  spectroscopic confirmation.}

The present LF also exhibits a higher volume density of bright sources
than the recent $z\sim7$ LF determinations by Schenker et al.\ (2013)
and McLure et al.\ (2013).  There are two likely contributing factors
that can account for this difference.  One contributing factor is that
the fact that the two wide-area fields used by these studies
(CANDELS-GS and CANDELS-UDS fields) appear to be systematically
underdense (by $\sim$1.5-2$\times$) in $z\sim7$ galaxies relative to
two other search fields also included here, i.e., the CANDELS-EGS and
CANDELS-GN fields (Figure~\ref{fig:f2f}).  A second contributing
factor is the HUDF12 team treating $z\sim7$ galaxy candidates as point
sources in measuring their fluxes.  The comparisons we present in
Appendix H also suggest that the luminosities that Schenker et
al.\ (2013) and McLure et al.\ (2013) derive for the brightest sources
are $\sim$0.25 mag too faint in the median (see
Figure~\ref{fig:compm13}).  McLure et al.\ (2013)'s treating $z\sim7$
galaxies as point sources in deriving selection volumes for their
$z\sim7$ LF could also contribute to differences between our two
studies (perhaps 10\% at the bright end).  Together these issues could
result in the HUDF12 team deriving a $UV$ LF that shows a
significantly fewer bright $z\sim6$-7 sources.

Finally, the present LF results are in excellent agreement with the
new LF determination at $z\sim7$ from Bowler et al.\ (2014), except
for their faintest LF bin (see Figure~\ref{fig:comp410}).  Bowler et
al.\ (2014) derived their LF based on 34 $z\sim7$ candidates they
identify over a 1.65 deg$^2$ search area within the UltraVISTA and
UKIDSS UDS search fields.  It is unclear why the faintest LF bin from
Bowler et al.\ (2014) is $\sim$0.8 dex lower than our own constraint
at this luminosity.  The three brightest candidates we find over the
CANDELS regions of the COSMOS and UDS fields are exactly the same
as Bowler et al.\ (2014) find, so differences in the derived LFs seem
unlikely to arise from our finding especially bright sources that
Bowler et al.\ (2014) miss.  We do however find three fainter sources
in the same magnitude interval that Bowler et al.\ (2014) do not find,
suggesting that Bowler et al.\ (2014) may suffer from more
incompleteness at the faint end than they estimate or the total
magnitudes we measure for sources may be slightly brighter ($\sim$0.1
mag) than what they recover.  For the three sources our probes have in
common, i.e., Bowler et al.\ (2014) $z=7$ candidates 211127 and 185070
and Himiko (Ouchi et al.\ 2009a), the total magnitudes we measure in
the $J_{125}$ band are 0.1$\pm$0.3, 0.3$\pm$0.2 mags, and 0.1$\pm$0.1
mag brighter, respectively.

\begin{figure}
\epsscale{0.55}
\plotone{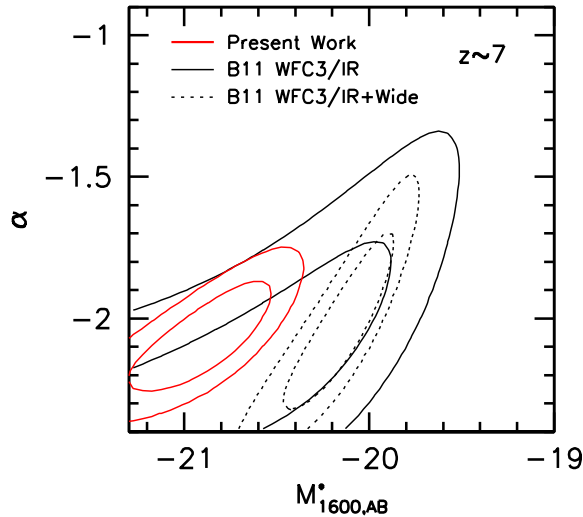}
\caption{Comparison of the current 68\% and 95\% confidence intervals
  on $M^*$ and $\alpha$ for the $UV$ LF at $z\sim7$ (\textit{solid red
    lines}) with what Bouwens et al.\ (2011) previously derived at
  $z\sim7$ based on their WFC3/IR search results alone (\textit{solid
    black line}) and combining their WFC3/IR search results with
  wide-area search results (\textit{dotted black line}: Ouchi et
  al.\ 2009b; Castellano et al.\ 2011; Bouwens et al.\ 2010c; see
  Figure 8 from Bouwens et al.\ 2011).  While our current constraints
  $M^*$ and $\alpha$ at $z\sim7$ differ from what we found in Bouwens
  et al.\ (2011), this is due to discrepancies between our new LF
  results using CANDELS and previous wide-area results (predominantly
  from Ouchi et al.\ 2009b).  The depth, area, and wavelength coverage
  of the CANDELS data set should make our new LF constraint robust
  (see Appendix F.3 and Appendix G).  In terms of our LF results using
  HST observations alone, our current constraints on $M^*$ and
  $\alpha$ agree quite well with what Bouwens et al.\ (2011) derived
  previously (\textit{solid black line}).\label{fig:agree7}}
\end{figure}

The best-fit value for the characteristic luminosity $M^*$ at $z\sim7$
($-20.87\pm0.26$) is brighter than what we presented in Bouwens et
al.\ (2011: $M_{UV,AB}=-20.14\pm0.26$).  The lower value for $M^*$
presented by Bouwens et al.\ (2011: and similarly for Grazian et
al.\ 2012) was largely driven by their use of the upper limits on
the volume densities of bright $z\sim7$ sources from Ouchi et
al.\ (2009b) based on their wide-area (1568 arcmin$^2$) search for
$z\sim7$ sources over the Subaru Deep Field and GOODS North.
Excluding the wide-area constraints from Ouchi et al.\ (2009b) and the
other wide-area searches (Castellano et al.\ 2010; Bouwens et
al.\ 2010b), Bouwens et al.\ (2011) would have found a characteristic
luminosity $M^*$ of $-20.6\pm0.4$ (see Figure~\ref{fig:agree7} and
also Table~\ref{tab:southvsnorth}).

\subsection{F.4  $z\sim8$ Results}

The present $z\sim8$ results are in broad agreement with previous
determinations of the LF at $z\sim8$ (Oesch et al.\ 2012; Yan et
al.\ 2012; Bradley et al.\ 2012; Schenker et al.\ 2013; McLure et
al.\ 2013; Schmidt et al.\ 2014), particularly at the faint end.
However, we do find a slight excess at the bright end relative to
other recent determinations, due to our discovery of three bright
$H_{160,AB}$ $z\sim8$ candidates over the CANDELS-EGS field.  These
bright candidates seem very likely to be at $z>7$ on the basis of
their robustly red $3.6\mu$m-$4.5\mu$m colors.

In terms of the characteristic luminosity $M^*$ we derive, we find a
much brighter value ($M^* = -20.63 \pm 0.36$) than essentially all
previous determinations: $M^* = -20.10\pm0.52$ (Bouwens et al.\ 2011),
$M^*=-19.80_{-0.57}^{+0.46}$ (Oesch et al.\ 2012b),
$M^*=-20.26_{-0.34}^{+0.26}$ (Bradley et al.\ 2012),
$M^*=-20.12_{-0.48}^{+0.37}$ (McLure et al.\ 2013),
$M^*=-20.44_{-0.33}^{+0.47}$ (Schenker et al.\ 2013),
$M^*=-20.15_{-0.38}^{+0.29}$ (Schmidt et al.\ 2014), $M^*=-19.5$
(Lorenzoni et al.\ 2011).  However, we note that the value of $M^*$ we
derive is consistent with what we would expect extrapolating from
lower redshift, i.e., $-20.94$ (\S5.1).  The bright value of $M^*$ we
derive is a direct consequence of our discovery of three bright $z>7$
candidates identified over the CANDELS EGS field.

\subsection{F.5  $z\sim10$ Results}

While there are still considerable uncertainties in the determinations
of the LF at $z\sim10$, the present results are in excellent agreement
with our earlier results as presented in Oesch et al.\ (2014), both in
term of the binned points and the best-fit Schechter parameters (see
Figure~\ref{fig:comp410}).  This is particularly clear if we adopt the
same shape for the LF as Oesch et al.\ (2014) use, i.e., $M^*=-20.12$
and $\alpha=-2.02$ and if we restrict ourselves to the same samples
and search fields.  The best-fit value for $\phi^*$ that Oesch et
al.\ (2014) find for these parameters is
$5.4_{-2.1}^{+3.3}\times10^{-5}$ Mpc$^{-3}$, while we find a best-fit
value of $6.2_{-2.4}^{+3.7}\times10^{-5}$ Mpc$^{-3}$.

\subsection{F.6  How can we reconcile current findings with previous claims for a dominant evolution of the $UV$ LF in $M^*$ at $z>4$?}

Over the past few years, a wide variety of conclusions have been drawn
regarding the evolution of the $UV$ LF at high redshift.  Some
analyses have argued that the primary evolution in the $UV$ LF is in
$\phi^*$ (e.g., Bouwens et al.\ 2004a; Beckwith et al.\ 2006; Capak
2008), while other analyses have argued that the observations provide
better support for a primary evolution in the characteristic
luminosity $L^*$ (Bouwens et al.\ 2006; Bouwens et al.\ 2007; Bouwens
et al.\ 2008; McLure et al.\ 2009; Lorenzoni et al.\ 2011) or the
faint-end slope $\alpha$ (Yan \& Windhorst 2004).

One particularly influential analysis has been that of Bouwens et
al.\ (2006) and Bouwens et al.\ (2007).  In those analyses, it was
found that the $UV$ LF showed much stronger evolution at the bright
end than it did at the faint end over the redshift interval $z\sim6$
to $z\sim3$.  The strong evolution Bouwens et al.\ (2006, 2007) found
at the bright end was very similar to the $6\times$ evolution found
earlier by Stanway et al.\ (2003) and Stanway et al.\ (2004), while
the weaker evolution Bouwens et al.\ (2006, 2007) found at the faint
end was in good agreement with the results of Giavalisco et
al.\ (2004) and Bouwens et al.\ (2003b).

The luminosity-dependent evolution that Bouwens et al.\ (2006, 2007)
observed could have been fit by an evolution in the faint-end slope
$\alpha$ of the $UV$ LF or the characteristic luminosity $M^*$.  Of
these two possibilities, Bouwens et al.\ (2006, 2007) found a better
fit to the observed surface density of sources adopting an evolution
in the characteristic luminosity.  Subsequent analyses of both similar
and even wider-area data sets (Su et al.\ 2011; McLure et al.\ 2009)
recovered approximately the same set of Schechter parameters as what
Bouwens et al.\ (2007) found.

The present LF determinations provide further evidence for such
luminosity-dependent evolution.  However, the large number of bright
$z\sim6$-7 galaxies and particularly $z\sim10$ galaxies found in the
new wide-area WFC3/IR observations (Oesch et al.\ 2014) have made it
clear that the general luminosity-dependent evolution can be better
fit through an evolution in the faint-end slope $\alpha$ and volume
density $\phi^*$, not exclusively with the characteristic magnitude
$M^*$ as was originally found by Bouwens et al.\ (2006).\footnote{Even
  though Beckwith et al.\ (2006) appear to have been generally correct
  in their use of $\phi^*$ to capture one aspect of the evolution of
  the LF, Beckwith et al.\ (2006) did not correctly capture the other
  aspect of the evolution of the $UV$ LF, which is the very strong
  luminosity-dependent evolution (Figure~\ref{fig:shapelf}).  Beckwith
  et al. (2006) found no difference in the rate of evolution at the
  bright and faint ends of the LF.}

Determining the exact form of the evolution of the $UV$ LF at high
redshift has been rather challenging for at least two reasons.  First
of all, the Schechter parameters become highly degenerate in cases of
a steep faint-end slope $\alpha$, i.e., $\alpha\lesssim-1.8$, due to
the limited contrast between the faint-end slope of the LF and the
effective slope of the LF at the bright end.  This makes it more
difficult to accurately measure the position of the knee of the LF
(making the Schechter parameters highly degenerate).  Second, accurate
measurements of the position of the knee of the LF are further
complicated by (1) field-to-field variations, (2) the large volumes
one needs to probe to accurately determine the bright end of the LF,
and (3) systematic errors.  Systematic errors can affect
determinations of the bright end of the LF differently than the faint
end, due to the different data sets involved.  Such errors can also
have a different impact on determinations of the LF at $z\sim4$-5 than
at $z\sim6$-8.

Of all of the above factors, perhaps the most challenging issue has
been the substantial field-to-field variations in the surface
densities of luminous sources.  As Table~\ref{tab:brightnumbers} from
Appendix G illustrates, the surface density of bright $z\sim6$-8
galaxies appears to vary substantially depending upon where one
happens to search.  If one searches for bright $z\sim6$-8 galaxies in
fields which are underdense (as appears to have been the case over the
CANDELS-GS), one would have inferred a faster evolution at the
bright end of the $UV$ LF (and hence $M^*$) than appears in fact to be
present (using all five CANDELS fields).  A comparison of the best-fit
Schechter parameters based on the CANDELS-GN+GS
(Table~\ref{tab:lfparm}: \textit{upper rows}) with the parameters
derived from all of our search fields (Table~\ref{tab:lfparm}:
\textit{lower rows}) suggests that this may have occurred.

Finally, if the $UV$ LF at $z>4$ in fact has a non-Schechter shape,
this could also have contributed to the past confusion regarding the
overall evolution of the $UV$ LF.  Thus far, however, we find no
evidence for such a non-Schechter form (\S4.4).

The recent discovery of four bright (apparently robust) $z\sim10$
galaxies over the CANDELS-GN and GS by Oesch et al.\ (2014) leaves
very little doubt as to how the $UV$ LF at high redshift evolves.
These bright $z\sim10$ galaxies simply cannot exist if the
characteristic luminosity $M^*$ is the dominant variable explaining
the evolution of the $UV$ LF from $z\sim10$ to $z\sim4$.

\begin{figure}
\epsscale{1.0}
\plotone{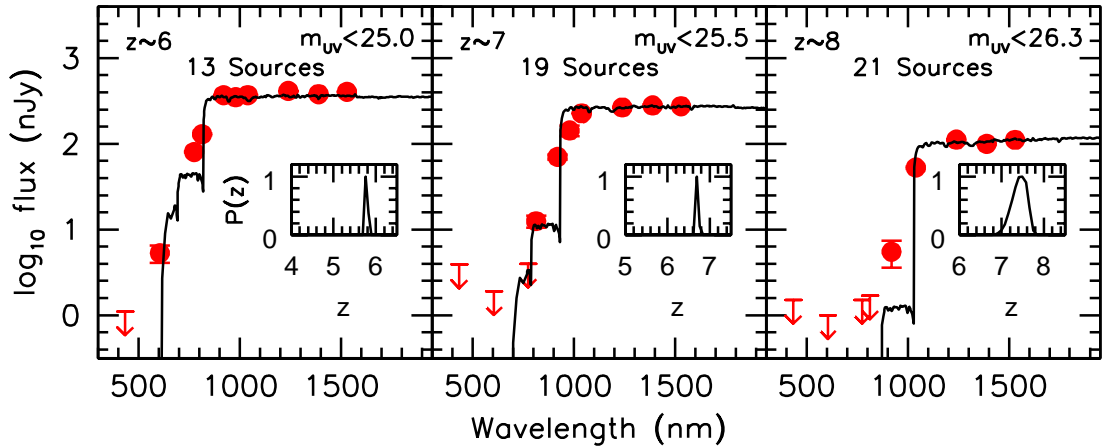}
\caption{Mean fluxes of the brightest $z\sim6$, $z\sim7$, and $z\sim8$
  galaxies identified over the five CANDELS fields (see Appendix G).
  The inset shows the redshift likelihood distribution we derive,
  using the photometric redshift code EAZY to estimate the probable
  redshift for the average source in our bright sample.  No
  significant flux is present in the stacked SED results blueward of
  the Lyman break, suggesting that the brightest $z\sim6$-8 candidate
  galaxies found over our search fields are almost all bona-fide
  $z\sim6$-8 galaxies.\label{fig:stack}}
\end{figure}

\begin{deluxetable*}{lrrrr}
\tablewidth{0pt}
\tablecolumns{11}
\tabletypesize{\footnotesize}
\tablecaption{Total number of especially bright sources\tablenotemark{a,b} in our $z\sim6$, $z\sim7$, $z\sim8$, and $z\sim10$ samples used in deriving the present high-redshift LFs.\label{tab:brightnumbers}}
\tablehead{
\colhead{} & \colhead{$z\sim6$} & \colhead{$z\sim7$} & \colhead{$z\sim8$} & \colhead{$z\sim10$}\\
\colhead{Field} & \colhead{\#} & \colhead{\#} & \colhead{\#} & \colhead{\#}}
\startdata
GOODS-S & 3 & 4 & 2 & 0\\
GOODS-N & 1 & 2 & 5 & 1\\
UDS & 0 & 2 & 3 & 0\\
COSMOS & 3 & 4 & 4 & 0\\
EGS & 4 & 7 & 5 & 0\\
Total & 13\tablenotemark{c} & 19 & 21\tablenotemark{d} & 1
\enddata
\tablenotetext{a}{See Appendix G}
\tablenotetext{b}{Included are candidate $z\sim6$ galaxies with
  $Y_{105,AB}<25.0$, $z\sim7$ galaxies with $J_{125,AB}<25.5$, 
  $z\sim8$ galaxies with $H_{160,AB}<26.3$, and $z\sim10$ galaxies with $H_{160,AB}<26.5$.}
\tablenotetext{c}{The other 2 bright $z\sim6$ candidates are found in the XDF and HUDF09-1 data sets.}
\tablenotetext{d}{The other 2 bright $z\sim8$ candidates are found in the XDF and HUDF09-2 data sets.}
\end{deluxetable*}

\section{G.  Robustness of our Constraints on the Bright End of the 
$z\sim6$-8 LFs}

Particularly central to many conclusions in this paper regarding the
shape of the $UV$ LF at $z\sim6$-8 concern the robustness of our
constraints on the volume density of bright $z\sim6$-8 galaxies.  Such
is an important question, given the tension between our results and
several previous $z\sim6$-7 results (though we note better agreement
with the new Bowler et al.\ 2015 $z\sim6$ results).

To ensure that our results are well-determined, it is useful for us to
look at the robustness of the redshift estimates we have on the
brightest $z\sim6$-8 sources and thus the contamination rate.  We
consider all $z\sim6$ candidates brighter than $Y_{105,AB}\sim25.0$
(13 sources), all $z\sim7$ candidates brighter than
$J_{125,AB}\sim25.5$ (19 sources), and all $z\sim8$ candidates
brighter than $H_{160,AB}\sim26.3$ (21 sources).  We combined the flux
measurements for all of the sources in these bright samples to produce
a mean SED for each sample.  The mean SED (presented in
Figure~\ref{fig:stack}) shows no evidence for flux blueward of the break
($<1\sigma$).  Moreover, using the photometric redshift code EAZY to
derive a redshift for the mean SED, we recovered $z=5.8$, $z=6.7$, and
$z=7.4$ for the redshifts.

As a second check on the robustness of the redshifts for bright
sources in our $z\sim6$-8 samples, we used the photometric redshift
code EAZY (Brammer et al.\ 2008) to compute their redshift likelihood
distributions.  Computing this distribution for all 57 individual
sources in our bright samples and averaging the results, the average
source showed just a 1.0\% probability of corresponding to a $z<4$
galaxy.  For the individual sources themselves, we found that all 53
bright candidates preferred a $z>4$ solution over a $z<4$ solution.

Second, we investigated how the measured volume density of the
brightest $z\sim6$-8 galaxy candidates varied from field to field.
Since all five CANDELS fields have approximately the same selection
volume for the brightest sources -- given their similar areas and
similar selectability of the brightest $z\sim6$-8 candidates -- the
number of bright candidates per CANDELS field should provide us with
an accurate estimate for the field-to-field variance in the volume
density of bright $z\sim6$-8 galaxies.

The total number of bright $z\sim6$, $z\sim7$, $z\sim8$, and $z\sim10$
candidates in each of our search fields is given in
Table~\ref{tab:brightnumbers}.  Interestingly enough, the number of
bright candidates per field appears to show an approximate Poissonian
distribution relative to the mean, with the most extreme upward
deviation from the mean being the number of bright $z\sim7$ galaxies
in the CANDELS-EGS field.\footnote{Of course, we should emphasize that
  we would expect to find at least 7 bright $z\sim7$ galaxies in at
  least one of the 5 CANDELS fields 38\% of the time -- even assuming
  simple Poissonian statistics and the mean number of bright galaxies
  found across all 5 CANDELS fields.}

Bootstrap resampling the number of bright candidates in each of the
CANDELS fields, we find that the number of bright $z\sim6$, $z\sim7$,
and $z\sim8$ candidates has a mean and $1\sigma$ uncertainty of
2.2$\pm$0.6 (0.12 dex), 3.8$\pm$1.0 (0.12 dex), and 3.8$\pm$0.6 (0.07
dex), respectively.  Since the $1\sigma$ uncertainty here includes
both the large-scale structure and Poissonian uncertainties, it
provides our best estimate on the uncertainties in the volume density
of the brightest $z\sim6$-8 candidates.

In summary, all of our tests indicate that the volume density of
bright $z\sim6$-8 galaxies we derive is extremely robust.

\begin{figure*}
\epsscale{0.9}
\plottwo{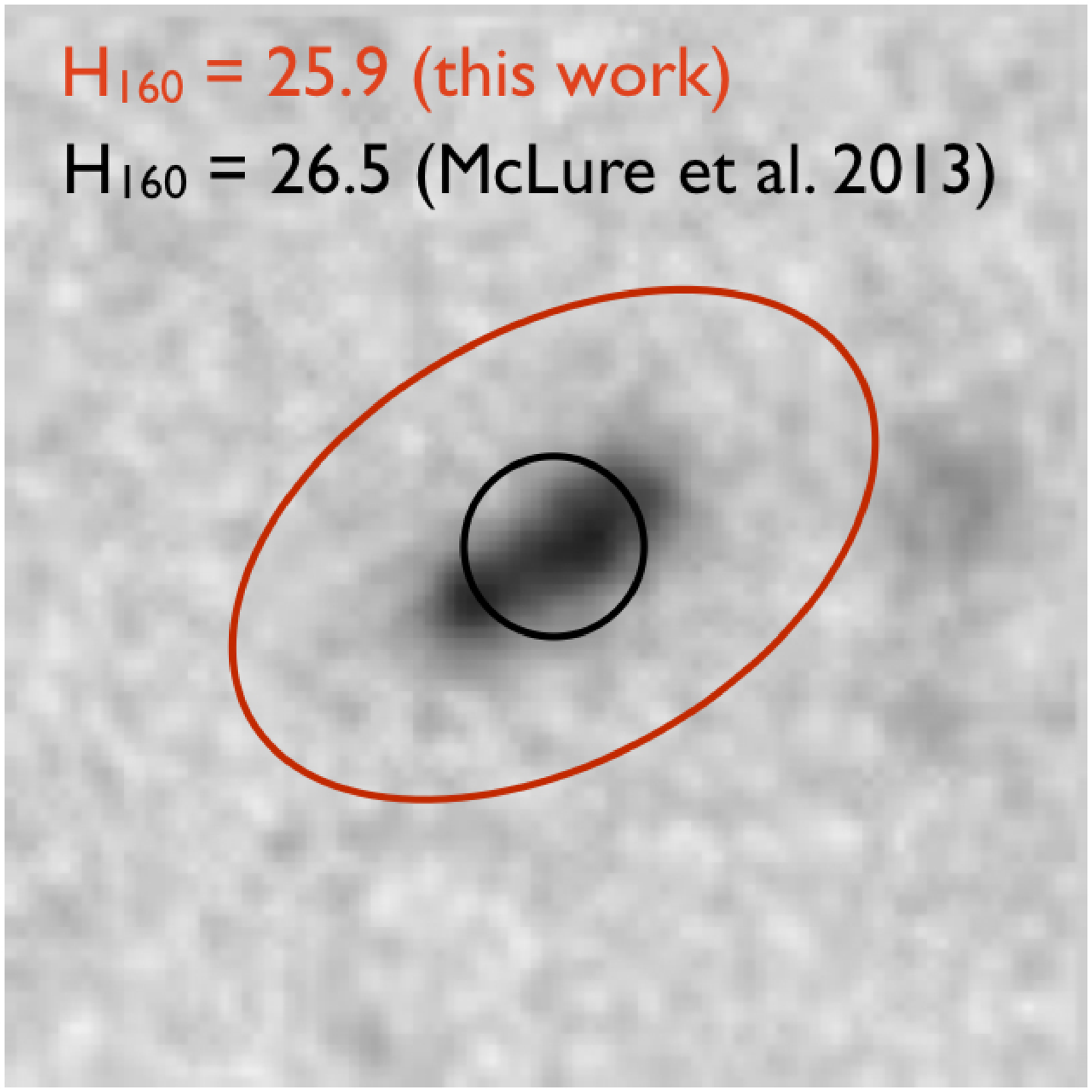}{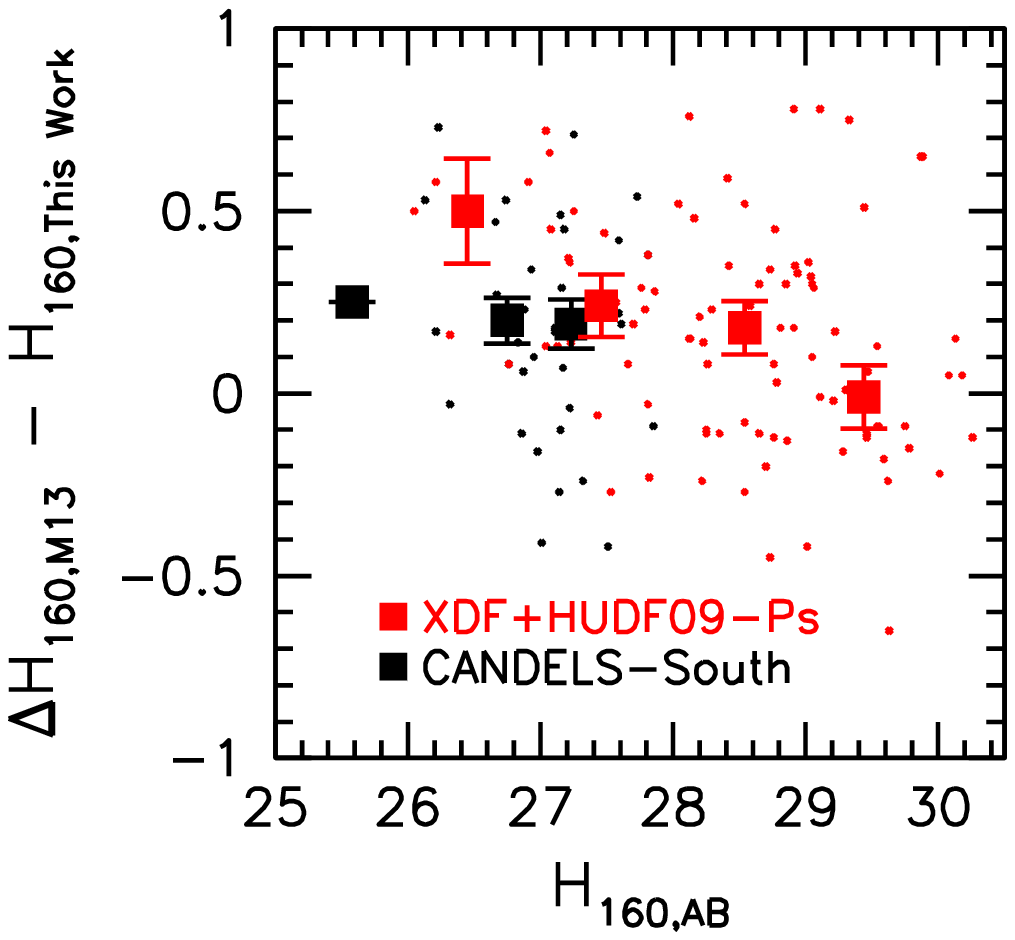}
\caption{(\textit{left}) Illustration of how the scalable Kron
  apertures used here to measure total magnitudes for galaxies
  (\textit{red ellipse}) compare with the fixed $0.50''$-diameter
  apertures McLure et al.\ (2013) use (\textit{black circle}).  See
  Appendix H.  The apertures are shown with respect to the
  $Y_{105}$-band image of one relatively large $z\sim7$ galaxy from
  the HUDF/XDF UDFz-42566566 (in a $3''\times3''$ box).  For sources
  like the one shown, the McLure et al.\ (2013) methodology will
  result in large biases in the measured magnitudes.  The total
  magnitude we measure for this source, i.e., 25.9 mag, is 0.6 mag
  brighter than what McLure et al.\ (2013) derive for the same source.
  (\textit{right}) Differences between the total magnitude
  measurements from McLure et al.\ (2013) in the $H_{160}$ band and
  those derived here for candidate sources at $z\sim7$-8.  The small
  red points show the observed differences for individual sources from
  the XDF, HUDF09-1, and HUDF09-2 fields, while the small black points
  show the observed differences for sources in the CANDELS-GS and ERS
  fields.  Magnitude differences are plotted as a function of the mean
  total magnitude measured in our two studies.  The large squares show
  the median differences for sources in 1-mag bins centered on
  $H_{160,AB}$ of 25.5, 26.5, 27.5, 28.5, and 29.5.  As illustrated in
  the left panel, we would expect a systematic bias in the total
  magnitude measurement by McLure et al.\ (2013) as a result of their
  treatment of $z\sim7$-8 galaxies as point sources, using fixed
  $0.50''$-diameter apertures to measure the magnitude of sources and
  correction to total magnitudes using the point-source encircled
  energy distribution.  This bias likely contributes to the deficit
  McLure et al.\ (2013) measure at the bright end of the $z\sim7$ LF
  relative to our own determination (see Figure~\ref{fig:comp410}).\label{fig:compm13}}
\end{figure*}

\section{H.  Comparisons against the Total Magnitude Measurements 
from McLure et al.\ (2013)}

One important difference between the methodology McLure et al.\ (2013)
use to determine the $UV$ LF at $z\sim7$-8 and the procedure used here
regard our procedures for measuring the total magnitudes of the
sources.  McLure et al.\ (2013) treat $z\sim7$-8 galaxies as point
sources, using fixed circular apertures enclosing 70\% of the expected
light for point sources and then applying a fixed 0.38-mag correction
to total.  We, however, derive total magnitudes for galaxies using the
light inside 2.5 Kron radii (ranging from $2''$ to $5''$ in radius for
$\sim$25 mag sources in CANDELS) and then applying an encircled energy
correction appropriate for point sources.

To determine whether these differences in methodology may have
resulted in any differences in measurements of the total magnitude, we
matched up sources from the McLure et al.\ (2013) and the present
catalogs and determined the difference in total $H_{160}$-band
magnitude.  We present the differences in Figure~\ref{fig:compm13} as
a function of the average of the total magnitude measurements.
Differences in the total magnitude measurements for sources from the
deepest data sets XDF, HUDF09-1, HUDF09-2 are shown in separate colors
from differences that occur for sources found in the CANDELS-GS and
ERS data set, due to the slight dependence total magnitudes can show
on the depth of a data set (when using variable apertures).

As is apparent from Figure~\ref{fig:compm13}, the total magnitude
measurements from McLure et al.\ (2013) appear to agree quite well
with our measurements for the faintest, lowest-luminosity $z\sim7$-8
galaxies.  However, for more luminous sources, the total magnitude
measurements from McLure et al.\ (2013) are offset (in the median) by
$\sim$0.25 mag faintward of our total magnitude measurements.  While
it might be surprising to see such large differences, biases would
clearly be expected in the McLure et al.\ (2013) photometry for the
largest, most extended sources (e.g., see the $z\sim7$ galaxy shown in
the left panel of Figure~\ref{fig:compm13}).  We verified that we
could reproduce the quoted magnitudes in McLure et al.\ (2013) using
similar $0.5''$-diameter aperture photometry and then aperture
correcting the results.

We expect similar systematic biases in the Schenker et al.\ (2013) LF
results due to their use of an identical photometric procedure.

\section{I.  Bouwens et al.\ (2008) Conditional Luminosity Function Model}

As an alternative to comparisons with the results from large
hydrodynamical simulations (e.g., Jaacks et al.\ 2012), we make use of
a much more simple-minded theoretical model using a conditional
luminosity function (CLF: Yang et al.\ 2003; Cooray \&
Milosavljevi{\'c} 2005) formalism where one derives the LF from the
halo mass function using some mass-to-light kernel:
\begin{equation}
\phi(L) = \int_M \phi(L|M) \frac{dN}{dM} dM
\end{equation}
For the kernel, we adopt the same functional form as Cooray \& Ouchi
(2006):
\begin{eqnarray*}
\phi(L|M) &= \frac{1}{\sqrt{2\pi}(\log_{e} 10)\sigma L} \times \\
           & ~ ~ \exp \left\{ - \frac{\log_{10} [L/L_c (M)]^2}{2\sigma^2} \right\}
\end{eqnarray*}
where $\frac{dN}{dM}$ is the Sheth-Tormen (1999) halo mass function, where
$\log_{e} 10\approx2.303$ and where $\phi(L|M)$ is the transfer
function that expresses the distribution of galaxies in luminosity at
a given halo mass.  $L_c (M)$ represents the $UV$ luminosity of the
central galaxy in some halo of mass $M$, while the parameter $\sigma$
expresses the dispersion in the relationship between the halo mass and
the $UV$ light of the central galaxy.  For convenience, we ignore the
contribution from satellite galaxies to the luminosity function in the
above equation since they appear to constitute $\lesssim10$\% of the
galaxies over a wide-range in luminosity (see, e.g., Cooray \& Ouchi
2006).

In Bouwens et al.\ (2008), we found that we could reproduce the
observed $UV$ LF at $z\sim4$ assuming that the luminosity $L_c$ of
galaxies depended on halo mass in the following way:
\begin{equation}
L_c = (2.51\times10^{22}\,\textrm{W}\, \textrm{Hz}^{-1})
  \frac{(M/m_c)^{1.24}}{(1+(M/m_c))} \left( \frac{1+z}{1+3.8} \right)
\label{eq:kernel}
\end{equation}
where $\sigma=0.16$ and $m_c = 1.2\times10^{12}\,M_{\odot}$.
$2.51\times10^{22}\, \textrm{W}\,\textrm{Hz}^{-1}$ is equivalent to
$-21.91$ AB mag.  Bouwens et al.\ (2008) included the
$(\frac{1+z}{1+3.8})$ factor in the above expression to approximately
match the apparent evolution in the mass-to-light ratio of dark matter
halos found in that study.  We make use of the same parameters in the
modeling we do here, with one exception.  We have modified the above
expression so that the $\left( \frac{1+z}{1+3.8} \right)$ factor was
taken to the 1.5 power to better fit the evolution of the $UV$ LFs
from $z\sim8$ to $z\sim4$.  The $(1+z)$ factor to the 1.5 power also
nicely matches the expected evolution in the dynamical time scales of
galaxies at early times.

\end{document}